\documentclass[showpacs,prb,twocolumn,floats,superscriptaddress]{revtex4}

\usepackage{graphicx}
\usepackage{amsmath}
\usepackage{amssymb}
\usepackage[colorlinks=true,citecolor=blue,linkcolor=blue]{hyperref}
\usepackage{amsfonts}
\usepackage[latin9]{inputenc}
\usepackage{color}

\begin{document}

\title{Quantum transport across van der Waals domain
walls in bilayer graphene}
\date{\today }
\author{Hasan M. Abdullah}
\email{g201002200@kfupm.edu.sa}
\affiliation{Department of Physics, King Fahd University of Petroleum and Minerals, 31261 Dhahran, Saudi Arabia}
\affiliation{Saudi Center for Theoretical Physics, 31261 Dhahran, Saudi Arabia}
\affiliation{Department of Physics, University of Antwerp, Groenenborgerlaan 171, B-2020 Antwerp, Belgium}
\author{B. Van Duppen}
\email{ben.vanduppen@uantwerpen.be}
\affiliation{Department of Physics, University of Antwerp, Groenenborgerlaan 171, B-2020 Antwerp, Belgium}
\author{M. Zarenia}
\affiliation{Department of Physics, University of Antwerp, Groenenborgerlaan 171, B-2020 Antwerp, Belgium}
\author{H. Bahlouli}
\affiliation{Department of Physics, King Fahd University of Petroleum and Minerals, 31261 Dhahran, Saudi Arabia}
\affiliation{Saudi Center for Theoretical Physics, 31261 Dhahran, Saudi Arabia}
\author{F. M. Peeters}
\affiliation{Department of Physics, University of Antwerp, Groenenborgerlaan 171, B-2020 Antwerp, Belgium}

\begin{abstract}
 Bilayer graphene can exhibit deformations such that the two graphene sheets  are locally detached from each other resulting in  a structure  consisting of domains with different inter-layer coupling. Here we investigate how the presence of these domains affect the transport properties of bilayer graphene. We derive analytical expressions for the transmission probability, and the corresponding conductance, across walls separating different inter-layer coupling domain. We find that the transmission can exhibit a valley-dependent layer asymmetry and that the domain walls have a considerable effect on the chiral tunnelling properties of the charge carriers. We show that transport measurements allow one to obtain the strength with which the two layers are coupled. We performed numerical calculations for systems with two domain walls and find that the availability of multiple transport channels in bilayer graphene modifies  significantly the  conductance dependence on  inter-layer potential asymmetry.
\end{abstract}

\maketitle
%%%%%%%%%%%%%%%%%%%%%%%%%%%%%%%%%%%%%%%%%%%%
\section{Introduction}
%%%%%%%%%%%%%%%%%%%%%%%%%%%%%%%%%%%%%%%%%%%%
A  decade ago,  researchers started investigating graphene and its associated multilayers  for use as a basis for  next  generation of fast and smart electronic logic gates. The absence of a band gap leads to different proposals for gap
generation\cite{14-0,14-1,14-2}.  For example, by changing the size of the graphene flakes into nanoribbons or quantum dots, one can control the energy gap through size quantization\cite{15,15-1,14}.  Important experimental advances were achieved in recent years which enabled the fabrication of graphene
based electronic devices at the nano scale\cite{16,16-1,intro-1}.

The increasing control over the structure of graphene flakes allowed for new devices that could constitute the building blocks for a fully
integrated carbon based electronics. An example of this is deformed  bilayer graphene, where the two layers are not aligned  due to a mismatch in orientation or stacking order resulting in e.g. twisted bilayer graphene. Its electronic structure is strongly different from normal bilayer graphene and exhibits very peculiar properties such as the appearance of additional   Dirac cones\cite{17,18,19,20, VanderDonck2016,VanderDonck2016b}.

Recent experiments have shown that epitaxial graphene can form step-like bilayer/single layer (SL/BL) interfaces or that it is possible to create bilayer graphene flakes that are connected to single layer graphene regions\cite{13-1,13-2,20-0}.
The appearance of these structures fueled  theoretical  and experimental investigations on  the behavior of massless and massive particles in such junctions. For example,  few  works have  investigated different domain walls   that separate, for instance, different type of stacking\cite{AB-BA-1,AB-BA-2,pelc}
or even different  number of layers\cite{20-1,22,26}. These theoretical investigations showed that the transmission probabilities through SL/BL interfaces exhibits a valley-dependent asymmetry which could be used  for  valley-based electronic applications\cite{23,24,25}. Other theoretical and experimental  works  focused on the emergence of Landau levels,  edge state properties and peculiar transport properties in such systems\cite{27-0,27-1,27-2,27,28,29,30,31,32,33,15}. Bilayer graphene flake sandwiched between two single  zigzag or armchair nanoribbons\cite{15,34} was also investigated and it was found that the conductance exhibits oscillations for energies larger than the inter-layer coupling. 

Most of these recent theoretical works considered domain walls separating patches of bilayer graphene with different stacking type or where only a single layer was connected to a bilayer graphene sheet. Very recently, however, a number of new bilayer graphene platforms have been synthesized. These consist of regions where the coupling between the two graphene layers is changed. For example in the case of folded graphene \cite{Wang2017,Rode2016} a part of the fold forms a coupled bilayer structure, while another part of it is uncoupled\cite{Schmitz2017,Hao2016,Yan2016}. One has also observed systems with domain walls separating regions of different Bernal stacking \cite{Yin2017,Yin2016}. In general, these systems can be modelled as being composed of two single
layers of graphene (2SL) which are locally bound  by van der Waals interaction into an AA- or AB-stacked bilayer structure.

 Here, we present a systematic study of electrical transport across domain walls separating  regions of different inter-layer coupling. We discuss  the dependance on the coupling between the graphene layers, on the distance between subsequent domain walls and on local electrostatic gating. For completeness, we also present all possible combinations of locally detached bilayer systems. Analytical expressions for the transport across a single domain wall are also obtained. These results can serve as a guide for future experiments. 

From a theoretical point of view, one can wonder how charge carriers will respond to transitions between systems that have completely different transport properties. For example, single layer graphene and AA-stacked bilayer graphene are known to feature Klein tunnelling at normal incidence while AB-stacked bilayer graphene shows anti-Klein tunnelling\cite{Katsnelson2006,Stander2009}. It is, therefore, interesting to investigate under which conditions these peculiar chirally-assisted tunnelling properties pertain in combined systems, as well as to investigate how the presence of multiple transport channels changes the transport properties. 

From our study we obtain useful analytical expressions for the transmission probability across a single domain wall. These results also show that the effect of local gating is to break the symmetry between the two layers and to introduce a valley-dependent angular asymmetry, which could be used for  a layer-dependent valley-filtering device. We show that the inter-layer coupling strength and stacking has a characteristic effect on the conductance across a domain wall which can be used  to measure structural deformations in bilayer graphene.  We find that the presence of multiple conductance channels in bilayer graphene can modify the  dependance of   the conductance  on an applied inter-layer potential difference from constructive to destructive. Finally, we show that transitions in-between AA-stacked and AB-stacked bilayer graphene systems largely conserve the parity of the transport channel.

The paper is organized as follows.
In Sec. \ref{Sec:Model}, we discuss the formalism, explain the geometry of the investigated domain walls, and define the possible scattering processes between the different transport modes.  In Sec. \ref{Symmetry}, we give analytical expressions for the transmission probabilities through one domain wall and analyze how the symmetry between the graphene layers can be broken by electrostatic potentials. An overview of the numerical results for more complex set-ups consisting of multiple domain walls and gates is presented in Sec. \ref{Results}.
Finally, in Sec. \ref{Concl} we  briefly summarize the main points of this paper and comment on possible experimental signatures of the presence of coupling domain walls in bilayer graphene.

%%%%%%%%%%%%%%%%%%%%%%%%%%%%%%%%%%%%%%%%%%%%
\section{Model}\label{Sec:Model}
%%%%%%%%%%%%%%%%%%%%%%%%%%%%%%%%%%%%%%%%%%%%
Single layer graphene consists of two inequivalent  sublattices, denoted as $\alpha$ and $\beta$, with interatomic distance $a=0.142$ nm and that are coupled in the tight binding (TB) formalism by $\gamma_{0}=3$  eV\cite{1}. It has a gapless energy spectrum with band crossings at the so-called Dirac points  $K$ and $K'$ that are located at the corners of the Brillouin zone. The energy dispersion around one of these points is depicted  in Fig. \ref{fig01}(a).

Bilayer graphene consists of two  single layers of graphene which can be stacked  in two stable configurations: AB-stacked bilayer  graphene (AB-BL) or AA-stacked bilayer graphene (AA-BL). In  AB-BL,  atom $\alpha_{2}$ is placed directly above  atom $\beta_{1}$ with inter-layer coupling $\gamma_1\approx0.4$ eV\cite{li2009band} as shown in Fig. \ref{fig01}(b). It has a parabolic dispersion relation with four bands. Two of them touch at zero energy, whereas the other two bands are split away by an energy $\gamma_1$. The skew hopping parameters $\gamma_3$ and $\gamma_4$  between the other two sublattices are negligible   since they have insignificant  effect on the transmission probabilities and band structure at high energies \cite{Ben}.

In AA-BL  two single layers of graphene are placed exactly on top of each other such that the structure becomes mirror-symmetric.  Atoms $\alpha_2$ and $\beta_2$ in the top layer are located directly above  atoms $\alpha_1$ and $\beta_1$ in the bottom layer,    with  direct inter-layer coupling $\gamma_1\approx0.2\ {\rm eV}$ \cite{AA-gamma1}, see Fig. \ref{fig01}(c). AA-BL has a linear energy  spectrum with two Dirac cones shifted  in energy by an amount of $\pm\gamma_1$ as depicted in Fig. \ref{fig01}(c) by the full curves.
%%%%%%%%%%%%%%%%%%%%%%%%%%%%%%%%%%%%%%%%%%%%%%%%%%%%%%%%%%%%%%%%
\subsection{Geometries}
\begin{figure}[t!]
\vspace{0.cm}
\centering\graphicspath{{./Figures1//Introd/}}
\includegraphics[width= 8.9 cm]{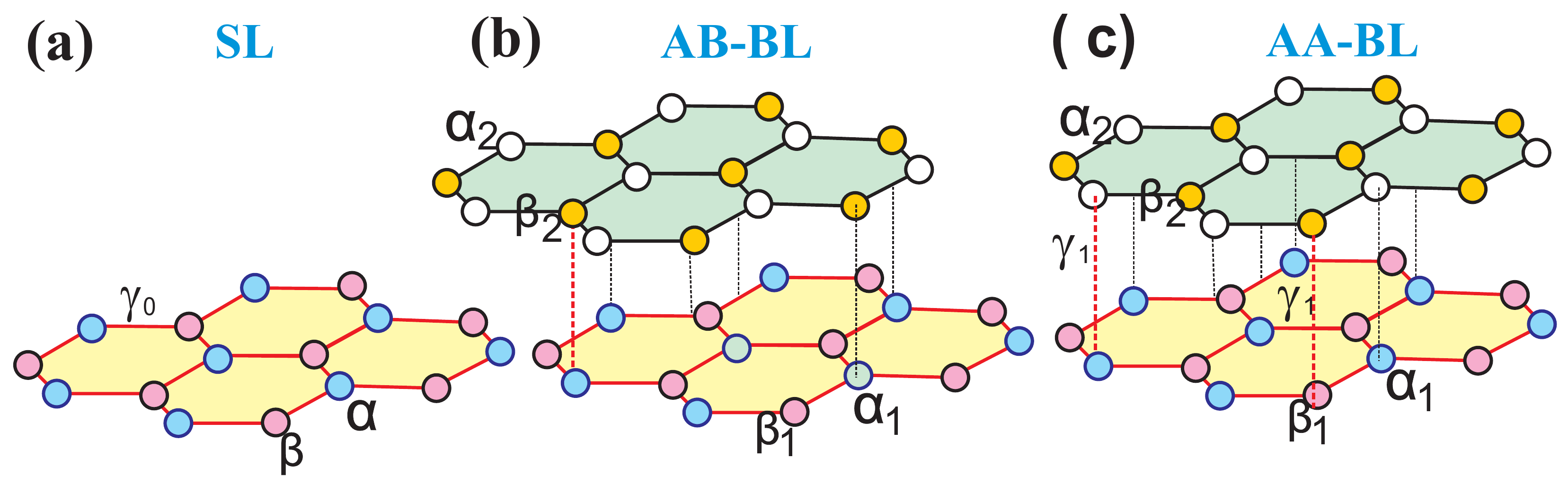}\\
\includegraphics[width= 2 cm]{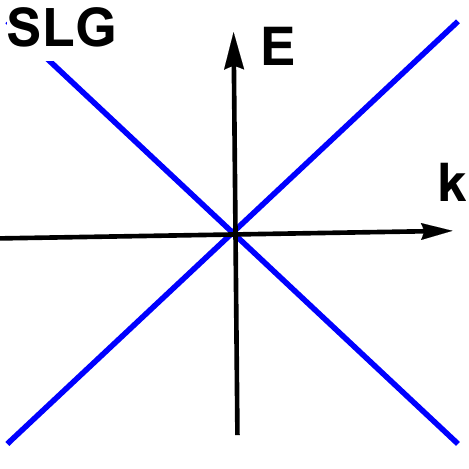}\ \ \ \ \ \ \ \ \
\includegraphics[width= 2 cm]{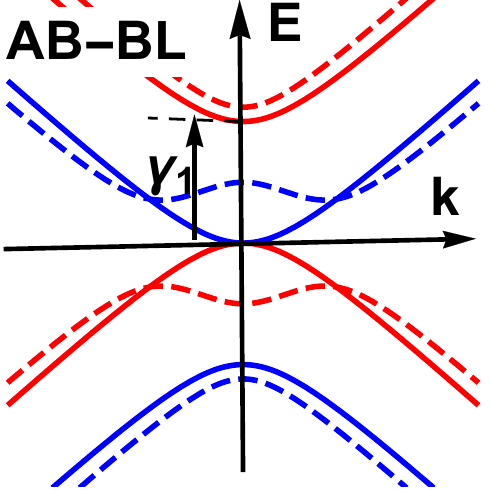}\ \ \ \ \ \ \ \ \
\includegraphics[width= 2 cm]{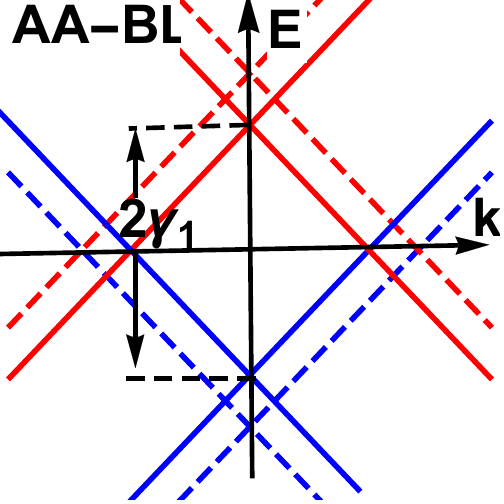}
\vspace{0.cm}
\caption{(Colour online)   Lattice structure with their corresponding energy spectrum of (a)  Monolayer graphene, (b) AB-stacked bilayer graphene, (c) AA-stacked bilayer graphene. The dashed curves correspond
to the spectrum in the presence of a finite bias. }
\label{fig01}
\end{figure}

%%%%%%%%%%%%%%%%%%%%%%%%%%%%%%%%%%%%%%%%%%%%%%%%%%%%%%%%%%%%%%%%
We consider four different junctions that can be made of from the building blocks depicted in Fig. \ref{fig01}: monolayer, AA- stacked and AB-stacked bilayer graphene. Without loss of generality, we assume that the charge carriers are always propagating from the left to the right hand side. Then we consider three different configurations: ($I$) a structure where the leads on the left ($x<0$) and on the right hand side ($x>d$) consist of two decoupled single layers while in between they are connected into an AB-BL (AA-BL) configuration. This is depicted in Fig. \ref{intro-fig02}(a). We will refer to such a structure  as  2SL-AB-2SL (2SL-AA-2SL). ($II$) A structure where the middle region is made up of two decoupled monolayers and whose leads are AB (AA) stacked bilayer graphene. This is depicted in Figs. \ref{intro-fig02}(b, d). Such a configuration henceforth will be refereed to  as AB-2SL-AB (AA-2SL-AA). ($III$) A structure where a domain wall separates an AB (AA) stacked structure from two decoupled single layers. We will assign the abbreviation 2SL-AB (2SL-AA) to this structure if the charge carriers are incident on one of the two separated layers or AB-2SL (AA-2SL) if the coupled bilayer structure is connected to the source. This is  depicted in Fig. \ref{intro-fig02}(c).
(IV) left  and right leads are bilayer graphene with AA- and AB  stacking, respectively, separated by a domain where the two layers are
completely decoupled (AA-2SL-AB), see Fig. \ref{intro-fig02}(e).
To describe transport in the above mentioned  structures, we allow for scattering between the layers as well as between the different propagating modes in an AB-BL or between the two Dirac cones in  AA-BL. In the next section, we describe the transport modes in 2SL and BL and how charge carriers can be scattered in-between them.

%%%%%%%%%%%%%%%%%%%%%%%%%%%%%%%%%%%%%%%%%%%%%%%%%%%%%%%%%%%%
\subsection{Scattering definitions}\label{Scat}
%%%%%%%%%%%%%%%%%%%%%%%%%%%%%%%%%%%%%%%%%%%%%%%%%%%%%%%%%%%%
In this section we define the model Hamiltonian that describes the different structures. For this purpose we use a suitable basis defined by $\mathbf{\Psi}=(\Psi_{\alpha 1},\Psi_{\beta 1},\Psi_{\alpha 2},\Psi_{\beta 2})^{T}, $ whose elements refer to the sublattices in each layer. The general form of the Hamiltonian near the K-point reads
\begin{equation}\label{eq01}
H=\left(
\begin{array}{cccc}
  V_{1} & v_{F}\pi^{\dag} & \tau\gamma_{1} & 0 \\
  v_{F}\pi & V_{1} &  \zeta\gamma_{1} & \tau\gamma_{1}\\
  \tau\gamma_{1} &   \zeta\gamma_{1} & V_{2}& v_{F}\pi^{\dag} \\
  0 & \tau\gamma_{1}& v_{F}\pi &V_{2} \\
\end{array}%
\right).
\end{equation}
The coupling between the two graphene layers is controlled by the parameters $\tau $ and $\zeta $ through which we can  ``\textit{switch on}" or ``\textit{switch off}" the inter-layer hopping between specific sublattices. This allows to model different stackings  by assigning different values to these parameters. For $\tau = \zeta = 0$, the two layers are decoupled and the Hamiltonian reduces to two independent SL sheets. To achieve AA-stacking we select $\tau =1$ and $\zeta =0$ while for AB-stacking we need $\tau =0$ and $\zeta =1$. In Eq. \eqref{eq01} $v_{F}\approx10^{6}$ m/s is the Fermi velocity\cite{1}  of charge carriers in each graphene
layer, $\pi=p_{x}+ip_{y}$ denotes the momentum, $V_1$ and $V_2$ are the potentials on layers 1 and 2. In the present study, we only apply these potentials in the intermediate region.
We assume that the domain wall is oriented in the $y$-direction and of infinite length. Therefore, the system is translational invariant and the momentum $p_{y}$  is  conserved. This enables us to write the wave function as $\mathbf{\Psi}(x,y)=e^{ik_{y}y} \mathbf{\Phi}(x)$.

\subsubsection{Decoupled graphene layers}

The eigenfunctions of the 2SL Hamiltonian are those of the isolated graphene sheet, \cite{1}
\begin{equation}\label{eq02}
\mathbf{\Phi}=\left(%
\begin{array}{cccc}
  \phi_{1}\   \\
   \phi_{2}\
\end{array}%
\right),\phi_j=\left(%
\begin{array}{cccc}
  \mu^{-}_j & -\mu^{+}_j  \\
  1 & 1
\end{array}%
\right)\left(%
\begin{array}{cccc}
  e^{ik_jx}   \\
  e^{-ik_jx}
\end{array}%
\right),
\end{equation}
where $j=1,2$ is the layer index, $ k_j=\sqrt{(\epsilon+s_{j}\delta)^2-k_y^2} $ with $s_j= $sgn$\left( j-1.5 \right)$, $\mu^\pm_ j=(k_j\pm\ ik_y)/(\epsilon+s_{j}\delta)$, $\epsilon=E-v_0,\ \delta=(V_1-V_2)/2,\ v_0=(V_1+V_2)/2$. Introducing the length
scale $l=\hbar v_{F}/\gamma_{1}$, which represents the inter-layer coupling length, allows
us to define the following dimensionless quantities:\
\begin{align}
\epsilon\rightarrow\frac{\epsilon}{\gamma_1},\ v_0\rightarrow\frac{v_0}{\gamma_1},\ \delta\rightarrow\frac{\delta}{\gamma_1},\ k_y\rightarrow lk_y,\ \text{and}\ \vec r\rightarrow\frac{\vec r}{l}.
\label{dimensionless}
\end{align}
Notice that for the two stacking configurations,   $\gamma_{1}$ was found to be different. For the AB-BL the value is $\gamma_1\approx0.4$ eV while for AA-BL it is $\gamma_1\approx0.2$ eV\cite{li2009band,AA-gamma1,AA-Yuehua2010}.

\begin{figure}[tb]
\vspace{0.4cm}
\centering \graphicspath{{./Figures1//introd/}}
\includegraphics[width=3.6 in]{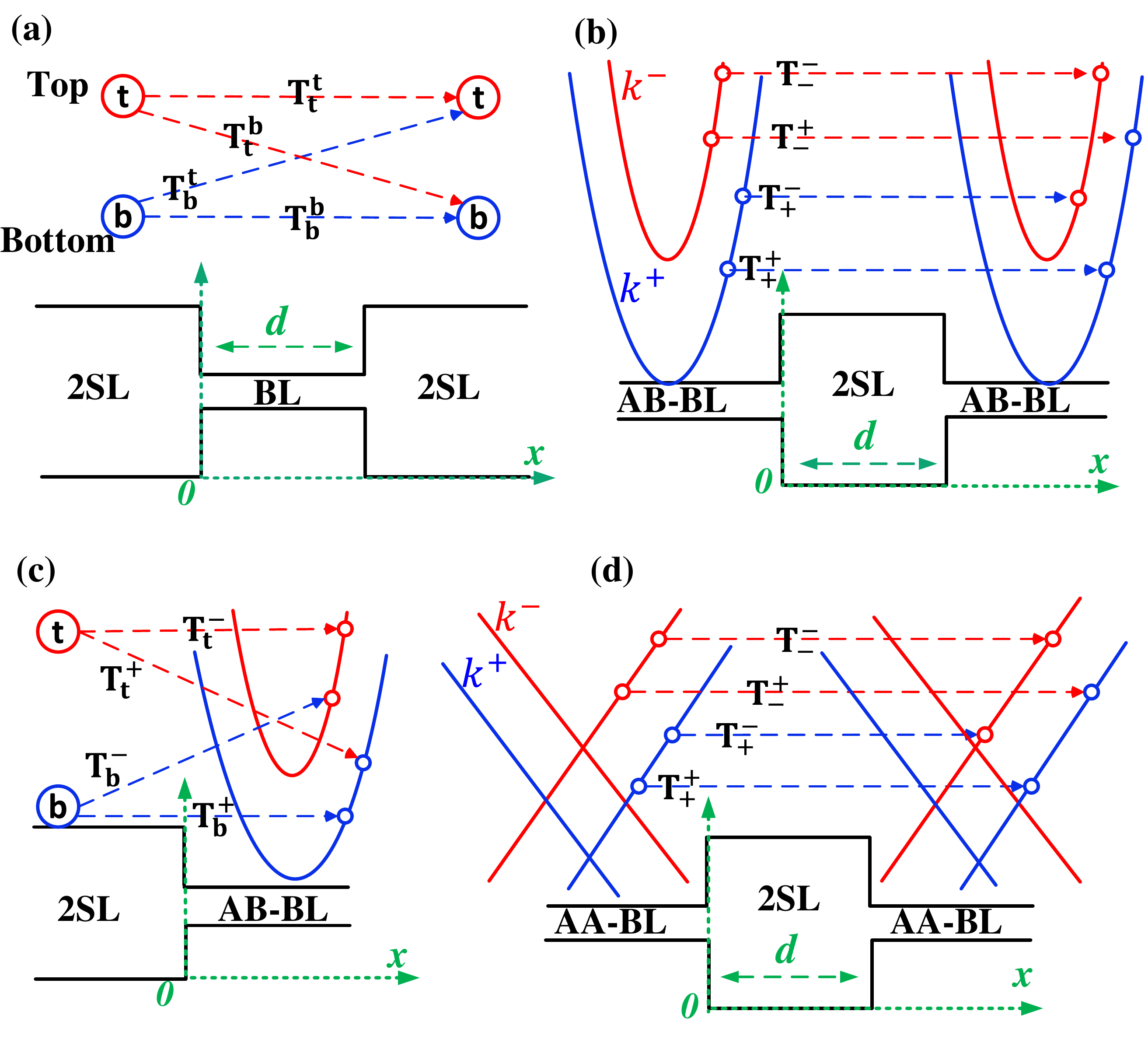}\\
\includegraphics[width=2.15 in]{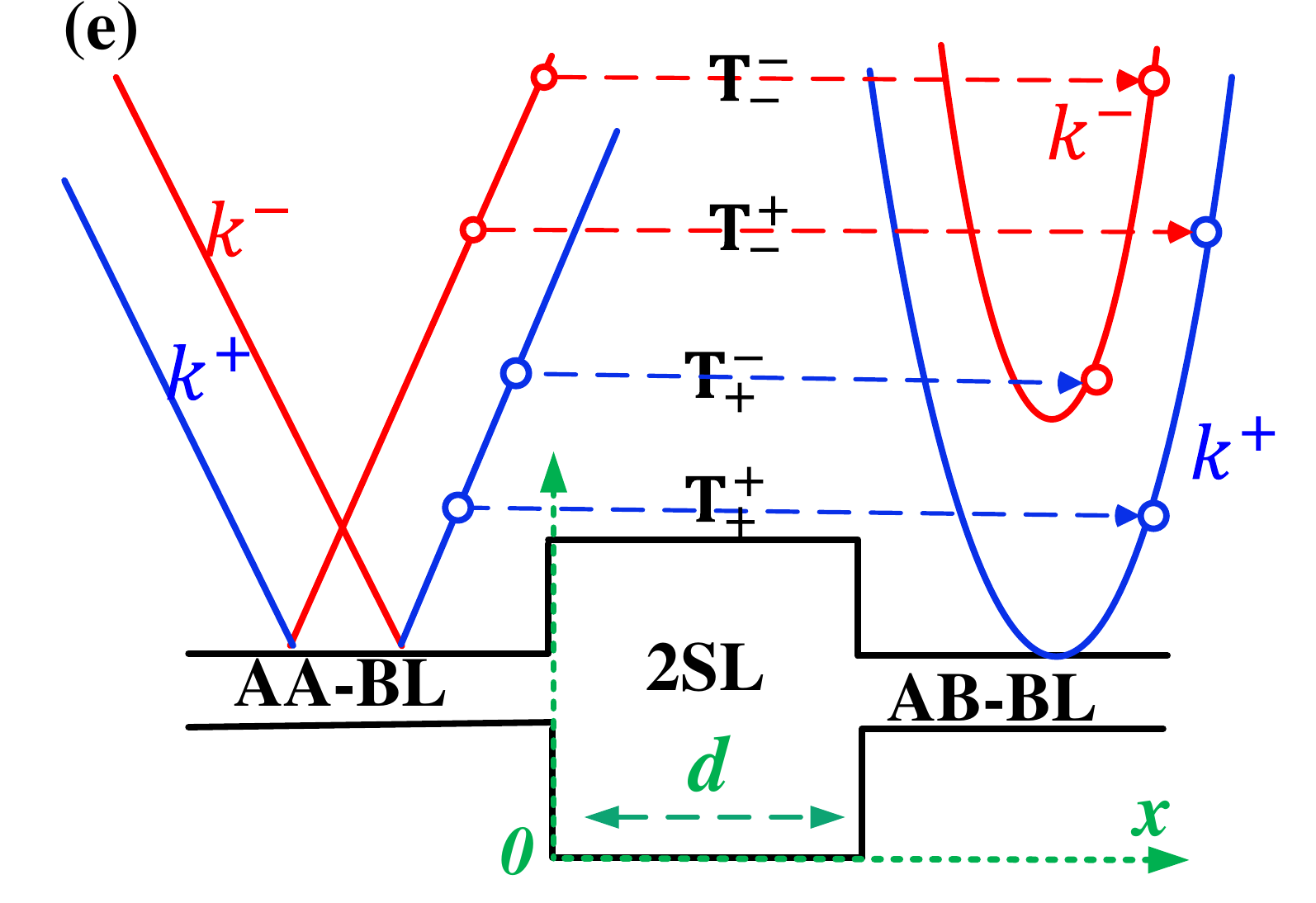}
\caption{(Colour online) Different geometries for bilayer
and two decoupled graphene layer interfaces with schematic representation of the transmission probabilities. (a) AA or AB stacking bilayer graphene  sandwiched between two SL garaphene layers (2SL-AA(AB)-2SL), (b) AB-BL leads with 2SL as intermediate region  (AB-2SL-AB), (c) two single  gaphene layers connected to AB-BL(2SL-AB), and  (d) similar to (b) but now with  AA-BL as the leads with two upper (red)-lower (blue) shifted Dirac cones (AA-2SL-AA). (e) left and right leads are bilayer graphene with different stacking connected to the two decoupled graphene sheets (AA-2SL-AB).  The possible transmission processes between the different conduction channel are indicated above the respective junctions. }\label{intro-fig02}
\end{figure}

In order to discuss the different scattering modes, we introduce the notation $A_{\rm incoming}^{\rm outgoing}$, where $A$ can stand for transmission ($T$) or reflection ($R$) probabilities and the indexes denote the mode by which the particles are incoming or outgoing. Fig. \ref{intro-fig02} depicts all possible transitions that are considered in the present work. Fig. \ref{intro-fig02}(a) shows all possible transmission processes in a 2SL-BL-2SL system where $t$ denotes the top layer on either side and $b$ the bottom layer. For example, $T^{b}_{t}$ denotes a particle coming through the top layer and exiting on the bottom layer. 

\subsubsection{AB-stacking}
For AB-BL  there are  two branches corresponding to propagating modes. These branches correspond to the wave vector $k^{\pm}$ given by
\begin{equation}\label{eq03}
k^{\pm}=\left[-k^{2}_{y}+\epsilon^{2}+\delta^{2}
\pm \sqrt{\epsilon^{2}(1+
4\delta^{2})-\delta^{2}}\right]^{1/2},
\end{equation}
The modes presented in Eq. (\ref{eq03}) labeled by ``$k^{+}$'' correspond to eigenstates that are odd under layer inversion, while the ``$k^{-}$''modes are even. These modes are shown, respectively, in blue and red in Fig. \ref{intro-fig02}(b). This means that there are two  available channels for transmission at a given energy, and an additional two for the reflection probabilities. Note that for energies $0<E<\gamma_1$,  there is only one propagating  mode and one transmission and reflection channel. Similarly, the wave function of  AB-BL can be written as \cite{Ben} 
\begin{equation}\label{eq04}
\Psi(x,y)=G M(x)C e^{ik_{y}y},
\end{equation}
where $M(x)$  corresponds to a $4 × 4$ diagonal matrix consisting of exponential terms, while the components of the constant vector  $C$ depend on the propagating region, and $G$ is given by 
\begin{equation}\label{eq05}
G=\left(%
\begin{array}{cccc}
  \xi^{+}_{-} & -\xi^{+}_{+} & \xi^{-}_{-} & -\xi^{-}_{+} \\
  1 & 1 & 1 & 1 \\
  \rho^{+} & \rho^{+} & \rho^{-} & \rho^{-} \\
  \zeta^{+}_{+} & -\zeta^{+}_{-} & \zeta^{-}_{+} & -\zeta^{-}_{-} \\
\end{array}%
\right),
\end{equation}
where $\xi^{\pm}_{\pm}=(k^{\pm}\pm
ik_{y})/E-\delta,\ \rho^{\pm}=(\epsilon-\delta)\left[1-((k^{\pm})^{2}+k^{2}_{y})/(\epsilon-\delta)^{2}\right]
$ and
$\zeta^{\pm}_{\pm}=(\epsilon-\delta)\rho^{\pm}\xi^{\pm}_{\pm}/(\epsilon+\delta)$. 

The use of the matrix notation will prove to be very useful to construct the transfer matrix as outlined below. 
%%%%%%%%%%%%%%%%%%%%%%%%%%%%%%%%%%%%%%%%%%%%%%%%%%
\subsubsection{AA-stacking}
%%%%%%%%%%%%%%%%%%%%%%%%%%%%%%%%%%%%%%%%%%%%%%%%%%
In the case of an AA-BL, the corresponding wave function can be written similar to Eq. \eqref{eq04} but now with the matrix $G$  given by
\begin{equation}\label{eq07}
G=\left(%
\begin{array}{cccc}
  \xi^{-}_{+} & \xi^{+}_{+} & \xi^{-}_{-} & \xi^{+}_{-} \\
  1 & 1 & 1 & 1 \\
  \zeta^{-}_{+} & \zeta^{+}_{+} & \zeta^{-}_{-} & \zeta^{+}_{-} \\
  \rho^{+} & \rho^{+} & \rho^{-} & \rho^{-} \\
\end{array}%
\right),
\end{equation}
where $\rho^{\pm}=\frac{1}{2\epsilon}\left[-(k_y^2+(k^\pm)^2)+(\epsilon^{}-\delta)^2+1)\right]
$, $\xi^{\pm}_{\pm}=(\rho^{\pm}+\delta+\epsilon)(ik_y\pm k^\pm)/(\delta^2-\epsilon^2+1) $ and $\zeta^{\pm}_{s}=(\xi^{\pm}_{\pm}-\rho^{\pm}(ik_y \pm k^\pm)/(\epsilon+\delta$).
To investigate when scattering between the Dirac cones of  AA-BL  is allowed or forbidden, one can apply a unitary transformation that forms symmetric and anti-symmetric combinations of the top and bottom layer. This yields a Hamiltonian in the basis $\mathbf{\Psi}=2^{-1/2}(\Psi_{\alpha 2}+\Psi_{\alpha 1},\Psi_{\beta 2}+\Psi_{\beta 1},\Psi_{\alpha 2}-\Psi_{\alpha 1},\Psi_{\beta 2}-\Psi_{\beta 1})^{T}$   of the form:
\begin{equation}\label{eq10}
H_{AA}=\left(
\begin{array}{cccc}
  \gamma_1+v_0 & v_{F}\pi^{\dag} & -\delta & 0 \\
  v_{F}\pi &  \gamma_1+v_0 &  0 & -\delta\\
  -\delta &   0 &  -\gamma_1+v_0& v_{F}\pi^{\dag} \\
  0 & -\delta& v_{F}\pi & -\gamma_1+v_0 \\
\end{array}%
\right).
\end{equation}
For $\delta=0$, this Hamiltonian is block-diagonal and represents two Dirac cones as shown in Fig. \ref{fig01}(c). The two cones correspond to modes with wave vector $k^{\pm}$ given by 
\begin{equation}
k^{\pm}=\left[-k^{2}_{y}+\left(\epsilon\pm \sqrt{(1+
\delta^{2})} \right)^2
\right]^{1/2}.
\label{k_vector_AA}
\end{equation}
In Fig. \ref{fig01}(c) the blue bands corresond to the odd $k^{+}$ modes and   red bands, denoting the even modes, are given by the $k^{-}$ wavevector. In these equations, $v_0$ denotes the energy shift of the whole spectrum. This shift can be chosen zero by assigning the same magnitude but different signs to the electrostatic potentials on both layers $V_1=-V_2$. Eq. \eqref{eq10} shows that for zero electric field ($\delta=0$) both cones are decoupled and the scattering between them is strictly forbidden. This was used before in  Ref. [\onlinecite{AA-cones}] to propose AA-BL as a potential candidate for \textit{\textquotedblleft cone-tronics\textquotedblright} based devices. However, this protected cone transport is broken for finite bias ($\delta\neq0$)  and hence  scattering between the cones is allowed. Furthermore, one might wonder if the charge carriers stay within their cone   transport through a domain consisting of two decoupled layers. 

\subsubsection{Scattering probability}

In order to calculate the scattering probability in the reflection and transmission channel, we use the transfer matrix method together with boundary conditions that require the eigenfunctions in each domain to be continious for each sublattice \cite{Barbier, Ben2}. To conserve probability current we normalize transmission probabilities $T$ and reflection probabilities $R$ such that
\begin{equation}
\sum_{i,j}\left( T_i^j+R_i^j \right)=1,
\end{equation}
where, the index $i$ refers to the incoming mode  while the index $j$ denotes the outgoing mode. For a coupled bilayer the different modes are labelled by ``$-$'' for the modes that are even under in-plane inversion and by ``$+$'' for odd modes. For a decoupled 2SL system, we employ the notation $t$ for the top layer and $b$ for the bottom layer. For example, for the system 2SL-AB-2SL and for an incident particle in the top layer of 2SL gives $T_t^t+T_t^b+R_t^t+R_t^b=1$. In Fig. \ref{intro-fig02} all possible transition probabilities are shown schematically. 

\subsubsection{Conductance}

To obtain measureable quantities, we finally calculate the zero temperature  conductance that can be obtained from the Landauer-B\"uttiker formula \cite{Landauer} where we have to   sum over all the transmission channels,
\begin{equation}\label{eq08}
{G_i^{j}}(E)=G_{0}\frac{L_y}{2
\pi}\int_{-\infty}^{+\infty}dk_{y} T_{i}^j(E,k_y),
\end{equation}
with $L_y$  the length of the sample in the $y$-direction and
$G_0=4\ e^2/h$. The factor $4$ comes from the valley and
spin degeneracy in graphene.
 The total conductance of any configuration  is the sum of all available channels $G_T=\sum_{i,j}G_i^j$. 
\begin{figure}[t!]
\vspace{0.cm}
\centering\graphicspath{{./Figures1//introd/}}
\includegraphics[width= 4.3 cm]{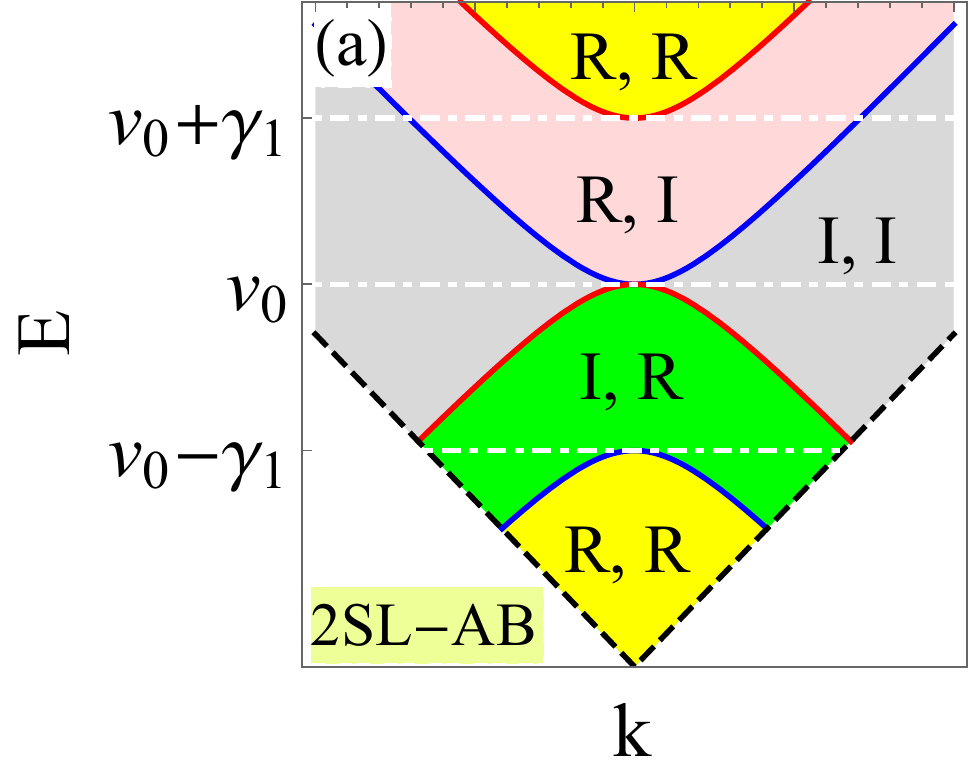}\ \ \ 
\includegraphics[width= 3.8 cm]{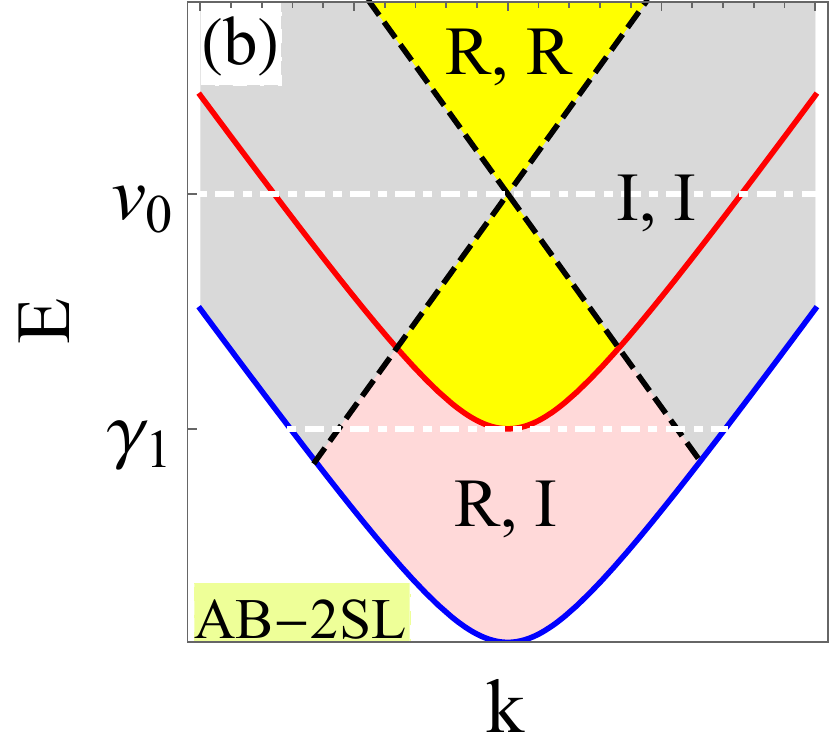}
\vspace{0.cm}
\caption{(Colour online)  Schematic diagrams, for one domain wall separating 2SL and AB-BL, showing the regions  where the modes ($k^+,\ k^-$) in AB-BL are either real (propagating) or imaginary (evanescent). (a)  shows the bands of pristine 2SL and gated AB-BL and vice versa in (b). In the yellow  region both modes are real (R, R),  while  one of them is real and the other is imaginary as in the green (I, R) and pink (R, I) regions. In the gray region  both modes  are imaginary (I, I). Blue, red and dashed black bands correspond  to $k^+$, $k^-$  and 2SL modes, respectively.}
\label{fig003}
\end{figure}
%%%%%%%%%%%%%%%%%%%%%%%%%%%%%%%%%%%%%%%%%%%%%%%%%%%%%%%%%%%%%%%%
\section{Transmission across a single domain wall}\label{Symmetry}
%%%%%%%%%%%%%%%%%%%%%%%%%%%%%%%%%%%%%%%%%%%%%%%%%%%%%%%%%%%%%%%%
\begin{figure}[t]
\vspace{0.4cm}
\centering \graphicspath{{./Figures1//SL-AA/}}
\includegraphics[width=1.25  in]{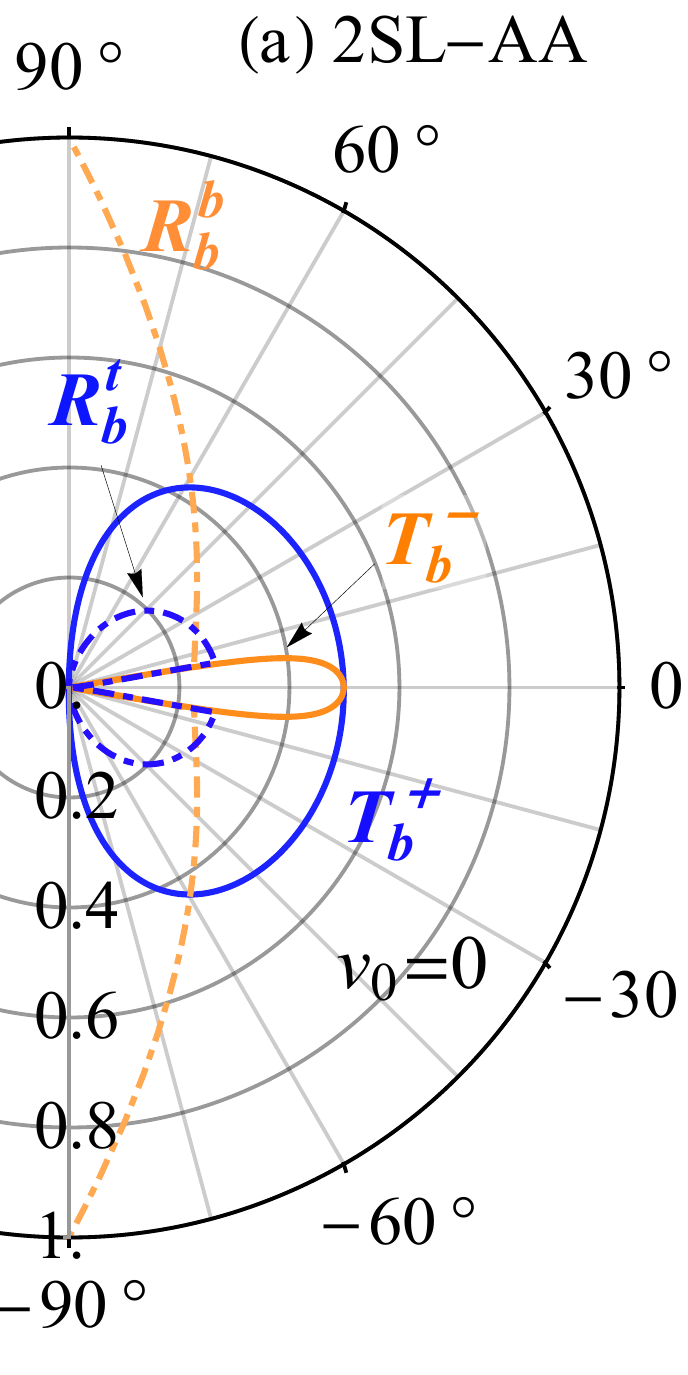} 
\includegraphics[width=1.25    in]{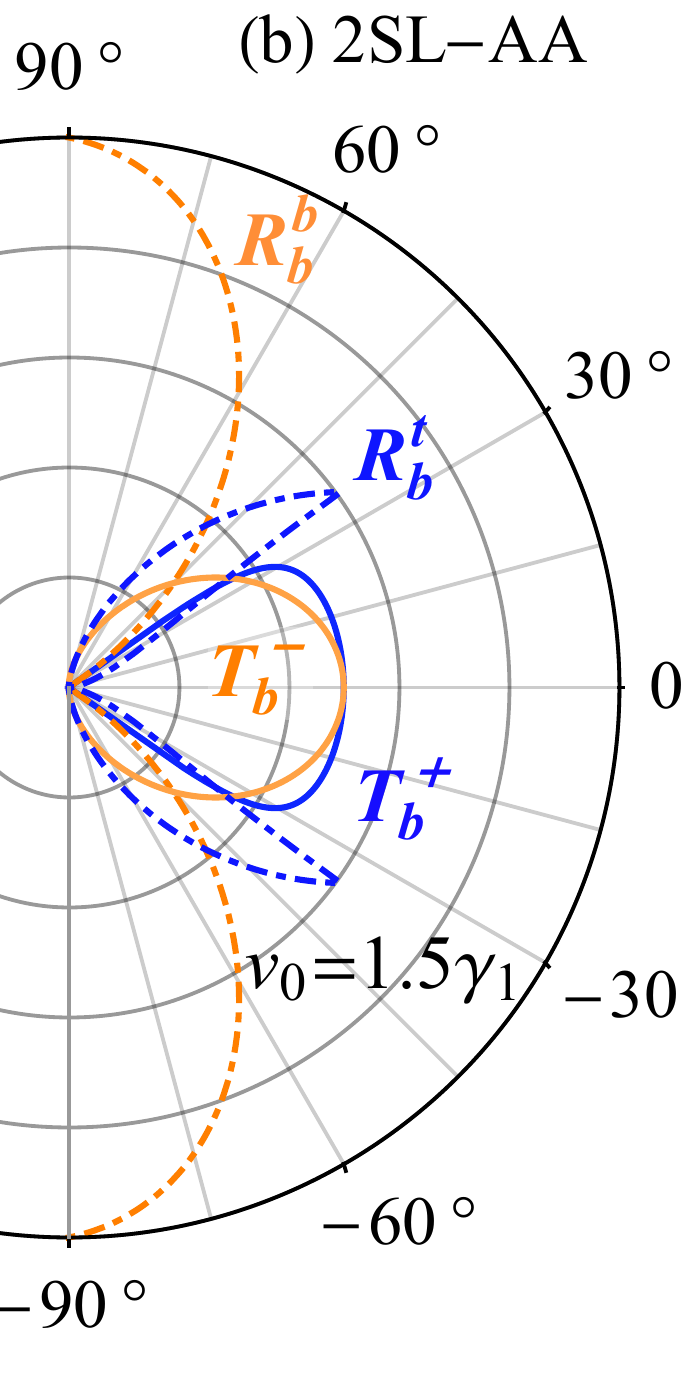}\\ 
\includegraphics[width=1.25 in]{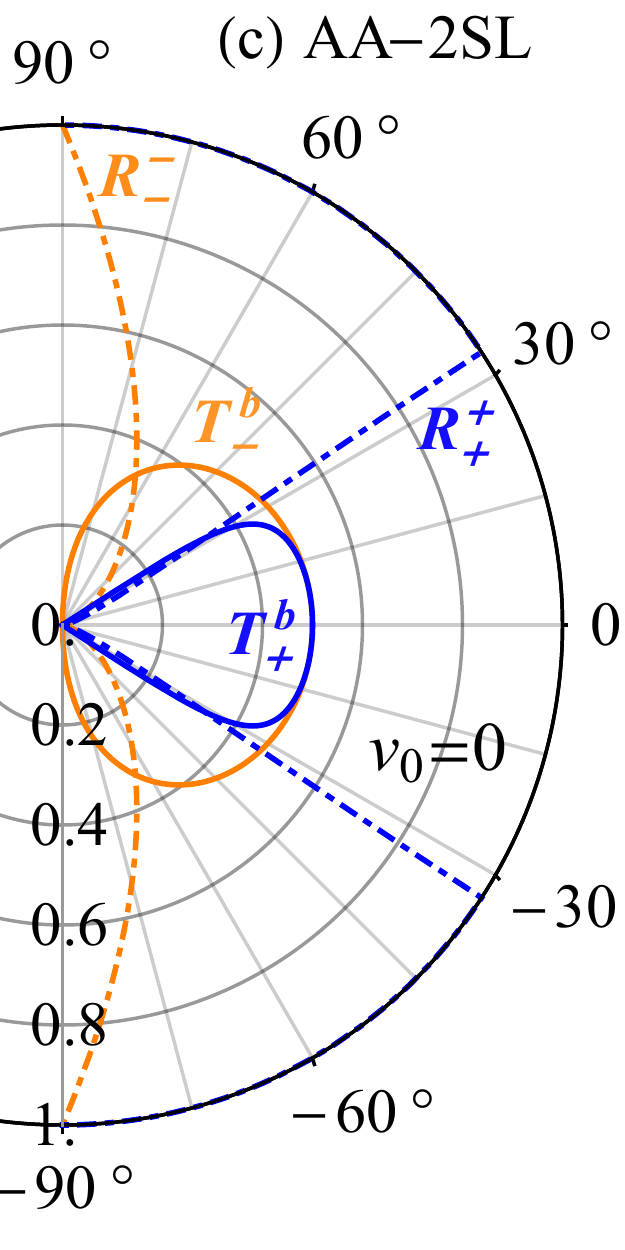}  
\includegraphics[width=1.25 in]{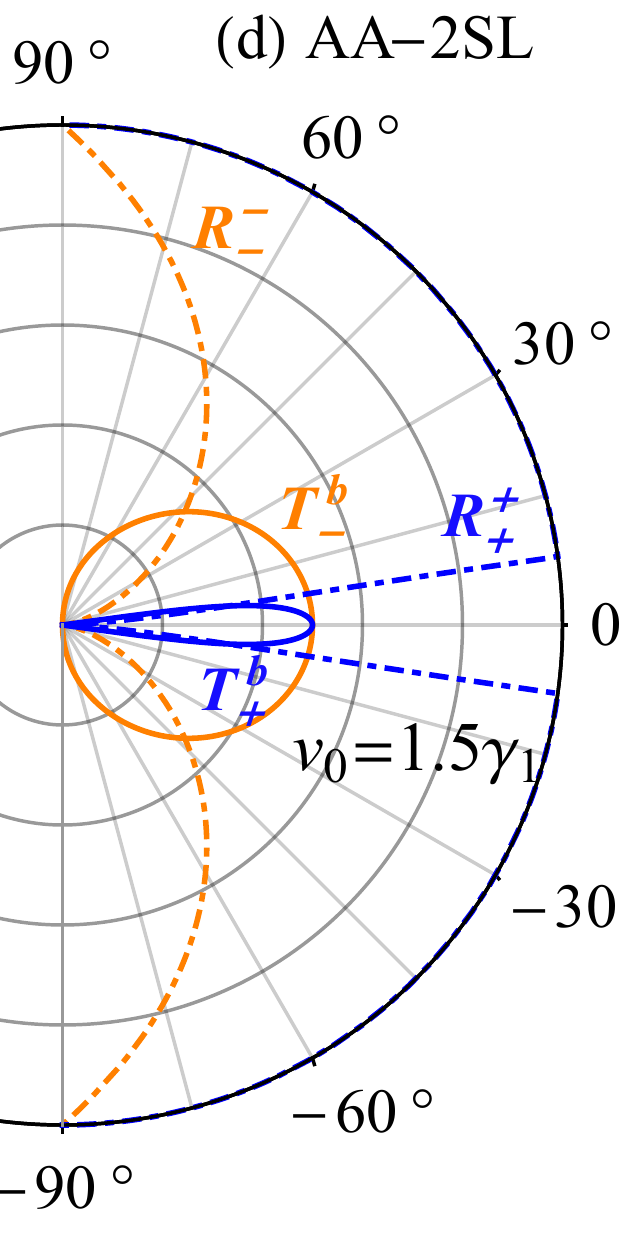}
\caption{(Colour online)   The angle-dependent transmission and reflection probabilities through (a, b) 2SL-AA and (c, d) AA-2SL systems. The systems in (b, d) are the same as in (a, c), respectively, but where now the right side of the junction is subjected to an electrostatic potential of strength $ v_0= 1.5\gamma_1.$ In the system 2SL-AA $R_b^{b(t)}=R_t^{t(b)}$ and $T_b^{\pm}=T_t^{\pm}$ while $R_+^{-}=R_-^{+}=0$ and $T_\pm^{b}=T_\pm^{t}$ in AA-2SL system. In all panels  $E=1.2\ \gamma_1$.}\label{polar-SL-AA}
\end{figure}

Here  we will present analytical expressions for the transmission probabilities of transport across a single domain wall. These analytical expressions will shed light on the requirements for transport across a domain wall and how local electrostatic gating can affect these transport properties. By doing so, we encounter that curiously, electrostatic gates can break the symmetry between the layers in the transmission probability if there are evanescent modes in the system. The breaking of the layer symmetry results in an asymmetric angular distribution of the transmission probability as will be shown further. 

We consider a situation where  two propagating modes exist in the AB-BL or AA-BL. This requires some caution in defining the incident angle  in the calculation of the transmission probabilities. Failing to do so may  result in erroneous results such as transmission exceeding unity or unexpected symmetry features\cite{kumar,Ben03}. Considering  one domain wall, the simplest configuration, separating 2SL and either AA or AB-BL allows  to obtain analytic expressions for the transmission probabilities. The incident angle for each propagating mode depends on the type of layer stacking in the incident region. Hence, for charge carriers incident from 2SL we define
\begin{equation}
k_j=E\cos\phi,\ k_y=E\sin\phi.
\end{equation}

On the other hand, when charge carriers are incident from AB-BL we need to define incident angle for each mode separately such that 
\begin{equation}
k^\pm=\sqrt{E^{2}\pm E}\cos\phi,\ k_y=\sqrt{E^{2}\pm E} \sin\phi.
\end{equation}
Finally, if charge carriers incident from AA-BL the associated angle is defined as
\begin{equation}
k^\pm=(E\pm 1)\cos\phi,\ k_y=(E\pm1) \sin\phi.
\end{equation}
 
A straightforward calculation  results in the transmission probability for charge carriers incident from 2SL and impinging on AA-BL 
\begin{equation}\label{eq15}
T_j^\pm=\frac{2(\epsilon+v_0)(\pm1+\epsilon)\textrm{Re}(k^{\pm})}{k_j\left[ \left( \pm1+\epsilon+k^\pm \sec\phi \right)^2+(\mp1+v_0)^{2}\tan^2\phi \right]}\ ,
\end{equation} 
while for the reverse configuration (AA-2SL) it is given by
\begin{equation}\label{eq16}
T_\pm^j=\frac{2\epsilon\textrm{Re}(k_{j})}{\cos\phi\left[ \left( \epsilon+k_{j} \sec\phi \right)^2+(\mp1+v_0)^{2}\tan^2\phi \right]}\ .
\end{equation}
Similar as performed for the AA-BL Hamiltonian, also the AB-BL Hamiltonian can be expressed in terms of symmetric and anti-symmetric combinations of the two layers. This manipulation allows to determine a closed-form expression for the transmission probability of the 2SL-AB structure. The derivation is outlined in Appendix \ref{Sec:Appendix} and results in
\begin{equation}\label{eq17}
T_j^\pm=4 \textrm{Re}(k^{\pm}) \frac{\eta\left[ \eta^2+\left(\textrm{Im}(k^{\mp})+\kappa_{j} v_0\sin\phi\right)^2\ \right]}{C_{0 }+\sum_{m=1}^4C_m\cos(m\phi)},
\end{equation}
with $\eta=\epsilon \cos\phi$ and $\kappa_{j}=+1(-1)$ for $j=b( t)$.  For the reverse configuration (AB-2SL) the transmission probabilities are 
\begin{equation}\label{eq18}
T_\pm^j=4 \textrm{Re}(k_j)k^{\pm} \frac{\lambda\left[  \mu^{\pm}+\kappa_{j} v_0 \sin\phi\ \textrm{Im}(k^{\mp}) \right] }{\left\vert Q^{\pm} \right\vert^2},
\end{equation}
where $ \lambda,\ C_m,\ \mu^\pm$, and$\ Q^\pm$ are functions defined in  Appendix \ref{Sec:Appendix}.

For a domain wall separating 2SL and AA-BL, the transmission probabilities are always symmetric with respect to  normal incidence as indicated in Eqs. (\ref{eq15},\ref{eq16}). In other words,  for the 2SL-AA   $T_b^\pm(\phi)=T_t^\pm(\phi) $ and similarly $T_\pm^b(\phi)=T_\pm^t(\phi) $ for AA-2SL configuration, and this symmetry still holds when the right side of the junction is gated ($v_0\neq0$). We will refer to this symmetry as \textquotedblleft\textit{layer symmetry\textquotedblright} since it is a consequence of the  equivalence of 2SL layers and the symmetric coupling of the AA-BL.
 
Notice that Klein tunnelling for normal incidence in SL and AA-BL is also conserved in the combined structure.
For example, in 2SL-AA and for normal incidence ($\phi=0$), the modes become $k_j=\epsilon+v_0,\ k^\pm=\pm1+\epsilon$ and hence Eq. \eqref{eq15} reads $T^\pm_j=1/2.$ Then, for charge carriers propagating in the bottom (top) layer it may be transmitted into $k^+$ or $k^-$ states and thus the total probability is $T_{b(t)}^++T_{b(t)}^-=1/2+1/2=1$.  As a result of Klein tunnelling at normal incidence, the corresponding reflection probabilities are zero such that $R_b^{b(t)}=R_t^{t(b)}=0$. In an analogous manner it can be shown that for normal incidence Eq. \eqref{eq16} gives $T_\pm^j=1/2.$  

Turning now to the 2SL-AB/AB-2SL case, one can infer from Eqs. (\ref{eq17},\ref{eq18}) that for $v_0=0$ the layer symmetry holds since the only term carrying asymmetric features is  proportional to $v_{0}$. However, for  $v_0\neq 0$ it is striking that despite the fact that a homogeneous electrostatic potential does not break any in-plane symmetry in the system, layer symmetry is broken. This leads to an angular asymmetry in the transmission channel, i.e. $T_b^\pm(\phi)=T_t^\pm(-\phi)$ for 2SL-AB and $T_\pm^b(\phi)=T_\pm^t(-\phi)$ for AB-2SL. Upon further analysis of Eqs. (\ref{eq17},\ref{eq18}), one notices that this asymmetric  feature is present in regions in the ($E,k_y$) plane where one of the two modes is propagating while the other  is evanescent. In Figs. \ref{fig003}(a,b) we show a diagram for these different regions associated with 2SL-AB  and AB-2SL, respectively.  The layer symmetry is broken  in the green and pink regions  while in the yellow regions layer symmetry holds.  

The mechanism for breaking the layer symmetry in configurations consisting of AB-BL is attributed  only to the evanescent modes. For example, in 2SL-AB (see Fig. \ref{fig003}) the transmission probability for charge carriers to be transmitted into $k^+$ from either bottom or top layers of 2SL is
\begin{equation}\label{eq19}
T_j^+=4 \textrm{Re}(k^{+}) \frac{\eta\left[ \eta^2+\left(\textrm{Im}(k^{-})+\kappa_{j} v_0 \sin\phi\right)^2 \right]}{C_{0 }+\sum_{m=1}^4C_m\cos(m\phi)},
\end{equation}
where $\kappa_{b(t)}=1(-1)$. The above equation shows that layer symmetry is broken, $T_b^+(\phi)=T_t^+(-\phi),$ only when $v_0\neq0$ and $\textrm{Im}(k^{-})\neq0$ which is satisfied  in the pink and gray regions in Fig. \ref{fig003}(a). However in the gray region there are no $k^+$ propagating states  and consequently the transmission probabilities $T_j^+$ are zero. The same analysis applies also to $T_j^-$ where the asymmetric  feature is preserved only when $\textrm{Im}(k^{+})\neq0$ as shown by  the green region in Fig. \ref{fig003}(a). For AB-2SL configuration, the layer asymmetry is only  reflected in the $T_+^j$ , see Eq. \eqref{eq18}, since $\textrm{Im}(k^{-})\neq0$  corresponds to the pink region in Fig. \ref{fig003}(b). While for $T_-^j$, the $k^-$ propagating states are only available for $E>\gamma_1$ (yellow region in Fig. \ref{fig003}(b)) which coincides with $\textrm{Im}(k^{+})=0$. Thus, the layer symmetry is always conserved in $T_-^j$ as it can be seen in  Eq. \eqref{eq18}. Now it is clear why  layer symmetry is not broken in the AA-BL configuration; because there are always  two propagating modes associated with any energy value. 

The breaking of angular symmetry in this situation is qualitatively similar to that obtained in AB-BL\cite{Ben} subject to an inter-layer bias. One can connect this layer asymmetry in the vicinity of the two valleys $K$ and $K'$ through  time-reversal symmetry. The Hamiltonian $H_{K'}$ can be related to the Hamiltonian $H_{K}$ through the transformation
\begin{equation}\label{eq20}
H_{K'}(\boldsymbol{k})=\Theta H_K(\boldsymbol{-k})\Theta^{-1},
\end{equation}    
where $\Theta$ is the time-reversal symmetry operator.
This implies, for example in the  $T_{b(t)}^+$ channel, that charge carriers moving from right to left and scattered from the bottom layer   to $k^+$ in $K$  valley are equivalent to those scattered from top layer to $k^+$ but moving in the opposite direction in  the vicinity of $K'$.  If layer symmetry  holds in the vicinity of one of the valleys, then the transmission  probabilities of charge carriers moving in the opposite directions must be the same. It is worth  pointing out here that  the layer asymmetry in the  $K$ valley is reversed in the  $K'$ valley and hence the overall symmetry of the system is restored.
Therefore, the macroscopic time reversal symmetry is preserved.
%%%%%%%%%%%%%%%%%%%%%%%%%%%%%%%%%%%%%%%%%%%%%%%%%%%%%%%%%%%%%%%%
\section{Numerical Results}\label{Results}
%%%%%%%%%%%%%%%%%%%%%%%%%%%%%%%%%%%%%%%%%%%%%%%%%%%%%%%%%%%%%%%%
We first present the results for transmission, and reflection probabilities  and for the conductance in the case of domain walls separating 2SL and AA-BL structures. The different regions as defined in Fig. \ref{fig003} are superimposed as dashed black and white curves. Moreover, in calculating  the transport properties we considered different magnitudes for  the electrostatic potential   $v_0$ and bias $\delta$ applied  to the drain structure.
%%%%%%%%%%%%%%%%%%%%%%%%%%%%%%%%%%%%%%%%%%%%%%%%%%%%%%%%%%
\subsection{AA-Stacking}
%%%%%%%%%%%%%%%%%%%%%%%%%%%%%%%%%%%%%%%%%%%%%%%%%%%%%%%%%%
\subsubsection{ 2SL-AA/AA-2SL}
%%%%%%%%%%%%%%%%%%%%%%%%%%%%%%%%%%%%%%%%%%%%%%%%%%%%%%%%%%
We consider charge carriers tunnelling through 2SL-AA and AA-2SL systems. In Fig. \ref{polar-SL-AA}(a) we show the transmission and reflection probabilities for charge carriers impinging  on pristine AA-BL as a function of  incident angle $\phi$. As a result of the layer symmetry, charge carriers  incident from bottom/top layer of 2SL and transmitted into the lower Dirac cone ($k^+$) in the AA-BL will have the same transmission probability $T_{b}^+=T_{t}^+$. Similarly, for those charge carriers transmitted into the upper cone, they will also have the same probability $T_{b}^-=T_{t}^-$ regardless which layer they are  incident from.  

This symmetry stems from the fact that the wavefunction in the 2SL are a superposition of two spinors corresponding  to the two sublattices while in AA-BL it is a superposition of four. For this reason, charge carriers incident from top or bottom layer of 2SL have the same dynamics and hence share their transmission probability. A partial reflection into the same layer, $R_{b}^{b}=R_{t}^{t}$ is shown in Fig. \ref{polar-SL-AA}(a), which  corresponds to evanescent modes  associated with the upper Dirac cone ($k^-$). As in transmission, charge carriers can be back scattered between the layers. However, the absence of the electrostatic potential  results in a small scattering current as depicted in Fig. \ref{polar-SL-AA}(a). In addition, scattering back from top to bottom layer or vice versa occurs also with the same reflection probabilities $R_{b}^{t}=R_{t}^{b}$.

Because of chiral decoupling of oppositely propagating waves in AA-BL and in SL, back-scattering  is forbidden for normal incidence ($\phi=0$) and thus the reflection probabilities for each channel are zero, i.e. $R_{b}^{b(t)}(0) = R_{t}^{t(b)}(0)=0$. This is associated  with perfect tunnelling $T_{b}^+(0) + T_{b}^-(0) = T_{t}^+(0) + T_{t}^-(0) = 1$. The effect holds for all forthcoming structures composed of AA-BL and 2SL. 

Fig. \ref{polar-SL-AA}(b) shows the numerical results of the same system, 2SL-AA, but now in the AA region, the potential is increased to $v_0=1.5 \gamma_1$. This shifts the two Dirac cones in energy to $ v_0\pm \gamma_1$. As a  result of the presence of the electrostatic potential, a strong scattered reflection $R_{b}^{t}/R_{t}^{b}$ takes place when there are no propagating modes in the AA section. 

In Figs. \ref{polar-SL-AA}(c,d), we show the reversed configuration, i.e. an AA-2SL system. The transmission and reflection probabilities for zero ($v_0=0$) and with nonzero ($v_0=1.5 \gamma_1$) electrostatic potentials applied to  2SL are reported in panels (c) and (d) respectively.
Similar to the 2SL-AA system, we can note that  layer symmetry still holds such that  $T_{+}^b = T_{+}^t$ and $T_{-}^b = T_{-}^t$. Furthermore, we find strong non-scattered reflection in the $R_{+}^+$ and $ R_{-}^-$ channels that is associated with evanescent modes on both sides of AA-BL and 2SL whereas the scattered reflection channels $R_{-}^+$ and $ R_{+}^-$ are always zero due to the protected cone transport discussed earlier.
%%%%%%%%%%%%%%%%%%%%%%%%%%%%%%%%%%%%%%%%%%%%%%%%%%%%%%%%%%
\begin{figure}[t]
\vspace{0.4cm}
\centering \graphicspath{{./Figures1//SL-AA-SL/}}
\includegraphics[width=1.5  in]{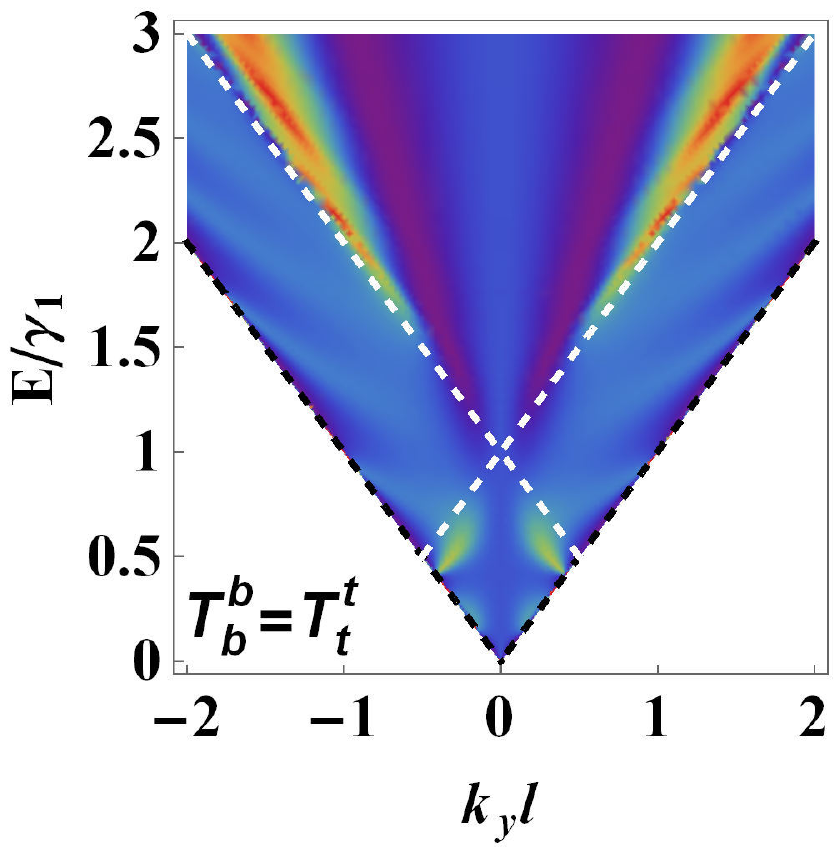}\ \
\includegraphics[width=1.73  in]{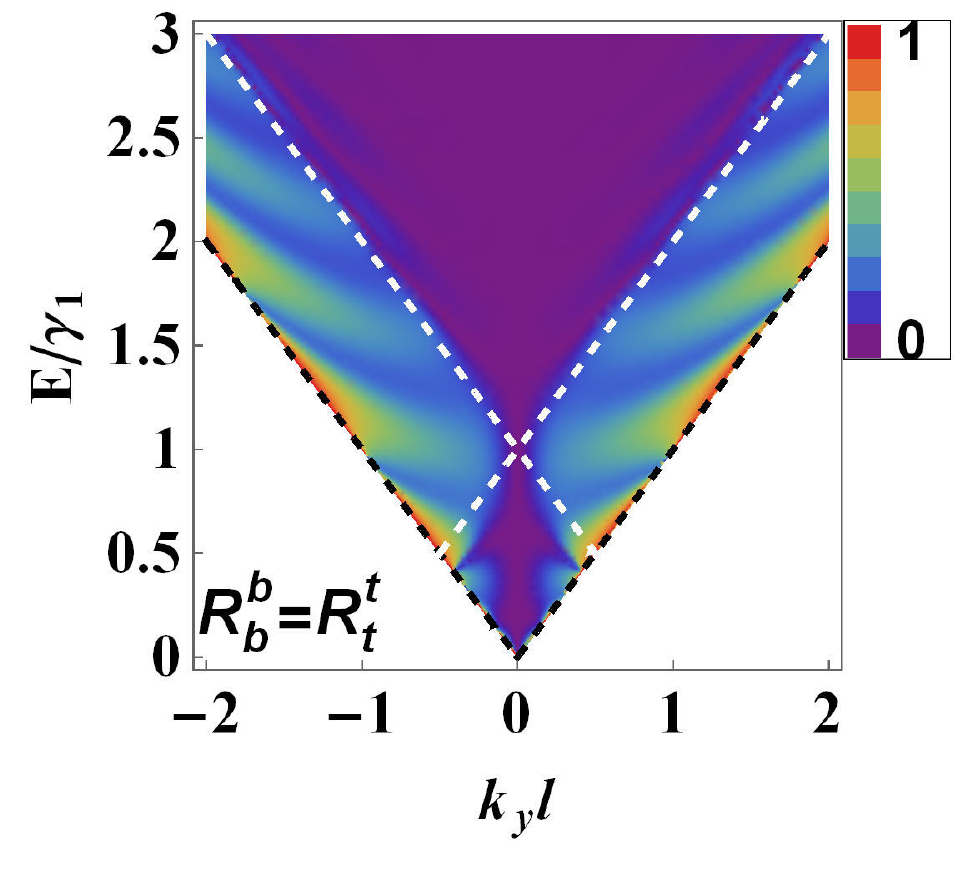}\\
\includegraphics[width=1.5 in]{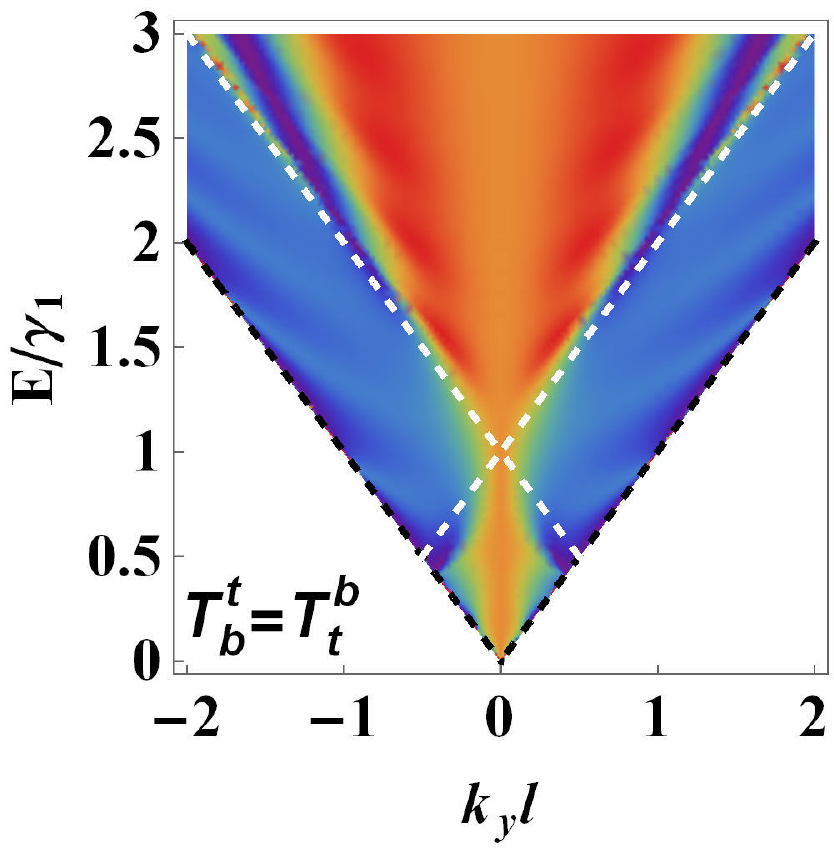}\ \
\includegraphics[width=1.73  in]{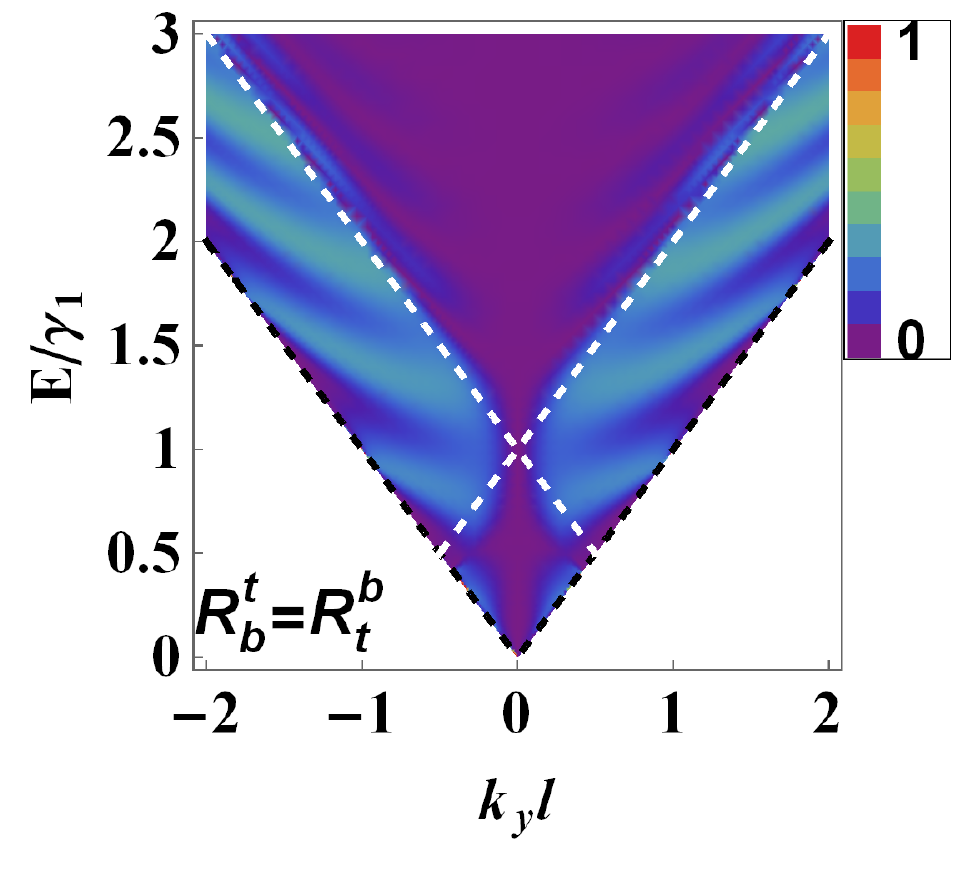}
\caption{(Colour online)   Density plot of the  transmission and reflection probabilities through 2SL-AA-2SL as a function of Fermi energy and
transverse wave vector $k_y$ with $v_0= \delta =0$ and width of the AA-BL $d=25$ nm. }\label{fig-SL-AA-SL}
\end{figure}
%%%%%%%%%%%%%%%%%%%%%%%%%%%%%%%%%%%%%%%%%%%%%%%%%%%%%%%%%%
\begin{figure}[t]
\vspace{0.4cm}
\centering \graphicspath{{./Figures1//SL-AA-SL/}}
\includegraphics[width=1.5  in]{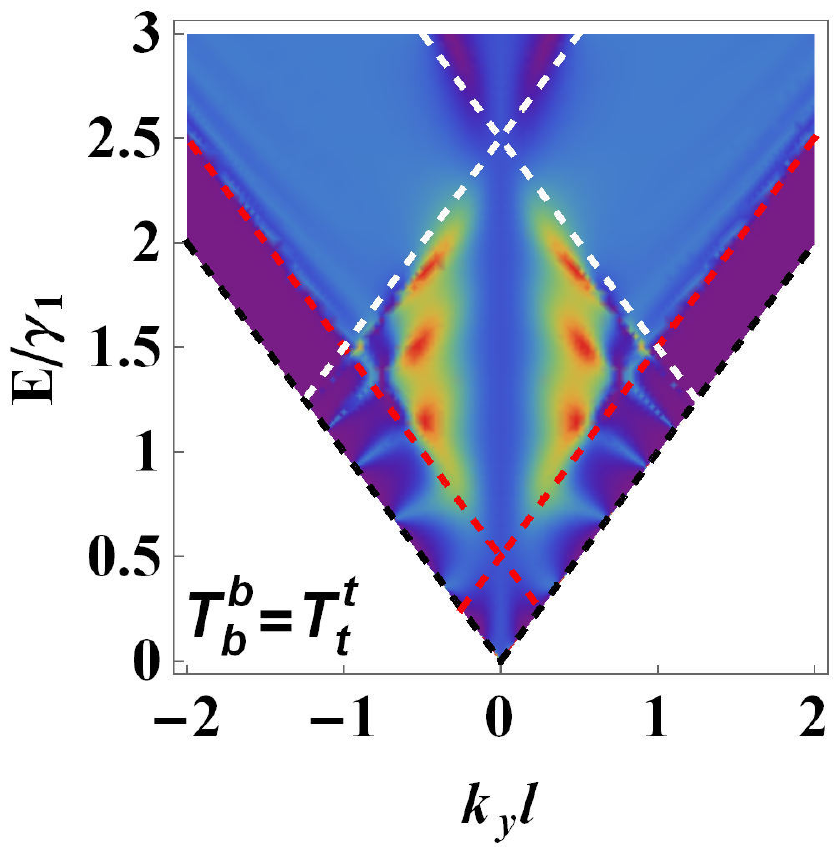}\ \
\includegraphics[width=1.73  in]{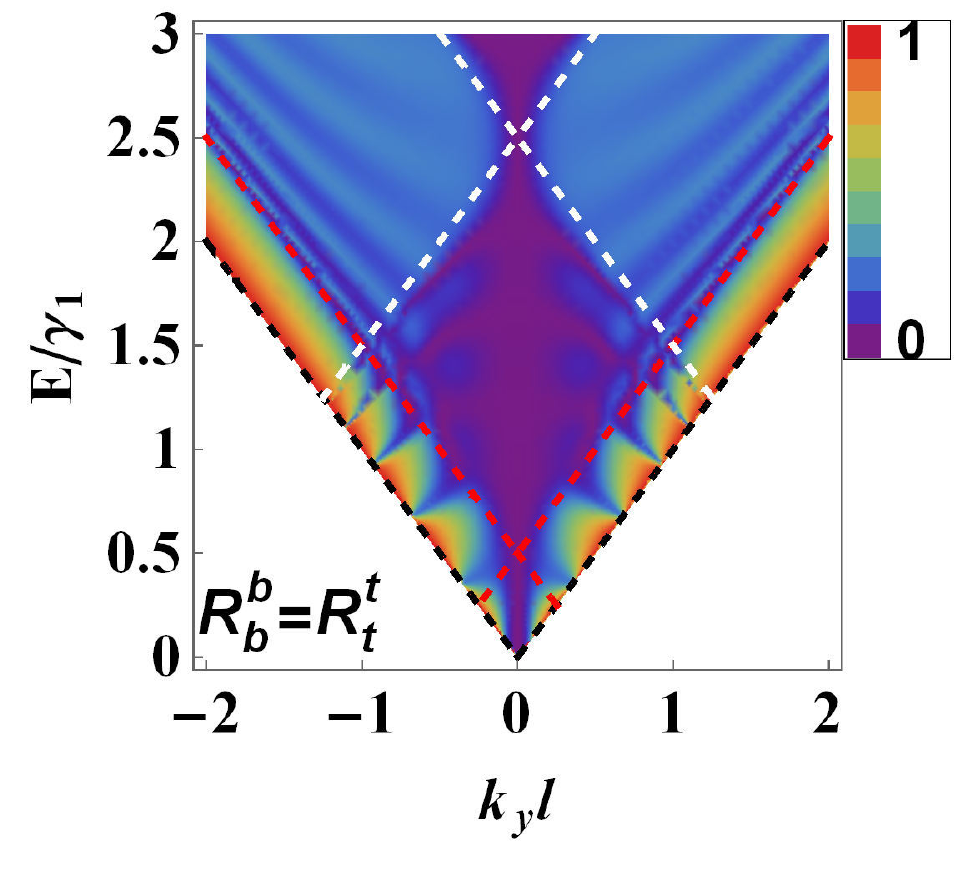}\\
\includegraphics[width=1.5 in]{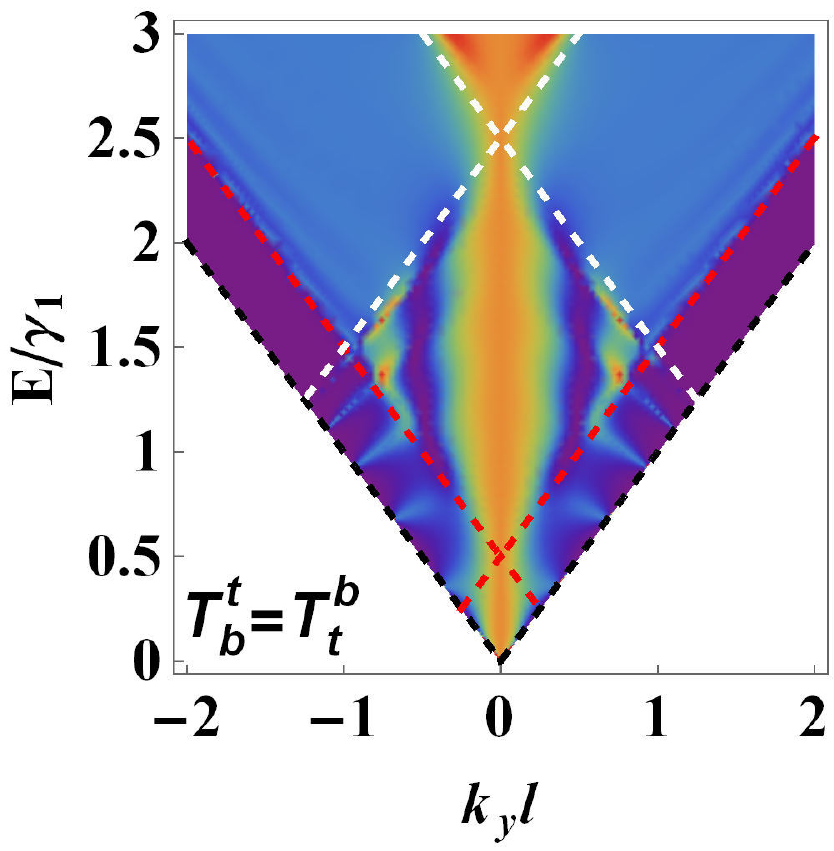}\ \
\includegraphics[width=1.73  in]{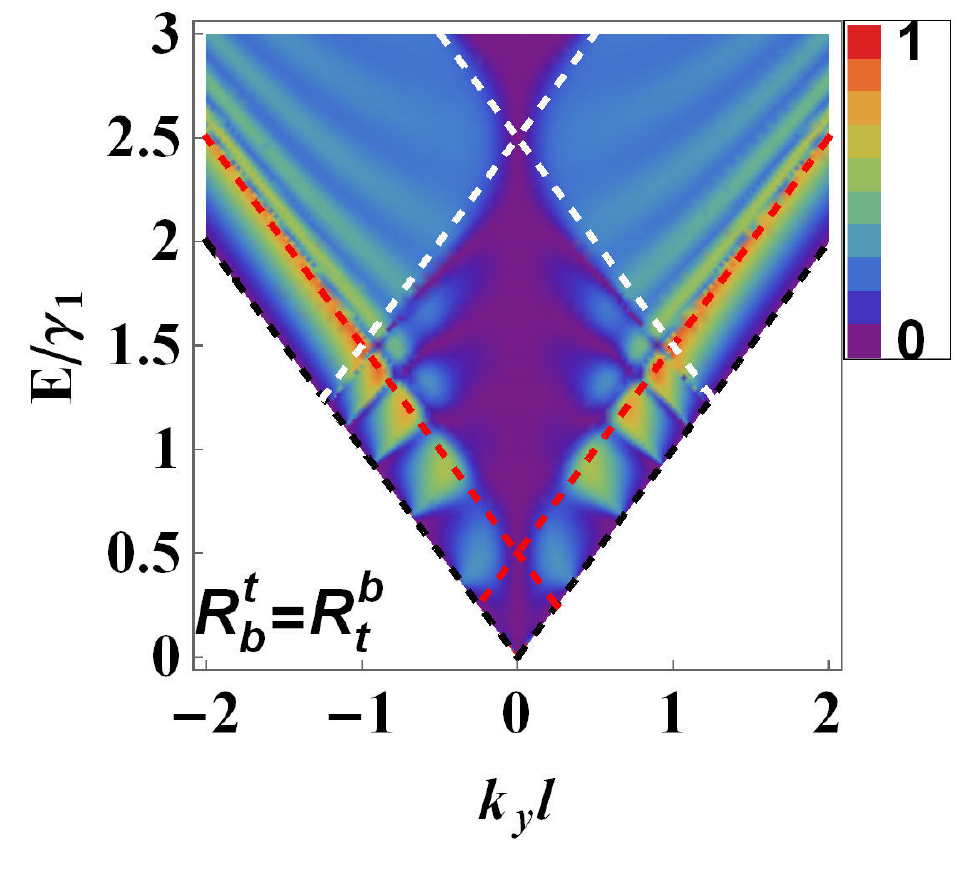}
\caption{(Colour online)   The same as in Fig. \ref{fig-SL-AA-SL}, but now with  $v_0=1.5 \gamma_1.$ Red and white dashed curves correspond to the lower and upper Dirac cones in AA-BL, respectively,  while
the black dashed curves are the bands of 2SL. }\label{fig-SL-AAv-SL}
\end{figure}
%%%%%%%%%%%%%%%%%%%%%%%%%%%%%%%%%%%%%%%%%%%%%%%%%%%%%%%%%%
\begin{figure}[t]
\vspace{0.4cm}
\centering \graphicspath{{./Figures1//SL-AA-SL/}}
\includegraphics[width=1.5  in]{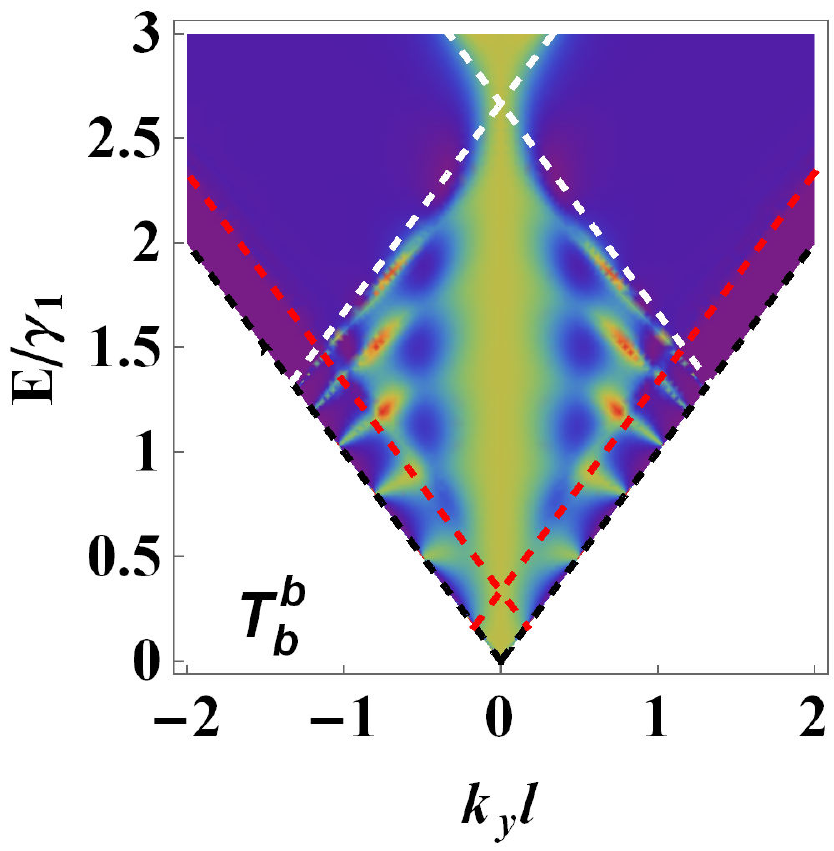}\ \
\includegraphics[width=1.73  in]{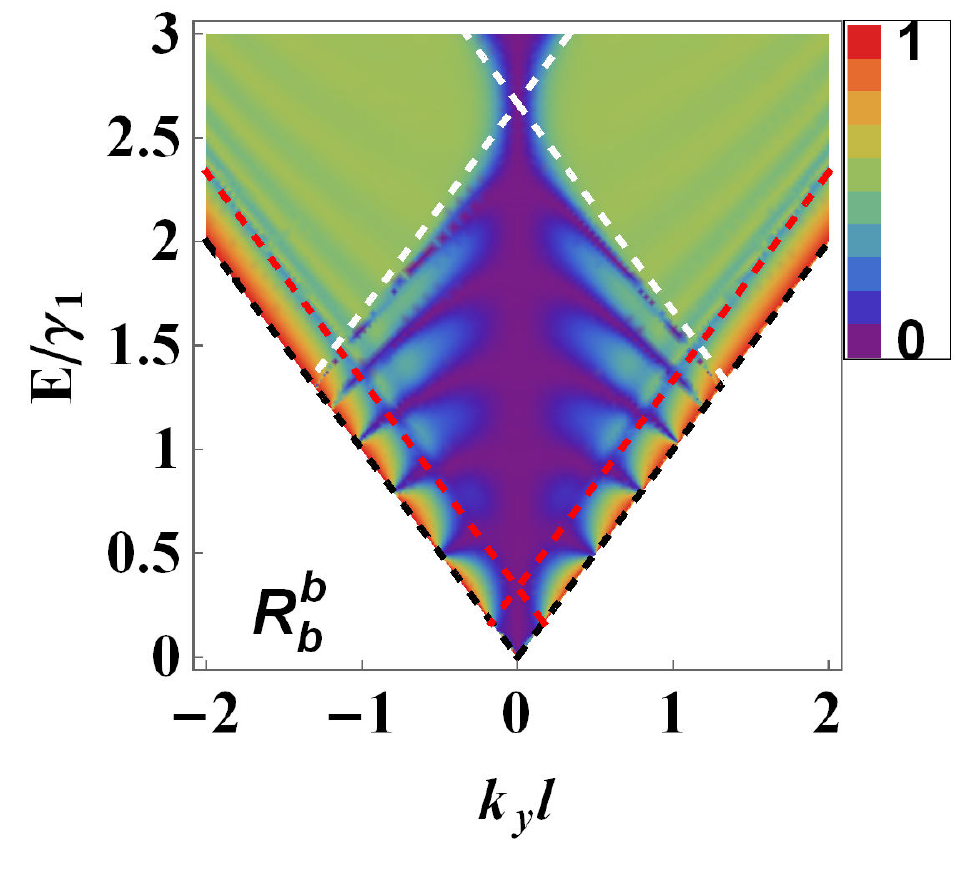}\\
\includegraphics[width=1.5 in]{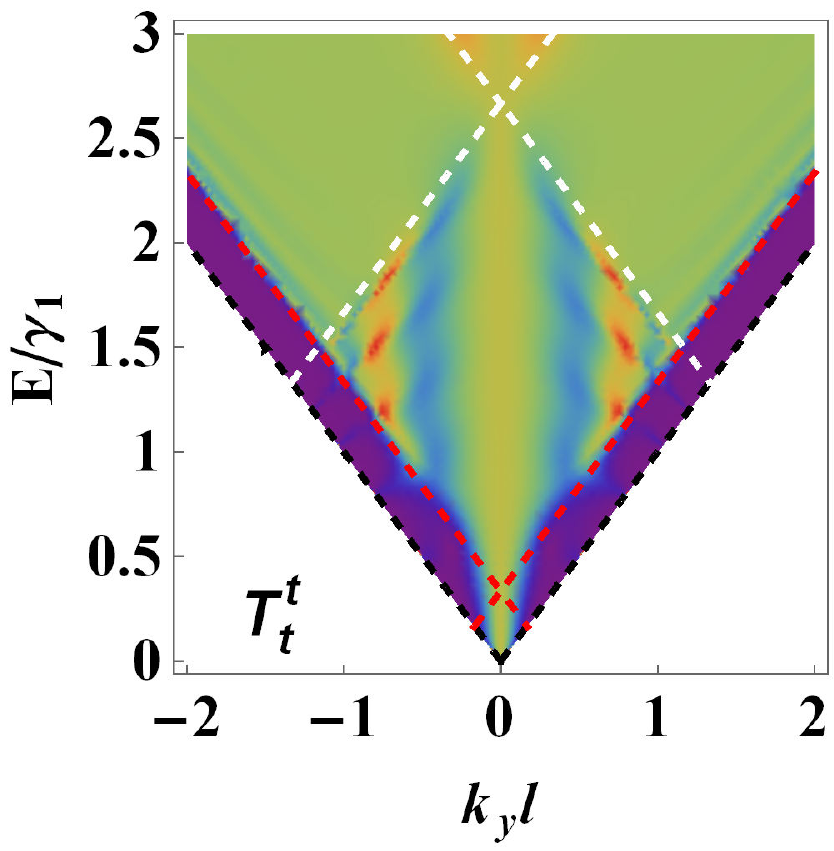}\ \
\includegraphics[width=1.73  in]{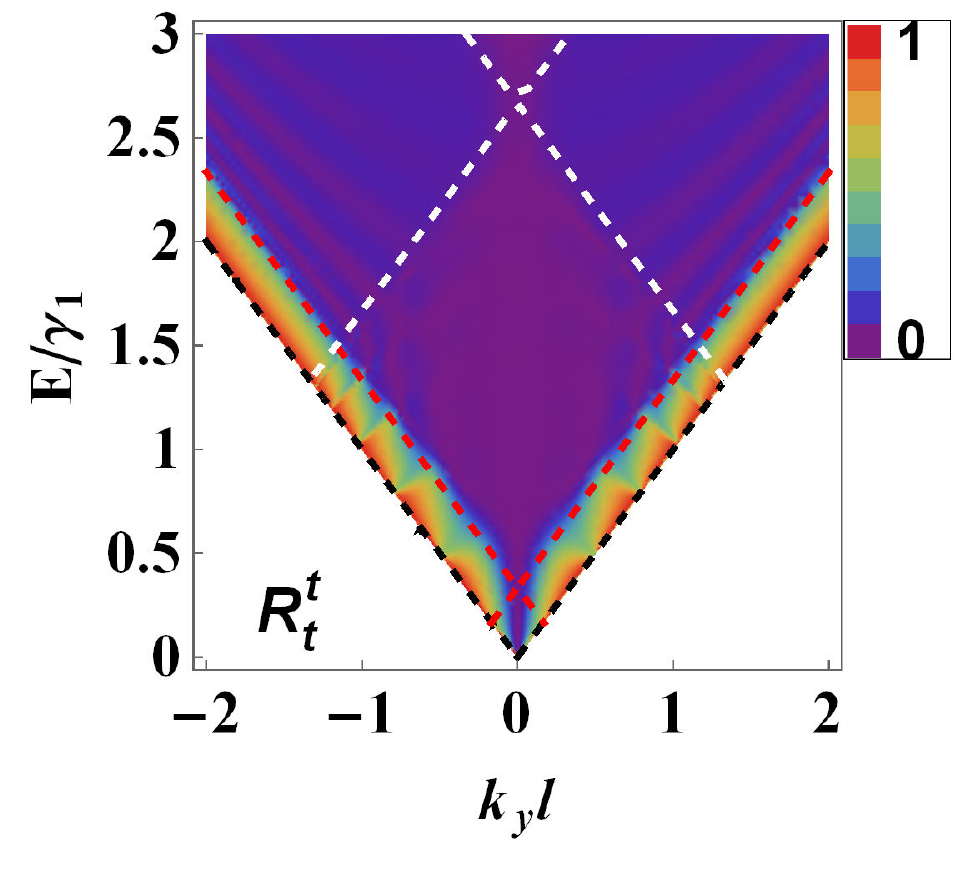}\\
\includegraphics[width=1.5 in]{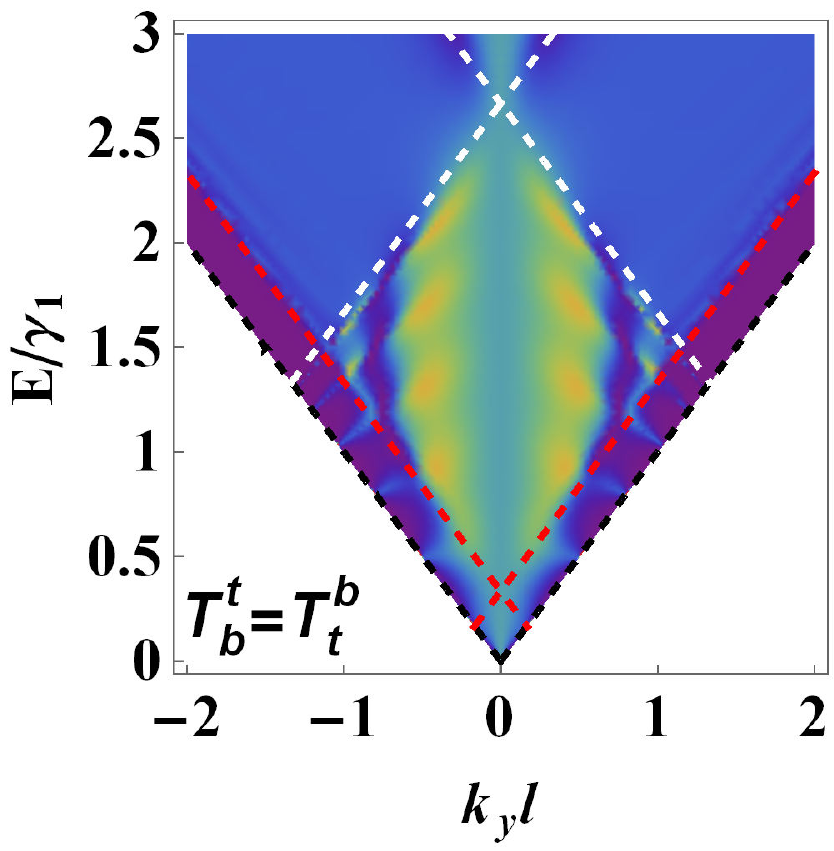}\ \
\includegraphics[width=1.73  in]{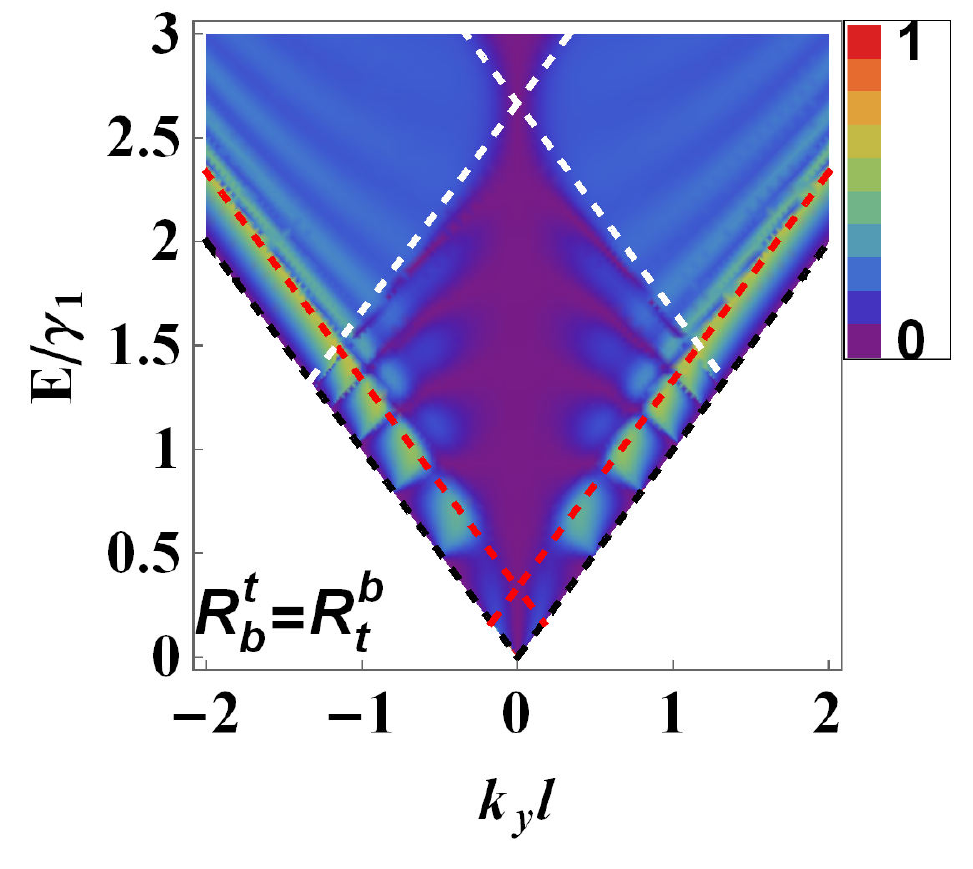}
\caption{(Colour online) The same as in Fig. \ref{fig-SL-AA-SL}, but now with     $v_0=1.5\gamma_1$ and $ \delta =0.6\gamma_1$. }\label{fig-SL-AAve-SL}
\end{figure}
%%%%%%%%%%%%%%%%%%%%%%%%%%%%%%%%%%%%%%%%%%%%%%%%%%%%%%%%%% 
%%%%%%%%%%%%%%%%%%%%%%%%%%%%%%%%%%%%%%%%%%%%%%%%%%%%%%%%%%
\subsubsection{ 2SL-AA-2SL }
%%%%%%%%%%%%%%%%%%%%%%%%%%%%%%%%%%%%%%%%%%%%%%%%%%%%%%%%%%
\begin{figure}[t]
\vspace{0.4cm}
\centering \graphicspath{{./Figures1//AA-SL-AA/}}
\includegraphics[width=1.5  in]{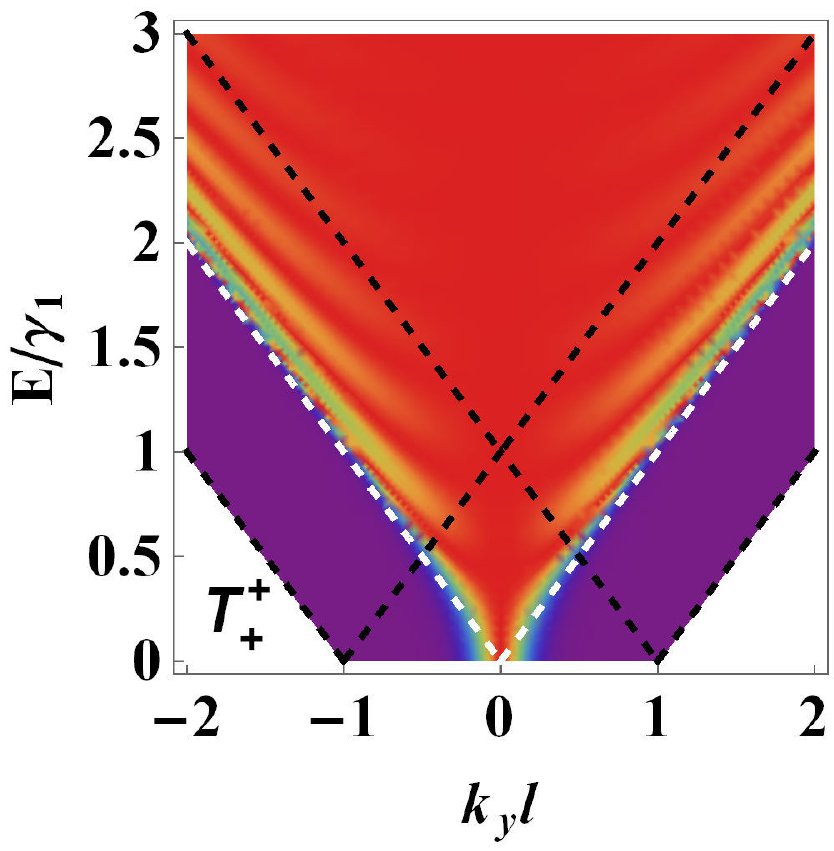}\ \
\includegraphics[width=1.73  in]{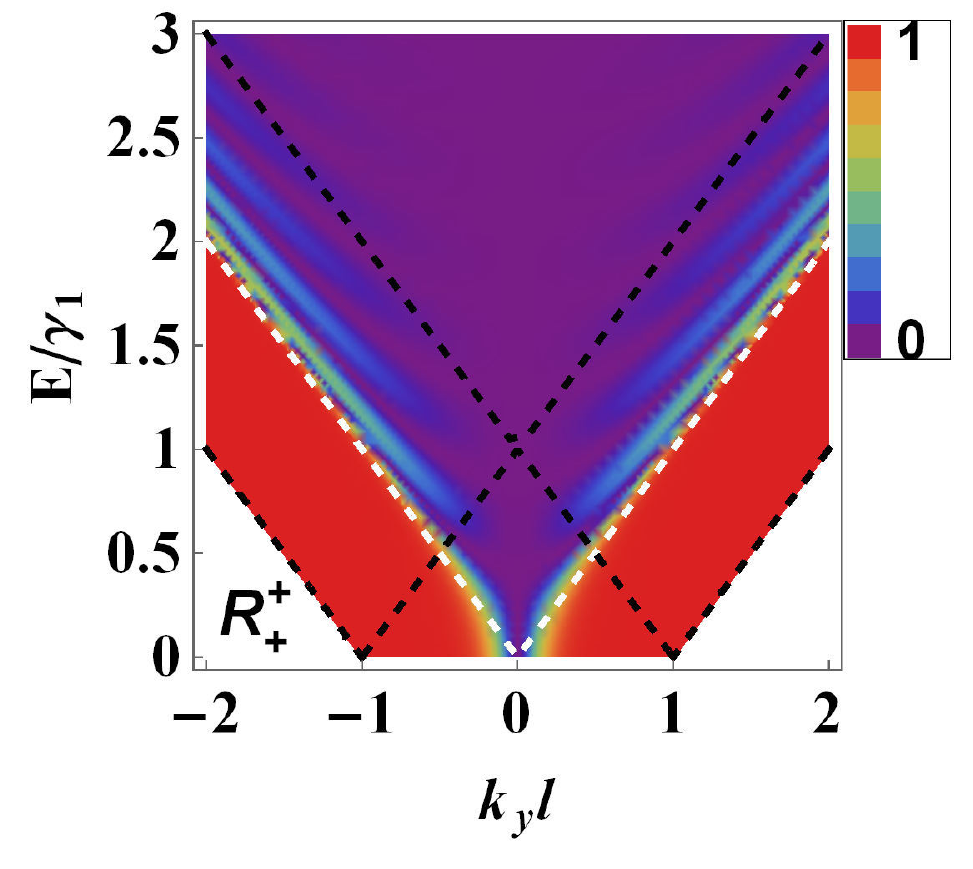}\\
\includegraphics[width=1.5 in]{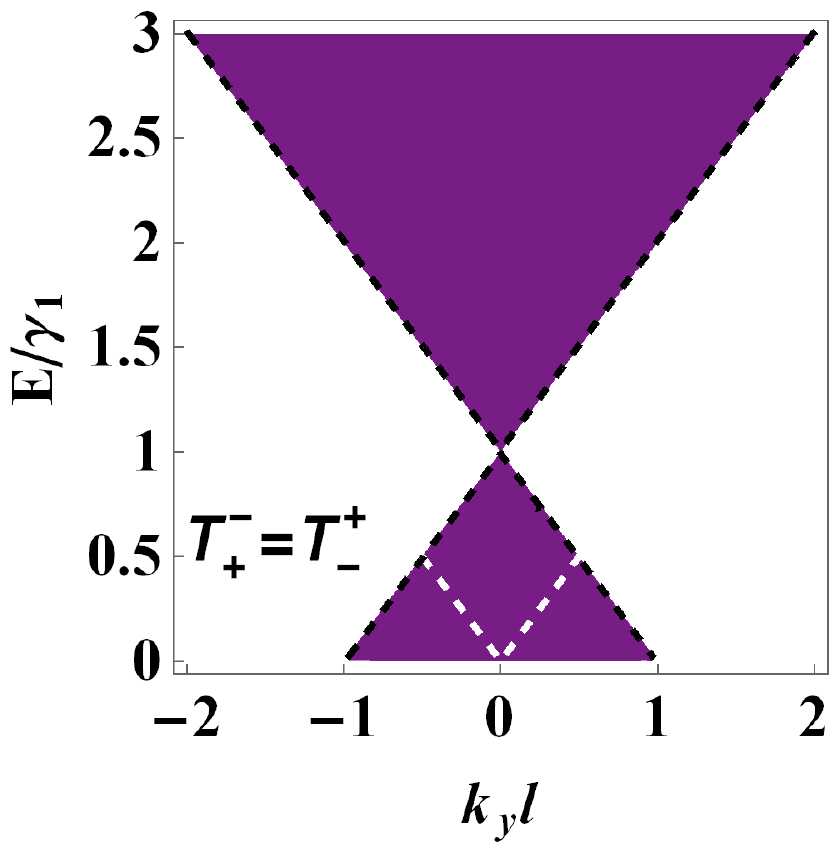}\ \
\includegraphics[width=1.73  in]{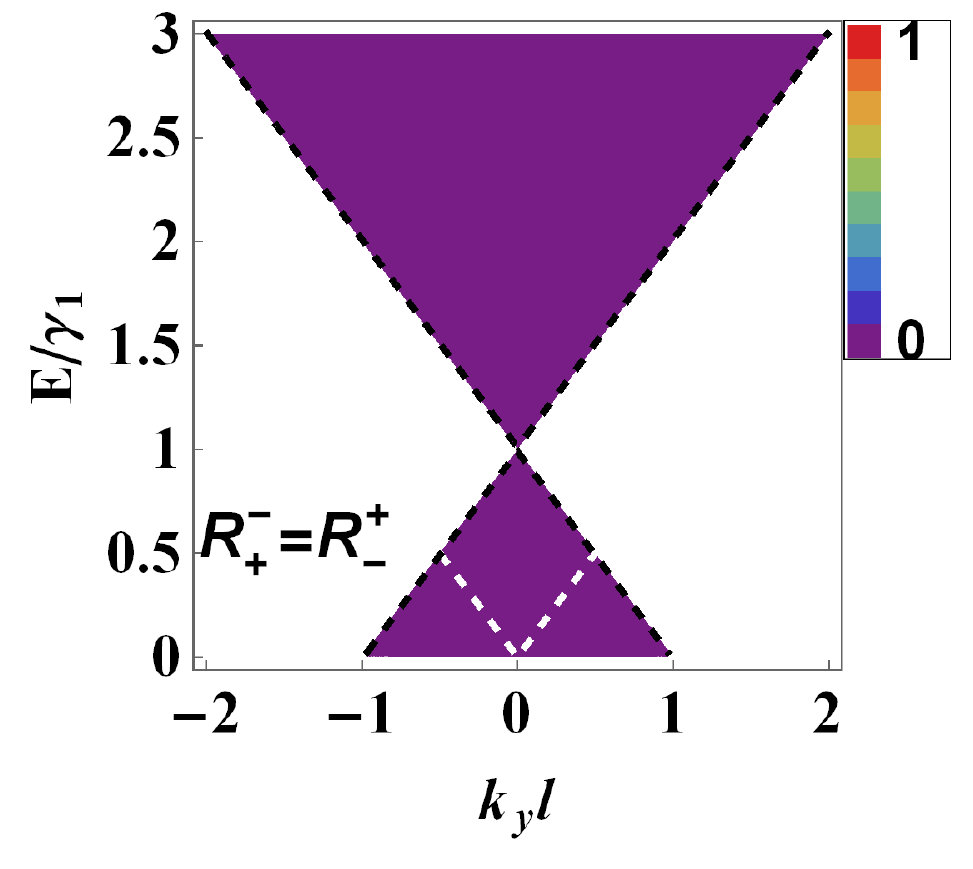}
\includegraphics[width=1.5  in]{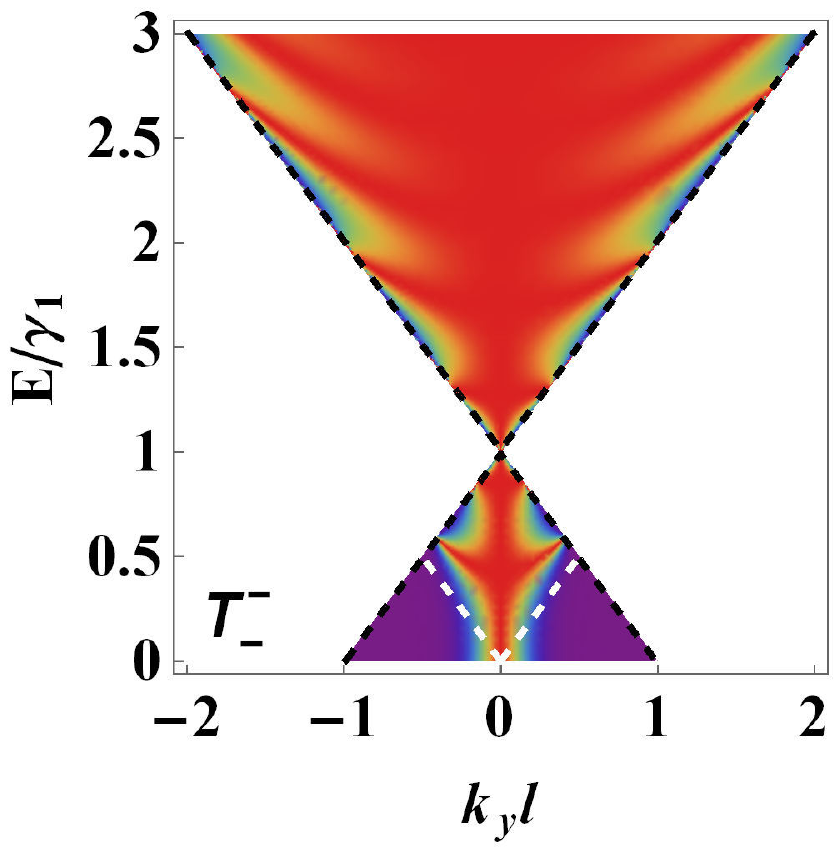}\ \
\includegraphics[width=1.73  in]{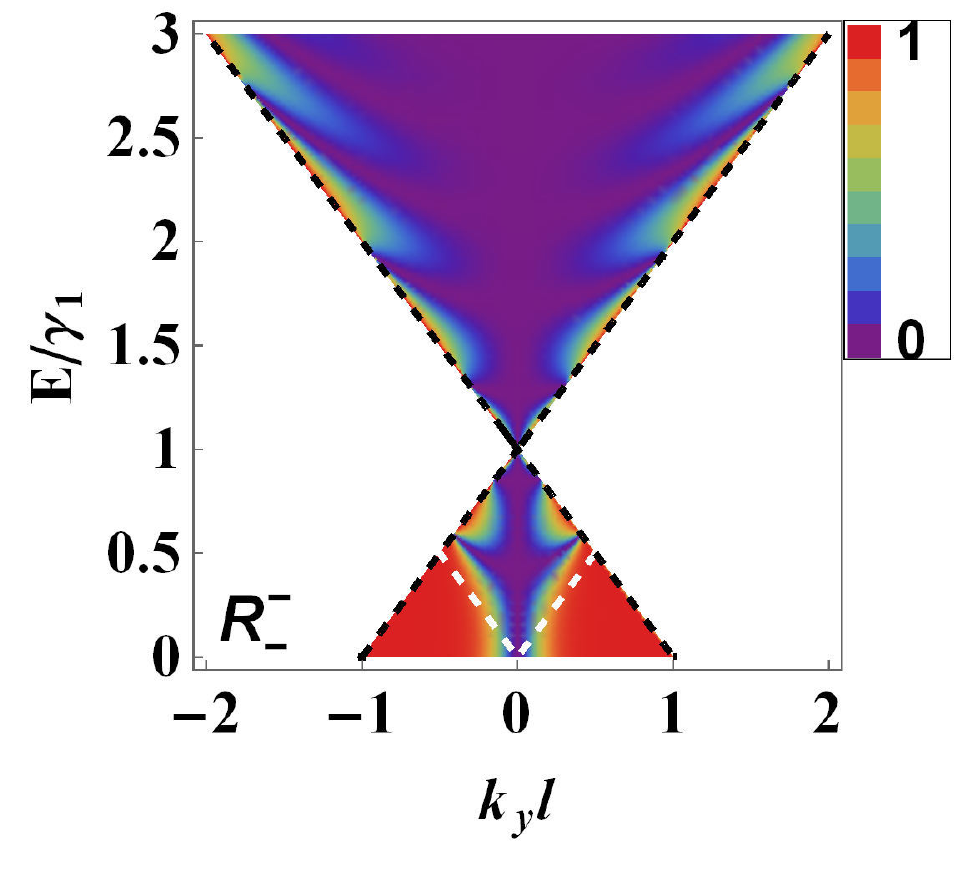}\\
\caption{(Colour online)   Density plot of the  transmission and reflection probabilities through AA-2SL-AA as a function of Fermi energy and
transverse wave vector $k_y$ with $v_0= \delta =0$ and width of the 2SL $d=25$ nm.  }\label{fig-AA-SL-AA}
\end{figure}
%%%%%%%%%%%%%%%%%%%%%%%%%%%%%%%%%%%%%%%%%%%%%%%%%%%%%%%%%%
\begin{figure}[t]
\vspace{0.4cm}
\centering \graphicspath{{./Figures1//AA-SL-AA/}}
\includegraphics[width=1.5  in]{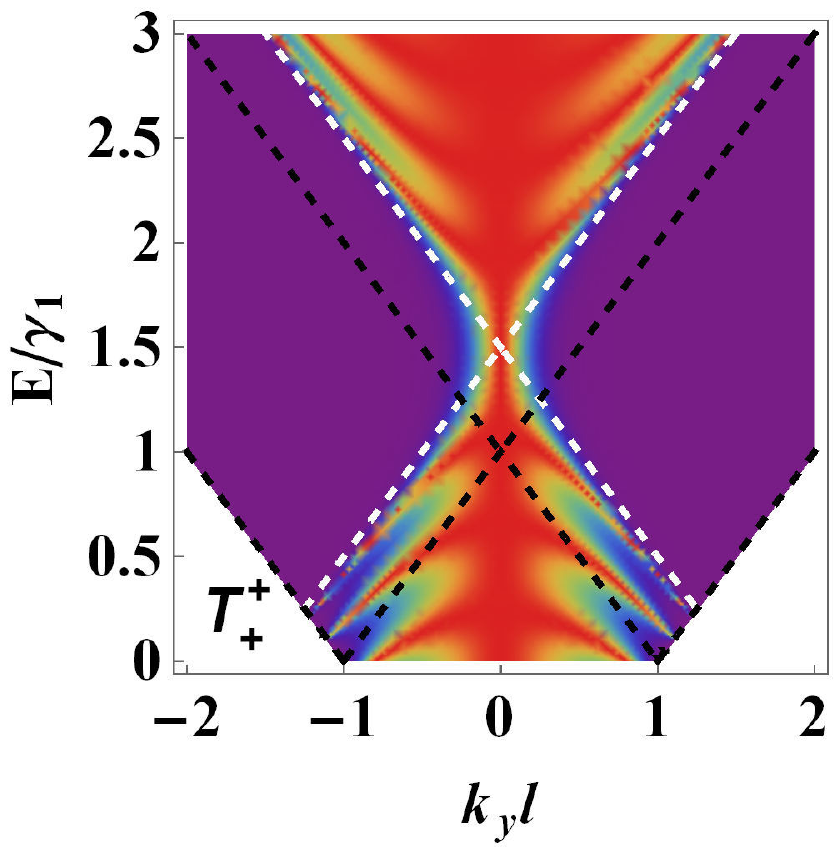}\ \
\includegraphics[width=1.73  in]{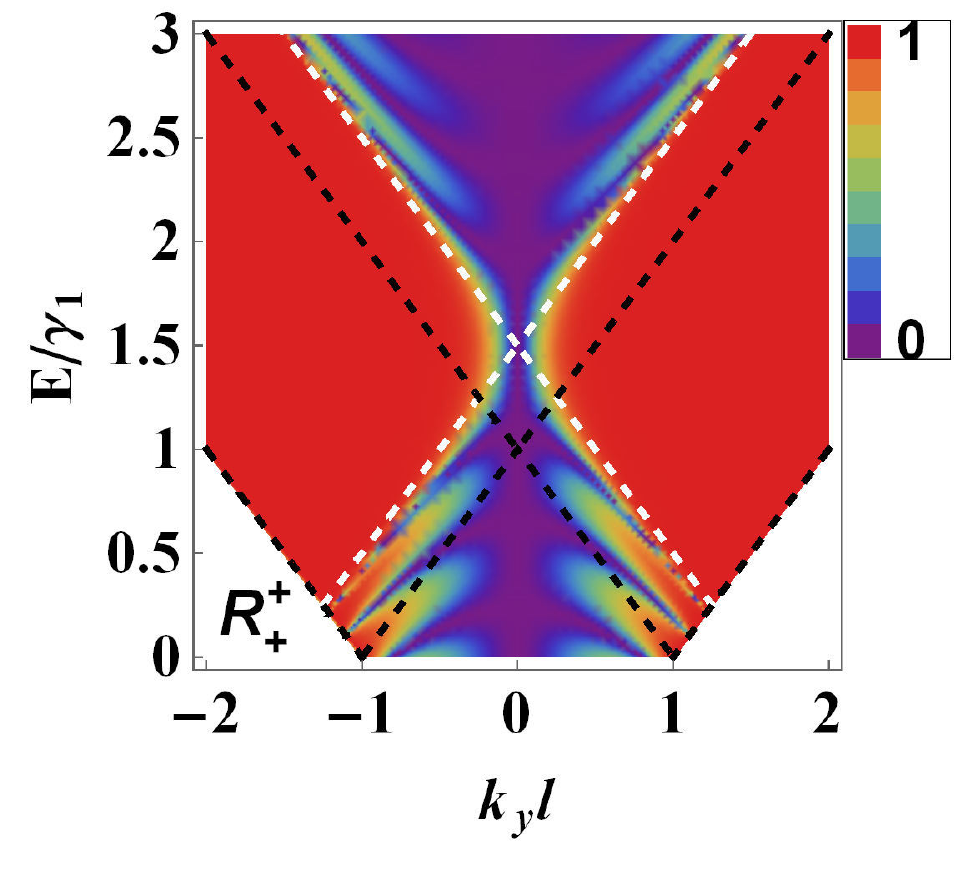}\\
\includegraphics[width=1.5 in]{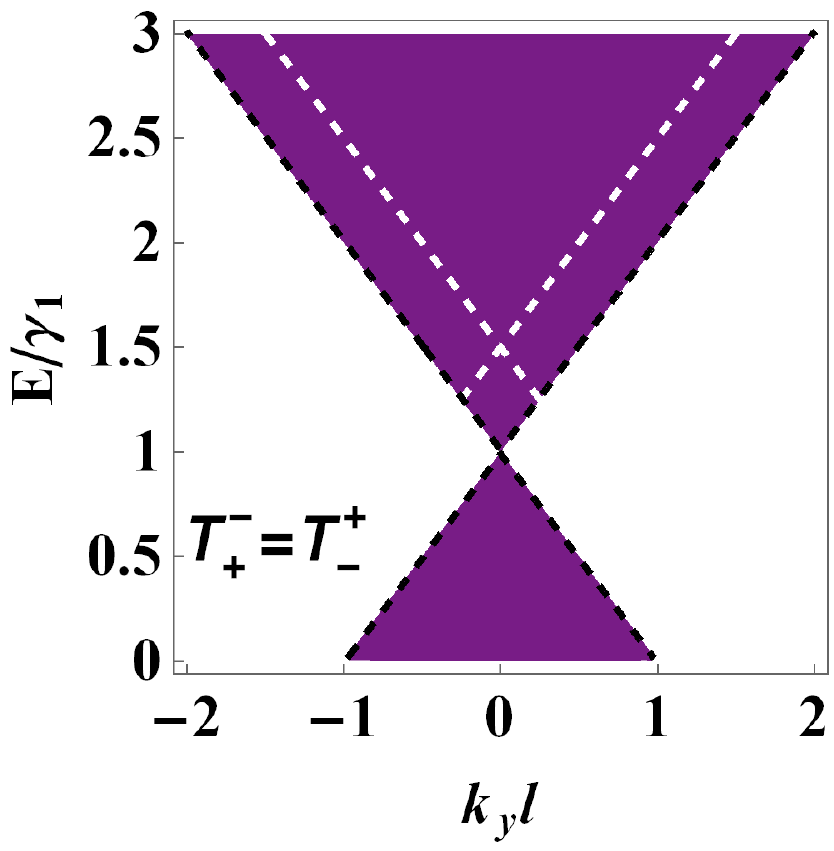}\ \
\includegraphics[width=1.73  in]{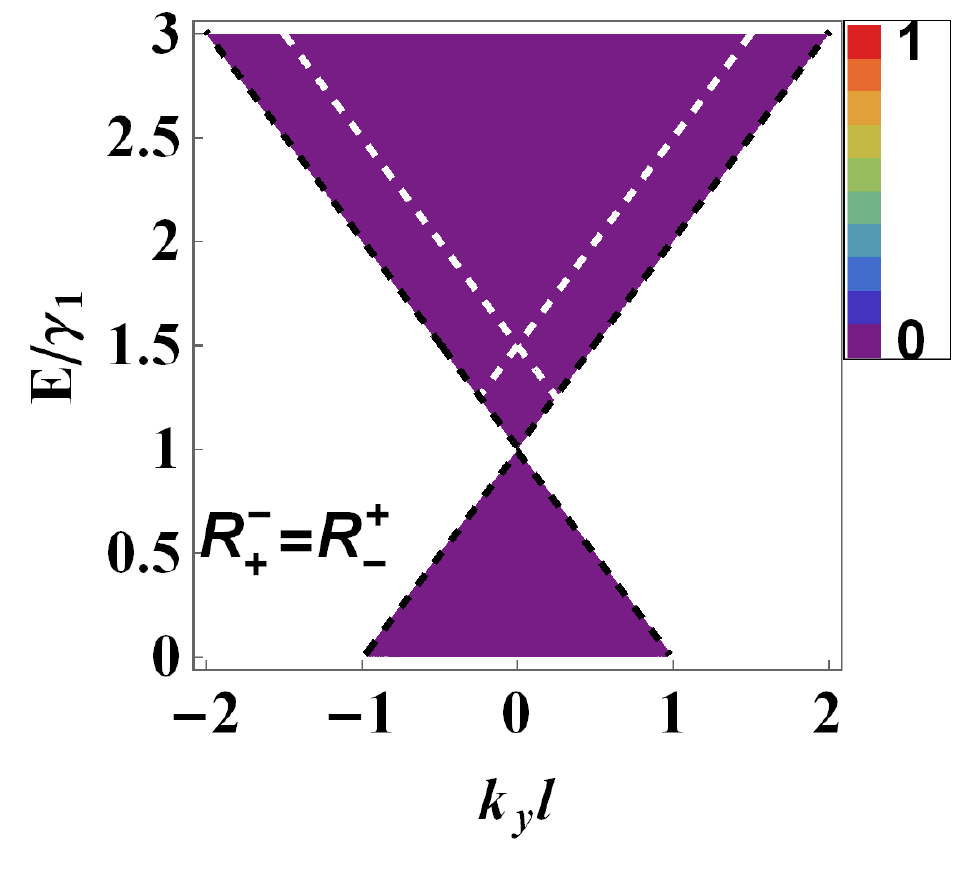}\\
\includegraphics[width=1.5  in]{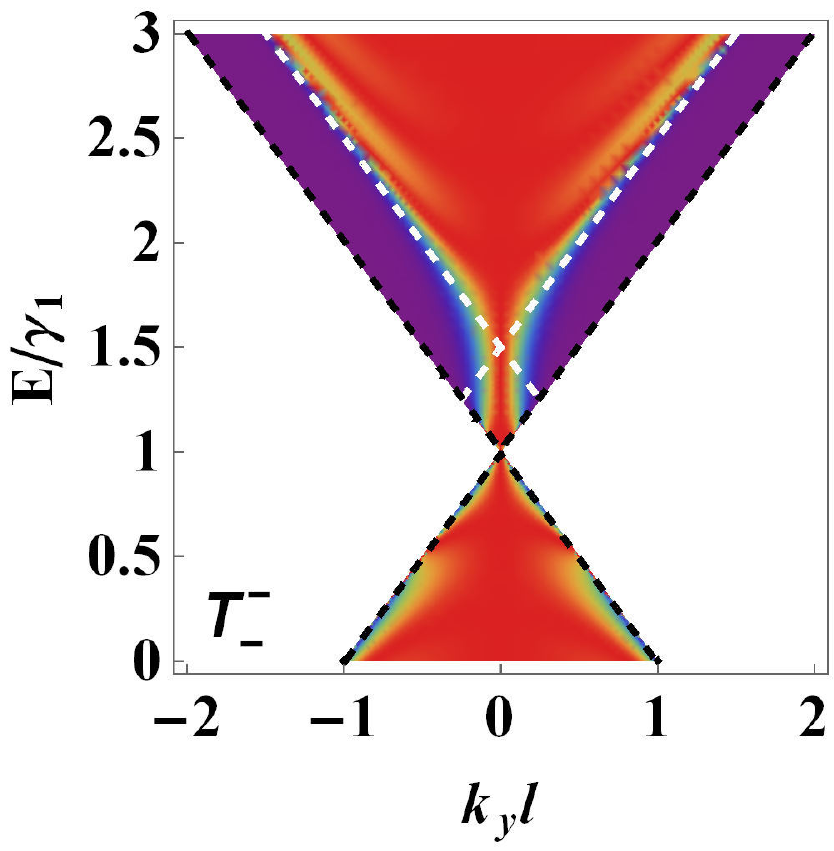}\ \
\includegraphics[width=1.73  in]{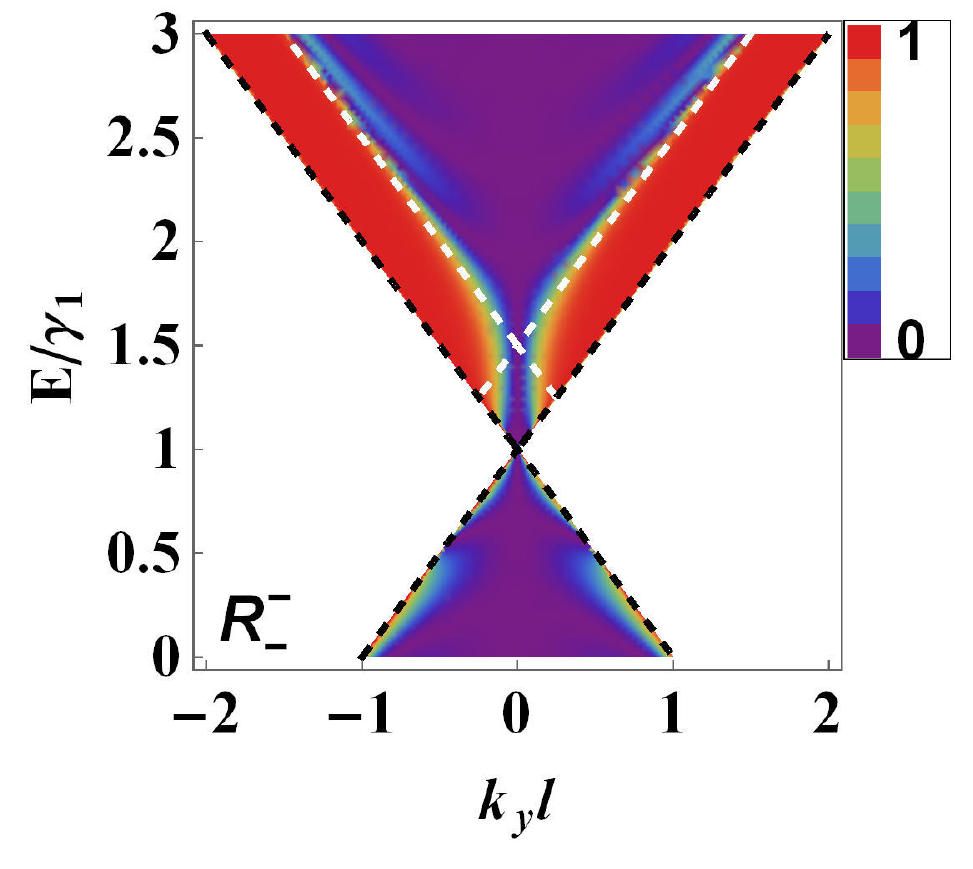}\\
\caption{(Colour online)   The same as in Fig. \ref{fig-AA-SL-AA}, but now with  $v_0=1.5\ \gamma_1.$ }\label{fig-AA-SLv-AA}
\end{figure}
%%%%%%%%%%%%%%%%%%%%%%%%%%%%%%%%%%%%%%%%%%%%%%%%%%%%%%%%%%
In this Section, we show the results of transport across two domain walls forming a system with three regions; where  AA-BL is sandwiched  between two regions of 2SL, see Fig. \ref{intro-fig02}(a). Such a system can exhibit a strong layer selectivity when current flows through the intermediate region , i.e.  AA-BL. This behaviour has already been investigated in Ref. [\onlinecite{Hasan1}]. Here, however, we go in much more detail to show how the different transmission and reflection channels are affected by the electrostatic potential or finite bias applied to the intermediate region.

In Figs. \ref{fig-SL-AA-SL} and \ref{fig-SL-AAv-SL} we show the scattered and non-scattered channels for transmission and reflection for pristine AA-BL and with electrostatic potential of strength $v_0=1.5 \ \gamma_1$, respectively. Layer symmetry is preserved in both reflection and transmission channels as clarified in Figs. \ref{fig-SL-AA-SL}.  and \ref{fig-SL-AAv-SL} also show strong scattered transmission, especially for normal incidence which can be altered depending on the width of the AA-BL. When an electrostatic potential is applied to the middle domain,  resonances appear in the transmission probabilities for $v_0+\gamma_{1}>E>v_0-\gamma_1$ as shown in Fig. \ref{fig-SL-AAv-SL}.  This is a consequence of the finite size of the AA-BL and the presence of charge carriers with  different chirality in the mentioned range of energies \cite{AA-cones}.  Introducing a finite bias $ \delta =0.6\gamma_1$ on AA-BL  breaks the layer symmetry of the system. As a result, $T_{b}^b \neq T_{t}^t$ and $R_{b}^b \neq R_{t}^t$. However, it is still preserved in the scattered channels  $T_{t}^b = T_{b}^t$ and $R_{t}^b = R_{b}^t$ ( see Fig. \ref{fig-SL-AAve-SL}). 

It is worth mentioning here that the finite bias does not break the angular symmetry with respect to normal incidence in the transmission and reflection probabilities as it does for normal AB-BL\cite{Ben}. This is a manifestation of the symmetric inter-layer coupling in  AA-BL.
%%%%%%%%%%%%%%%%%%%%%%%%%%%%%%%%%%%%%%%%%%%%%%%%%%%%%%%%%%
\subsubsection{ AA-2SL-AA }
%%%%%%%%%%%%%%%%%%%%%%%%%%%%%%%%%%%%%%%%%%%%%%%%%%%%%%%%%%
In this system we interchange the AA-BL and 2SL as shown in Fig. \ref{intro-fig02}(d). In this case, scattering is defined between the two cones in the AA-BL regions. In Figs. \ref{fig-AA-SL-AA} and \ref{fig-AA-SLv-AA} we show the transmission and reflection probabilities between the two Dirac cones through the pristine 2SL and in the presence of an electrostatic potential, respectively. The first and the last rows of Figs. (\ref{fig-AA-SL-AA}) and (\ref{fig-AA-SLv-AA}) show the non-scattered transmission and reflection probabilities corresponding to the lower and upper Dirac cones, respectively. We notice that Klein tunnelling is preserved at normal incidence. This shows that Klein tunnelling in AA-stacked bilayer graphene is a robust feature that is insensitive to local changes in the inter-layer coupling.  On the other hand we see that scattering between two different Dirac cones remains strictly forbidden even with a local decoupling of the two layers. Therefore, these devices could be used for conetronics. As a result, in the second row of Figs. \ref{fig-AA-SL-AA} and \ref{fig-AA-SLv-AA} the scattered transmission and reflection channels are zero $T_{+}^- = T_{-}^+=R_{+}^- = R_{-}^+=0$.  

\begin{figure}[t]
\vspace{0.4cm}
\centering \graphicspath{{./Figures1//AA-SL-AA/}}
\includegraphics[width=1.5  in]{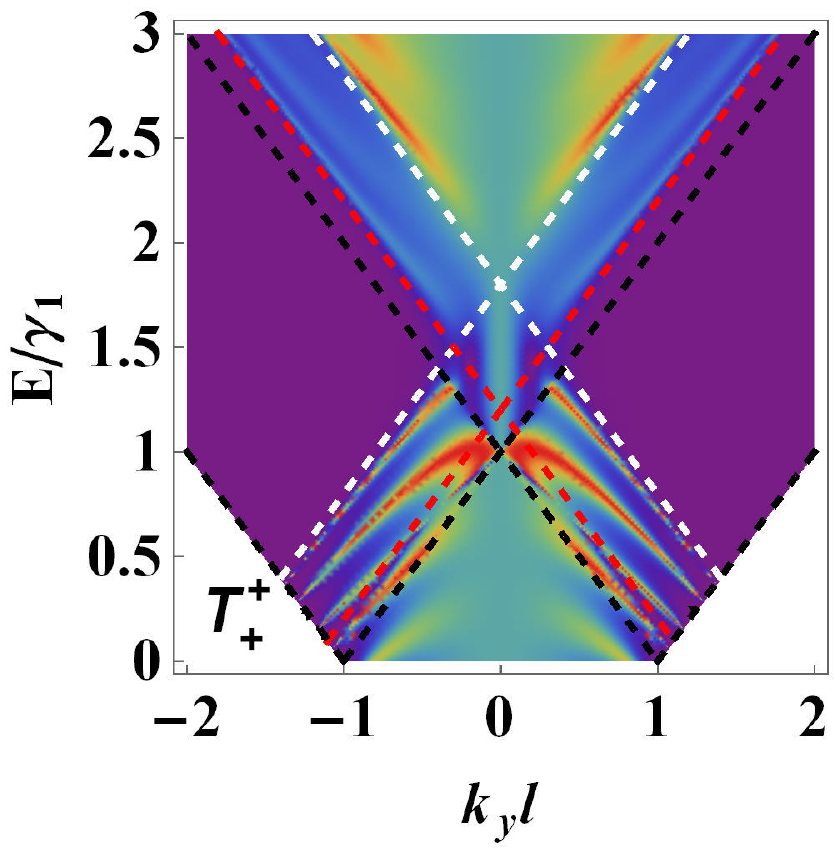}\ \
\includegraphics[width=1.73  in]{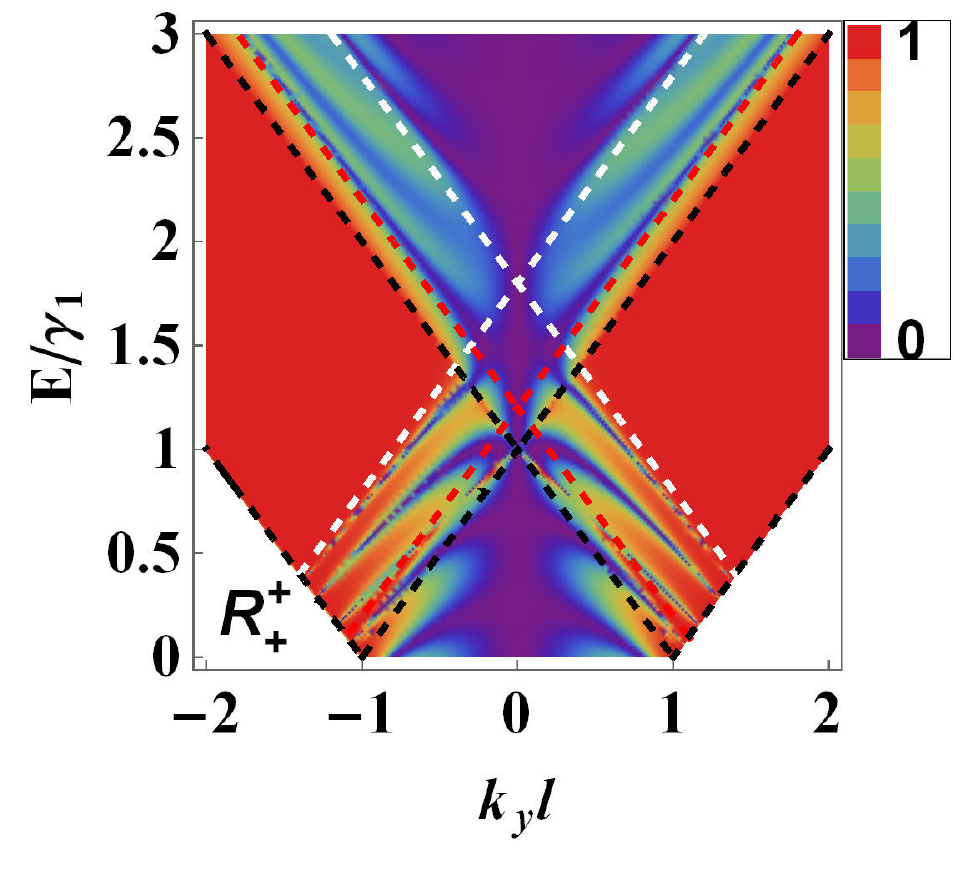}\\
\includegraphics[width=1.5 in]{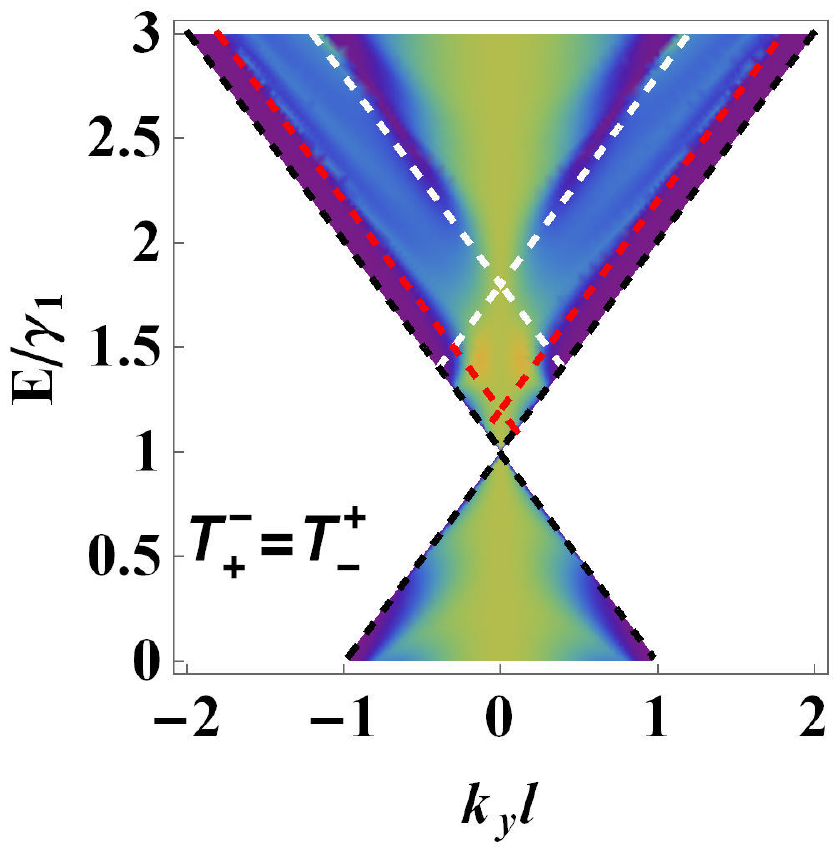}\ \
\includegraphics[width=1.73  in]{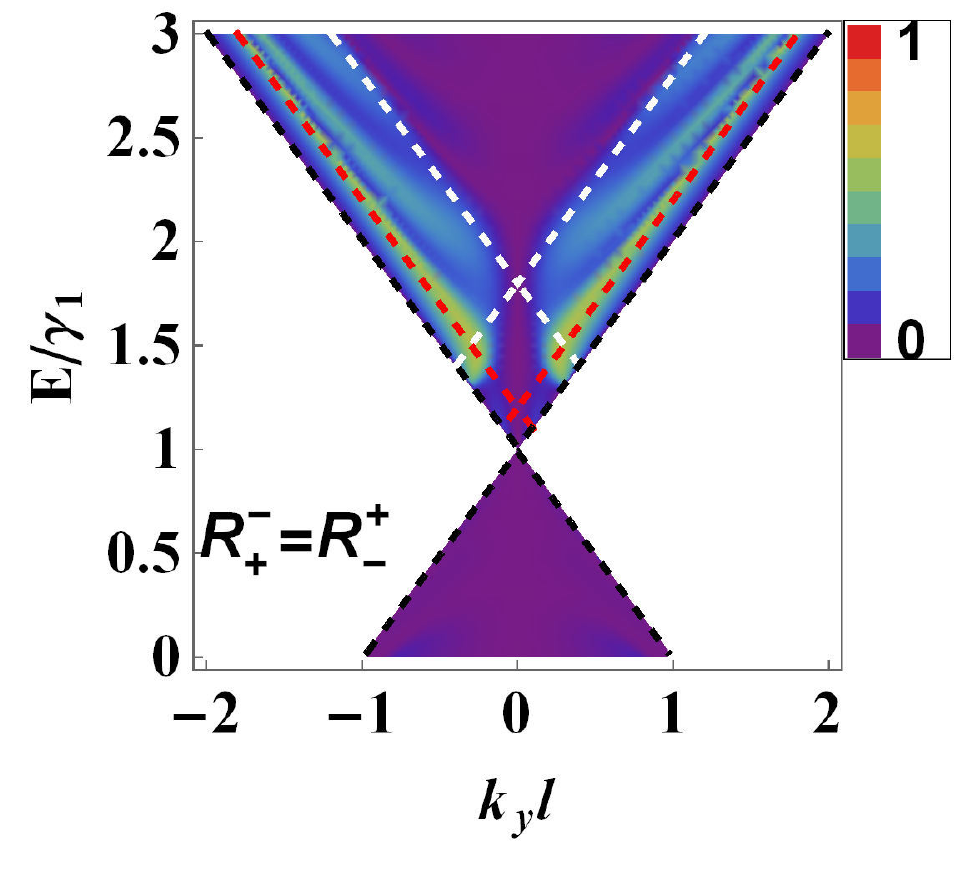}
\includegraphics[width=1.5  in]{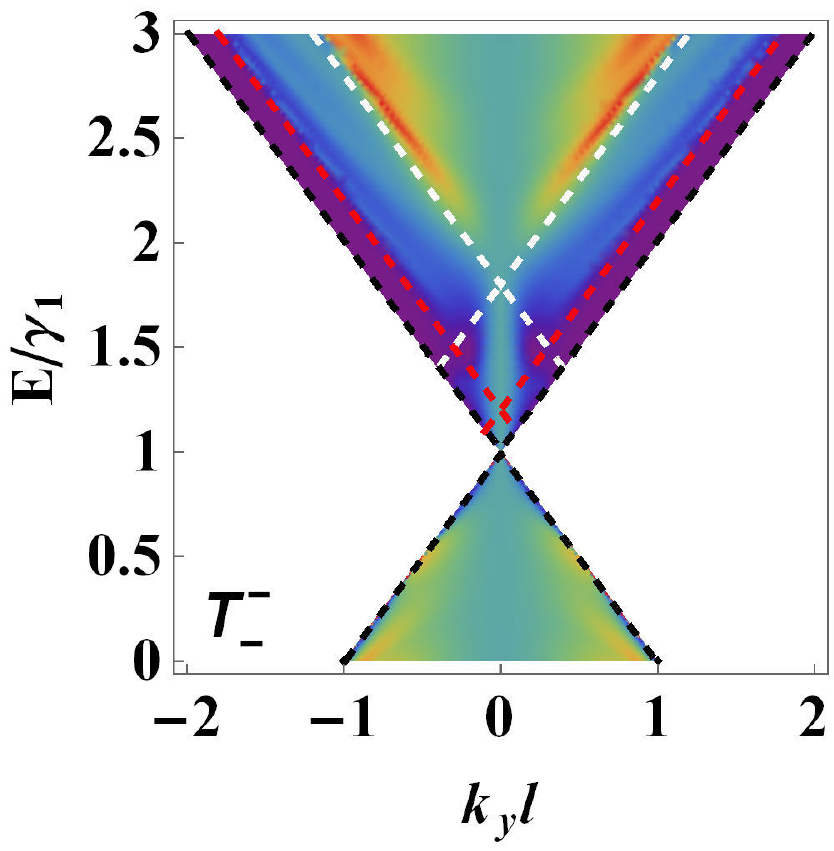}\ \
\includegraphics[width=1.73  in]{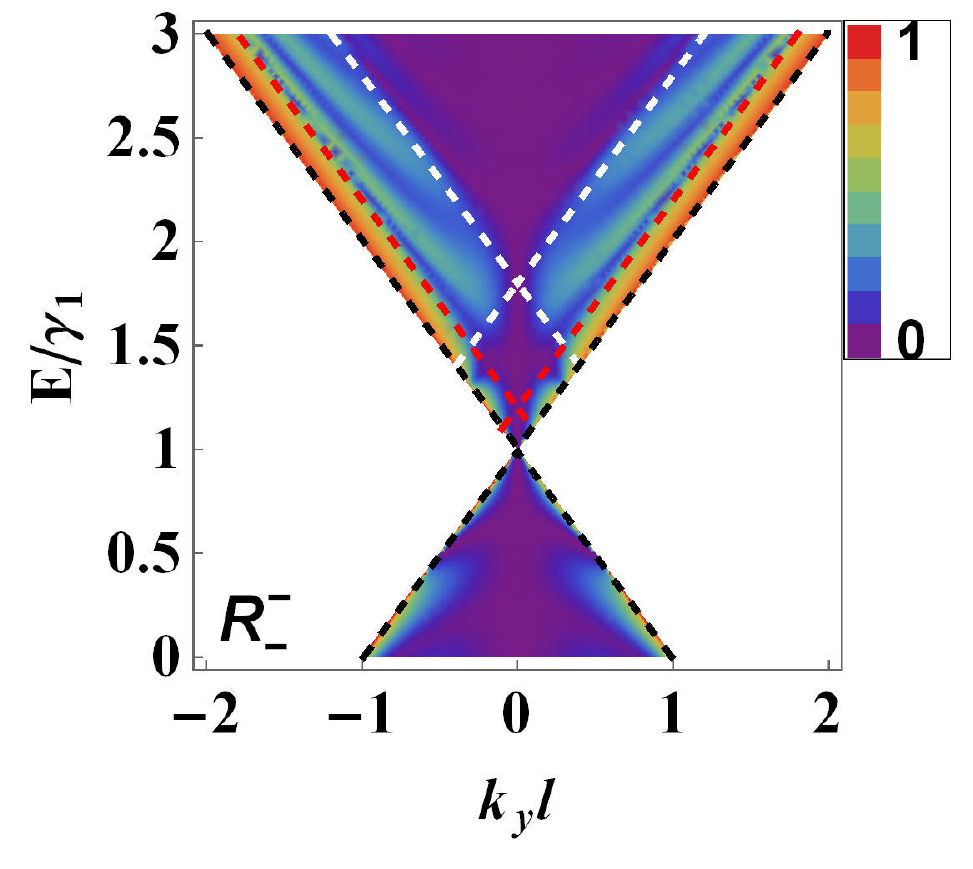}
\caption{(Colour online)   The same as in Fig. \ref{fig-AA-SL-AA}, but now with   $v_0=1.5\gamma_1$ and $ \delta =0.3\gamma_1$. Red and white dashed curves correspond to the bands of bottom and top layers of 2SL, respectively,  while
the black dashed curves are the AA-BL bands.}\label{fig-AA-SLve-AA}
\end{figure}

In Fig. \ref{fig-AA-SLve-AA} we plot the transmission and reflection probabilities for a potential strength $v_0=1.5\ \gamma_1$ and inter-layer bias $\delta = 0.3\ \gamma_1$.  The shift in the bands of the top (white) and bottom (red) layer of 2SL is due to   the inter-layer bias which couples the two Dirac cones as shown in Eq. \eqref{eq10}.
%%%%%%%%%%%%%%%%%%%%%%%%%%%%%%%%%%%%%%%%%%%%%%%%%%%%%%%%%%
Therefore, the suppression of the scattering transmission and reflection probabilities due to the protected cone transport does not hold anymore. It is, therefore, possible that scattering between different cones takes place as clarified in  the second row of Fig. \ref{fig-AA-SLve-AA}.
%%%%%%%%%%%%%%%%%%%%%%%%%%%%%%%%%%%%%%%%%%%%%%%%%%%%%%%%%%
\subsubsection{ Conductance }
\begin{figure}[t]
\vspace{0.4cm}
\centering \graphicspath{{./Figures1//SL-AA/}}
\includegraphics[width=1.65  in]{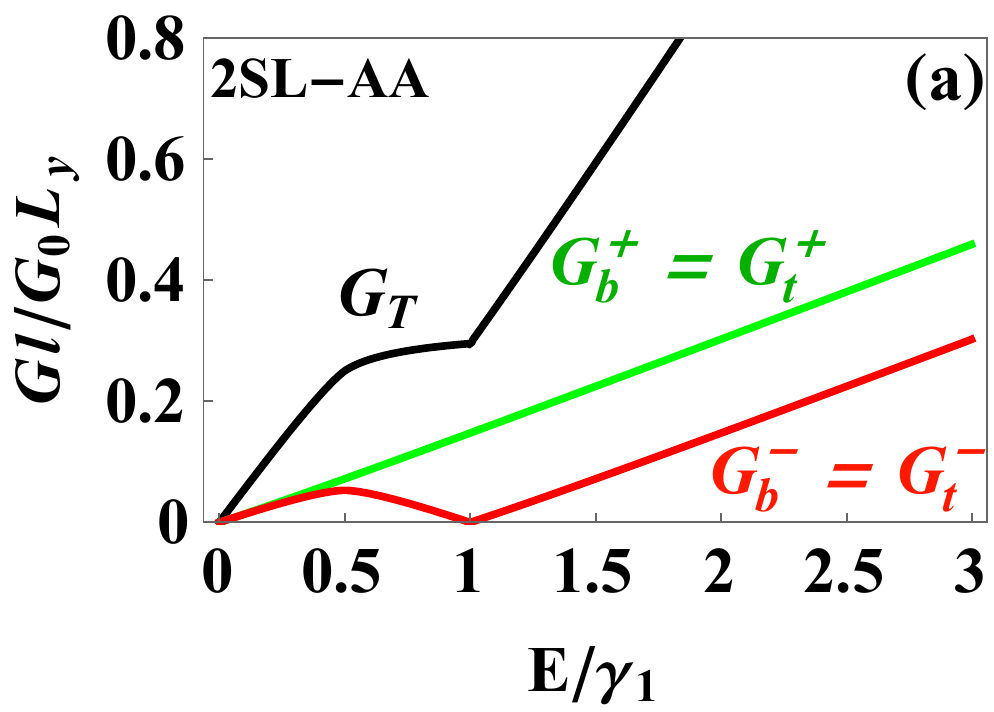}\ \
\includegraphics[width=1.65  in]{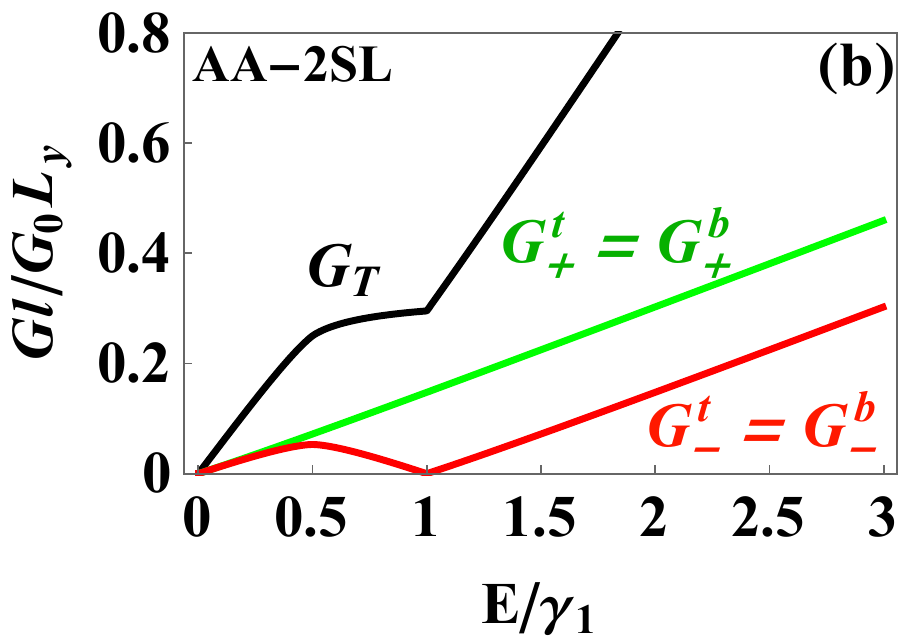}\\
\includegraphics[width=1.65 in]{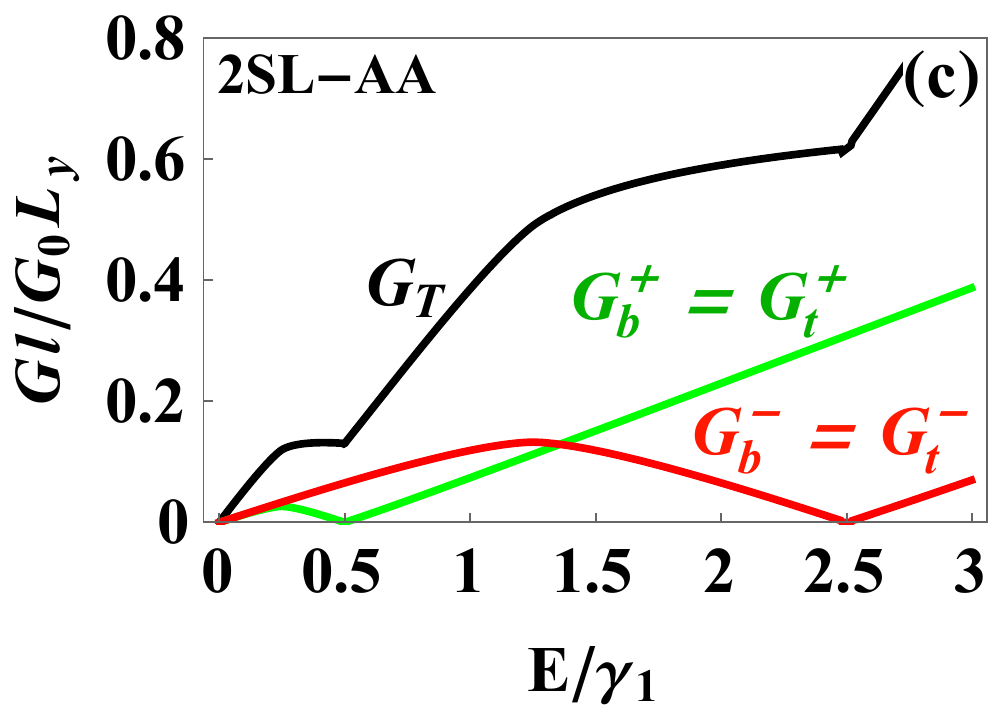}\ \
\includegraphics[width=1.65  in]{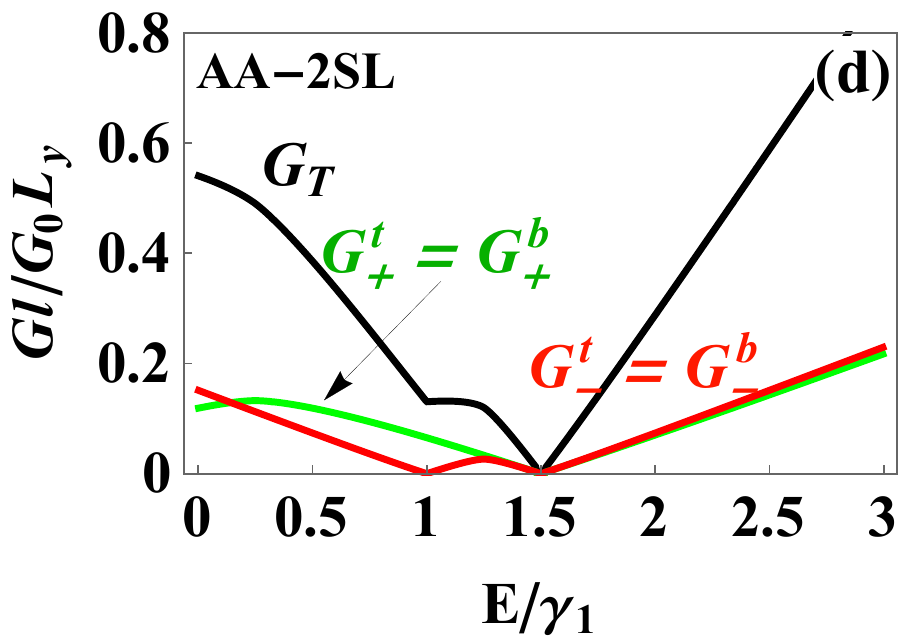}\\
\includegraphics[width=1.65  in]{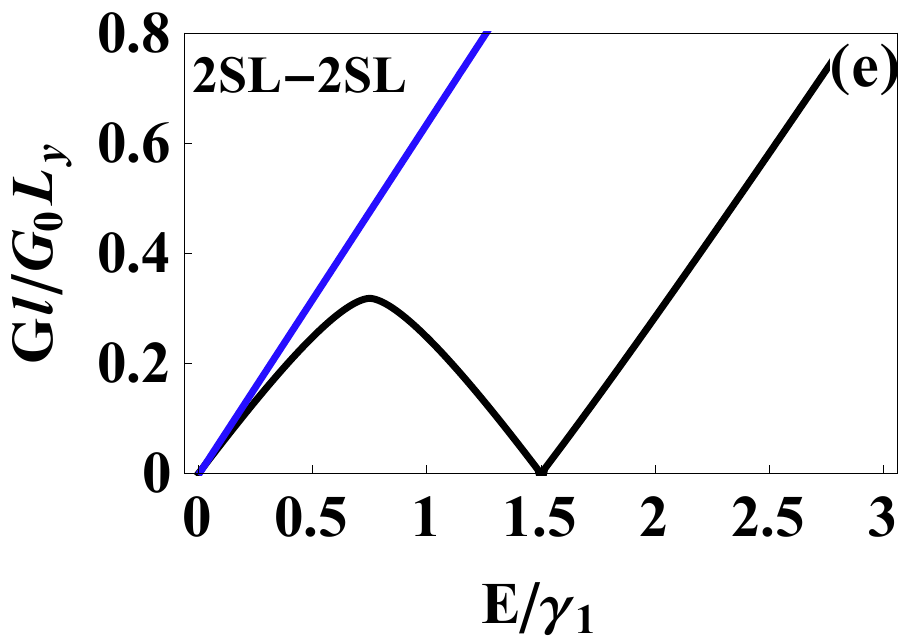}\ \
\includegraphics[width=1.65  in]{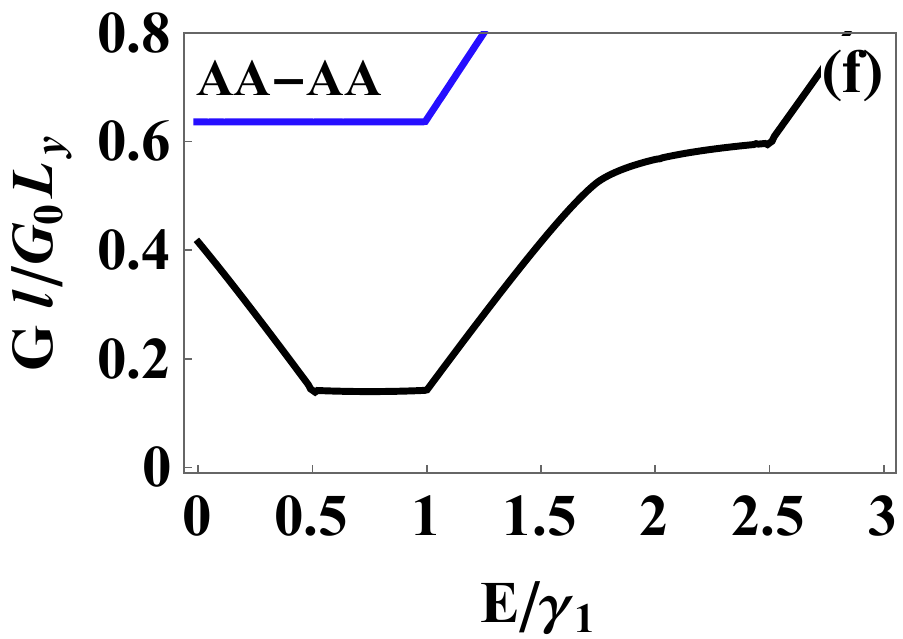}
\caption{(Colour online) Conductance of two-block system for different magnitudes of the applied gate: (a, b) $v_0= \delta =0$, (c, d) $v_0=3\gamma_1/2$, $\delta =0$ . $G_T$ is the total conductance  obtained by  summation of all possible channels,  (e, f) the total conductance for 2SL-2SL and AA-AA junctions, respectively, with $v_0=0$ (blue curves) and $v_0=1.5\gamma_1$ (black curves).}\label{SL-AA-G}
\end{figure}

\begin{figure}[t]
\vspace{0.4cm}
\centering \graphicspath{{./Figures1//AA-SL-AA/}}
\includegraphics[width=1.65  in]{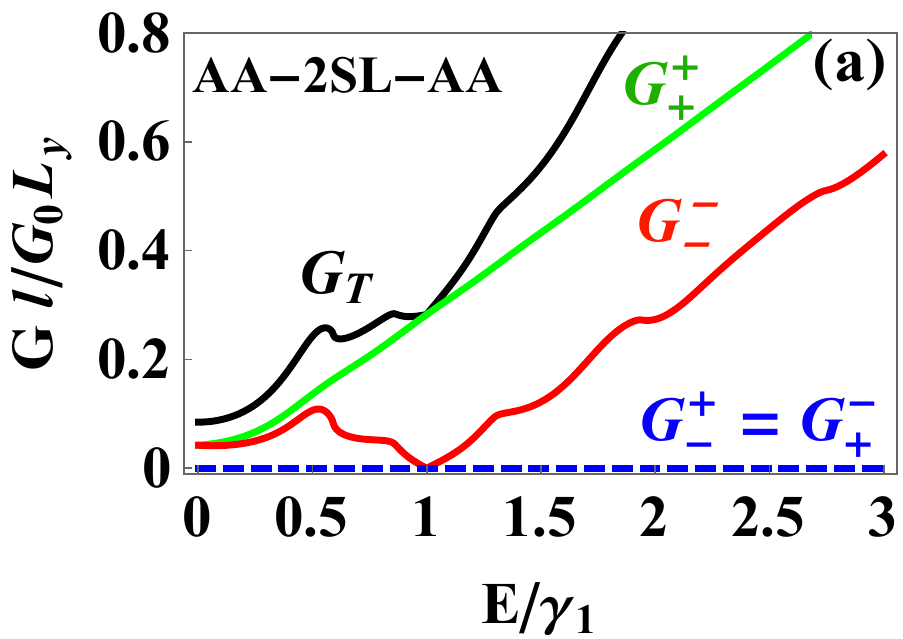}\ \
\includegraphics[width=1.65  in]{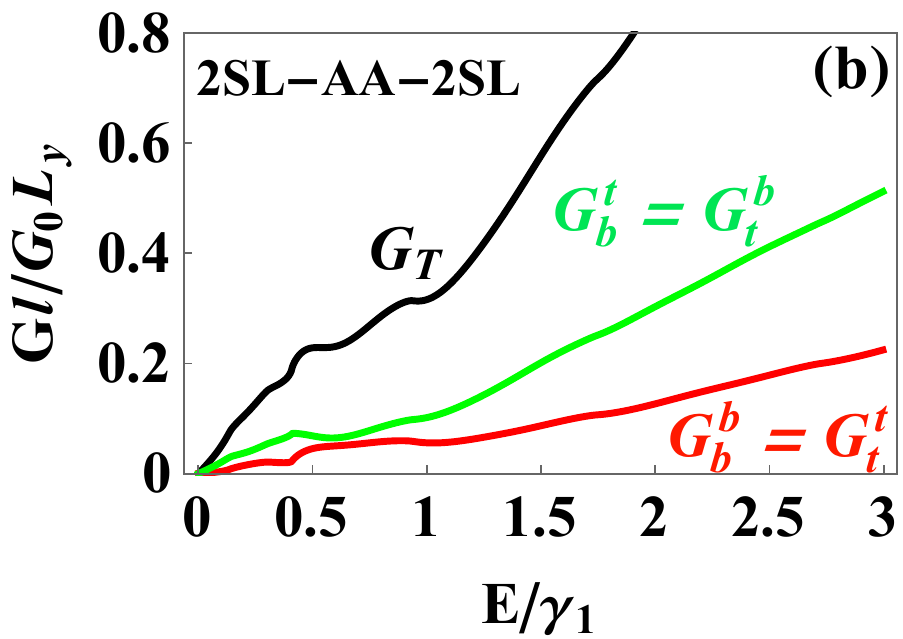}\\
\includegraphics[width=1.65 in]{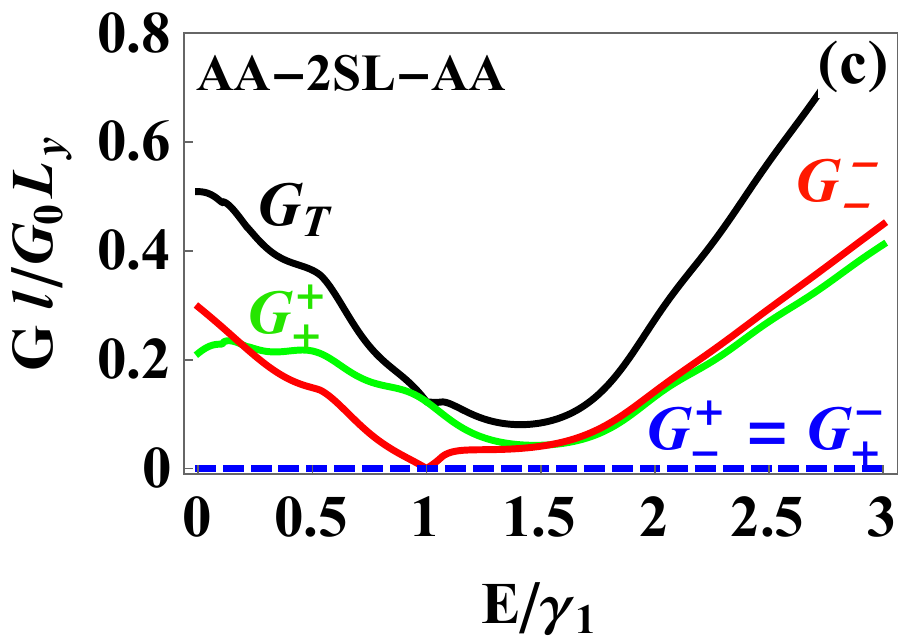}\ \
\includegraphics[width=1.65  in]{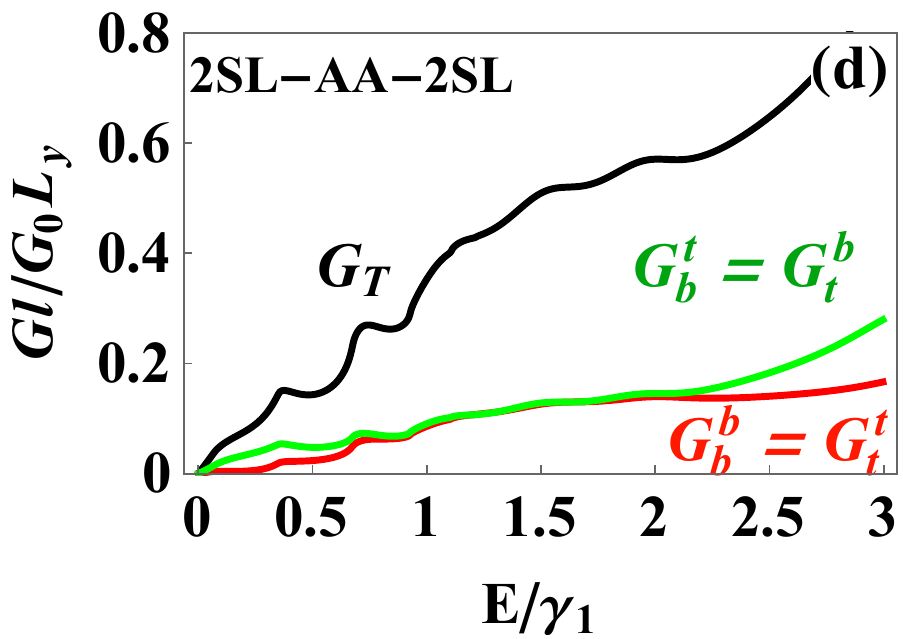}\\
\includegraphics[width=1.65  in]{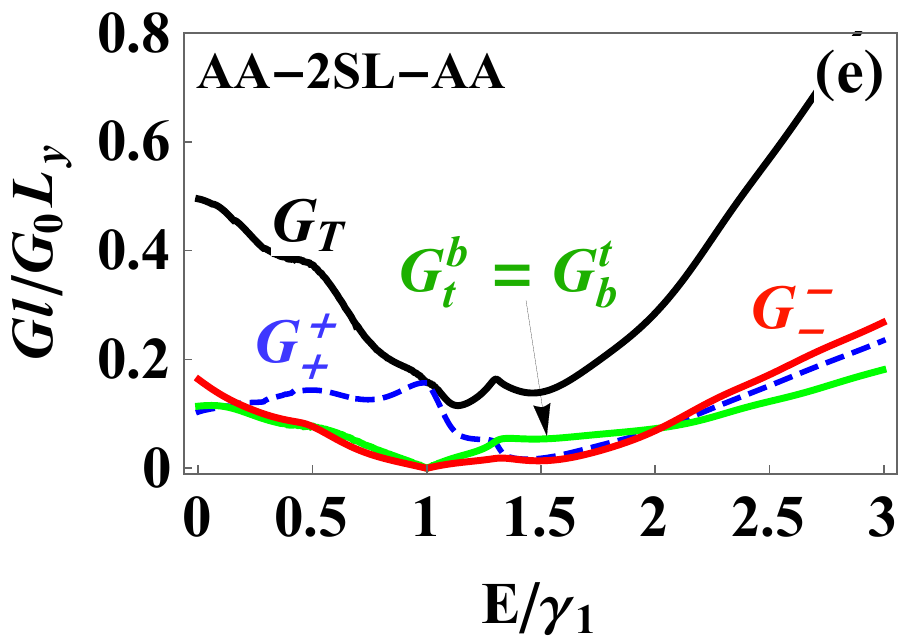}\ \
\includegraphics[width=1.65  in]{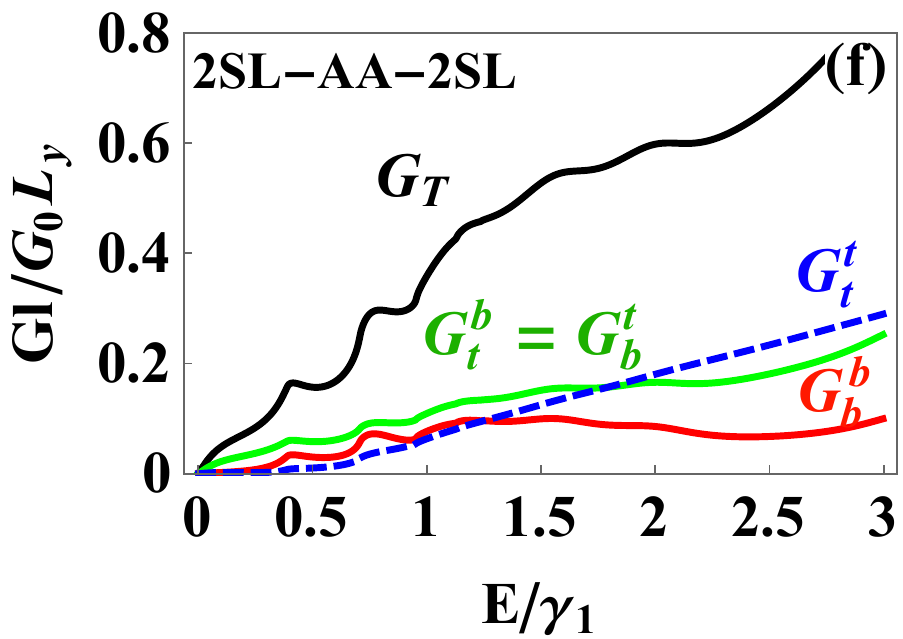}\\
\caption{(Colour online) Conductance of three-block system with different magnitudes of the applied gate: (a, b) $v_0=\ \delta =0$, (c, d) $v_0=3\gamma_1/2$, $\delta =0$ and (e, f) $v_0=3\gamma_1/2$, $\delta =0.6\gamma_1$. $G_T$ is the total conductance  obtained by  summation of all possible channels.   }\label{SL-AA-SL-G}
\end{figure}

%%%%%%%%%%%%%%%%%%%%%%%%%%%%%%%%%%%%%%%%%%%%%%%%%%%%%%%%%%
The conductance of two and three-block systems is shown in Figs. \ref{SL-AA-G} and \ref{SL-AA-SL-G}, respectively.
For the two systems 2SL-AA and AA-2SL with pristine AA-BL and 2SL, the conductance for different channels is shown in Figs. \ref{SL-AA-G}(a, b). It shows that the conductance of these two systems are identical. Referring to Figs. \ref{polar-SL-AA}(a, c)  we notice that the transmission probabilities for pristine 2SL-AA and AA-2SL are quite different. However, the corresponding conductances (see Fig. \ref{SL-AA-G}) exhibit  time reversal symmetry in spite of the fact that the domain wall separates two different systems. This is a strong point which  can be verified experimentally even in the case of zero electrostatic potential.     

Adding an electrostatic potential to one of the two sides leads to different behavior in the conductance of the above mentioned two systems as depicted in Figs. \ref{SL-AA-G}(c,d). In Fig. \ref{SL-AA-G}(c) the charge carriers  incident from 2SL and impinging on AA-BL whose bands are  shifted by $v_0$. Each conductance channel gives zero at $E=0$ due to the absence of propagating states in the 2SL at this energy, even though there are propagating states available in AA-BL corresponding to two cones. We note also that $G_{b}^\pm = G_{t}^\pm$ are almost zero at upper and lower cones  $v_0\pm \gamma_1$ as a result of the absence of states at these points as seen in Fig. \ref{SL-AA-G}(c).
In Fig. \ref{SL-AA-G}(d) we see that the conductance of different channels is not zero in contrast to the previous   case because here at $E=0$ there are propagating states  available in both AA-BL and 2SL.
Furthermore, all channels  have one minimum, due to the lack of states, at $E = v_0$ which corresponds to the Dirac cone in  2SL shifted by $v_0$ while $G_-^{t/b}$ has also another minimum at the upper cone $E=\gamma_1$ as shown in Fig. \ref{SL-AA-G}(d).
Finally,   for  comparison we add in Figs. \ref{SL-AA-G}(e, f) the conductance that will be measured in the absence of a domain wall  for 2SL-2SL and AA-AA junctions with $v_0=0$ (blue curves). Our results indicate   that domain walls are experimentally identifiable channels even in the absence of a gate. As a reference we also calculate the total conductance in the presence of an electrostatic potential   ($v_0=1.5\gamma_1$) as shown with black curves  in Figs. \ref{SL-AA-G}(e, f) which corresponds, in this case, to the usual p-n junctions in single-layer graphene  and AA-BL, respectively.   

The conductance of three-block systems is shown in Fig. \ref{SL-AA-SL-G} where left and right panels  correspond to AA-2SL-AA and  2SL-AA-2SL structure, respectively.
Protected cone transport leads to zero conductance in the scattered channels $G_-^+ = G_+^-=0$ as shown in Fig. \ref{SL-AA-SL-G}(a). A close inspection also reveals that  $G_-^- = G_+^+$ at $E=0$ with finite and non-zero values, regardless of the fact that in the 2SL region there are no available propagating states. This is attributed to the evanescent modes in 2SL at $E=0$ which are responsible for  ballistic transport in graphene\cite{Snyman}. We thus also expect that $G_-^-$ (red curve in Fig. \ref{SL-AA-SL-G}(a)) should be exactly zero at the Dirac cone $E = \gamma_1$ as a result of the absence of propagating states in the leads at this energy.

By shifting the bands of 2SL  using a local potential with strength $v_0 = 1.5\gamma_1$, a local minimum appears in the conductance $G_T$ at $E = v_0$ which corresponds  to the position of the charge-neutrality point in 2SL as shown in Fig. \ref{SL-AA-SL-G}(c). This minimum can be obtained by aligning the upper  cone in AA-BL and the Dirac cone in 2SL such that they are located at the same energy, this can be achieved by choosing $v_0=\gamma_1$. The main difference introduced by applying an inter-layer bias is the broken  protected cone  transport where now $G_+^-=G_-^+\neq0$ as depicted in Fig. \ref{SL-AA-SL-G}(e). For completeness, we performed similar calculations but now with 2SL as  the leads (2SL-AA-2SL) and the results for  the conductance with pristine, gated and biased AA-BL are shown in Figs. \ref{SL-AA-SL-G}(b, d, f), respectively. Here, all conductance channels are zero at $E=0$   such that $G_t^t=G_b^b$ and $G_t^b=G_b^t$ as shown in Figs. \ref{SL-AA-SL-G}(b, d). Similarly, the main features in  Fig. \ref{SL-AA-SL-G}(f) are in qualitative  agreement with those shown in Figs. \ref{SL-AA-SL-G}(b, d) but now the tunnelling equivalence through the same channel is broken so that $G_t^t\neq G_b^b$. This is a direct consequence of the perpendicular electric field which leads to the breaking  of the inter-layer sublattice equivalence. The peaks appearing in the total conductance are due to the finite size of the AA-BL region.
%%%%%%%%%%%%%%%%%%%%%%%%%%%%%%%%%%%%%%%%%%%%%%%%%%%%%%%%%%%%%%%%
\subsection{ AB-Stacking }
%%%%%%%%%%%%%%%%%%%%%%%%%%%%%%%%%%%%%%%%%%%%%%%%%%%%%%%%%%%%%%%%
\subsubsection{ 2SL-AB/AB-2SL}
\begin{figure}[t]
\vspace{0.4cm}
\centering \graphicspath{{./Figures1//SL-AB/}}
\includegraphics[width=1.25  in]{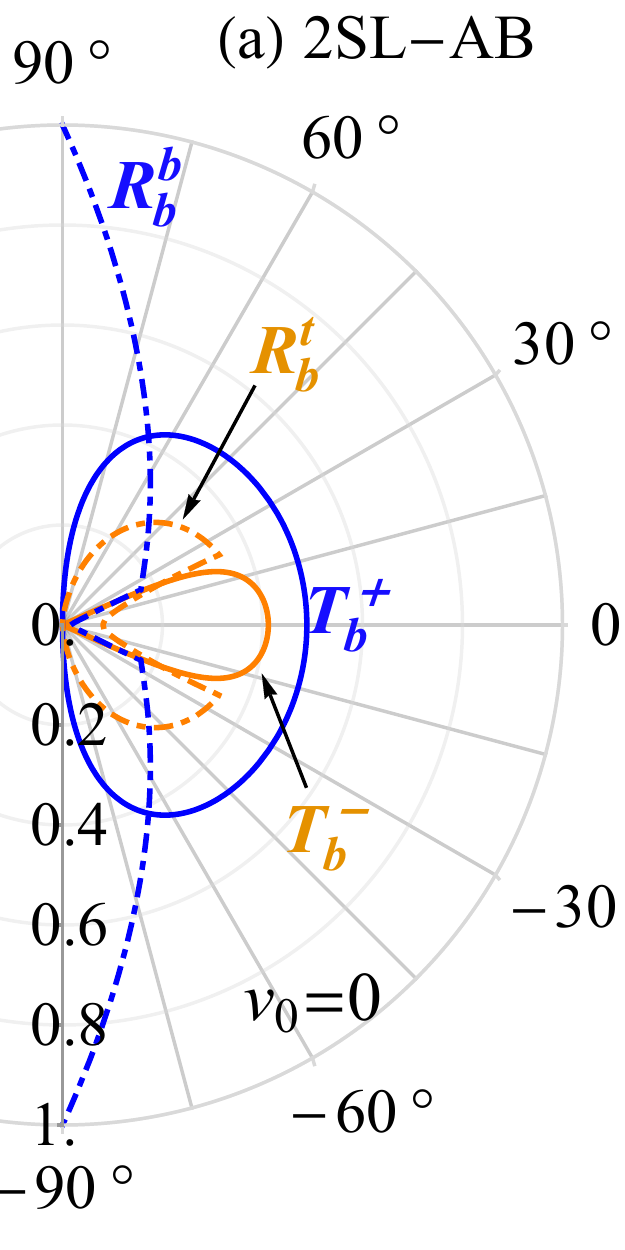}\ \ 
\includegraphics[width=1.25    in]{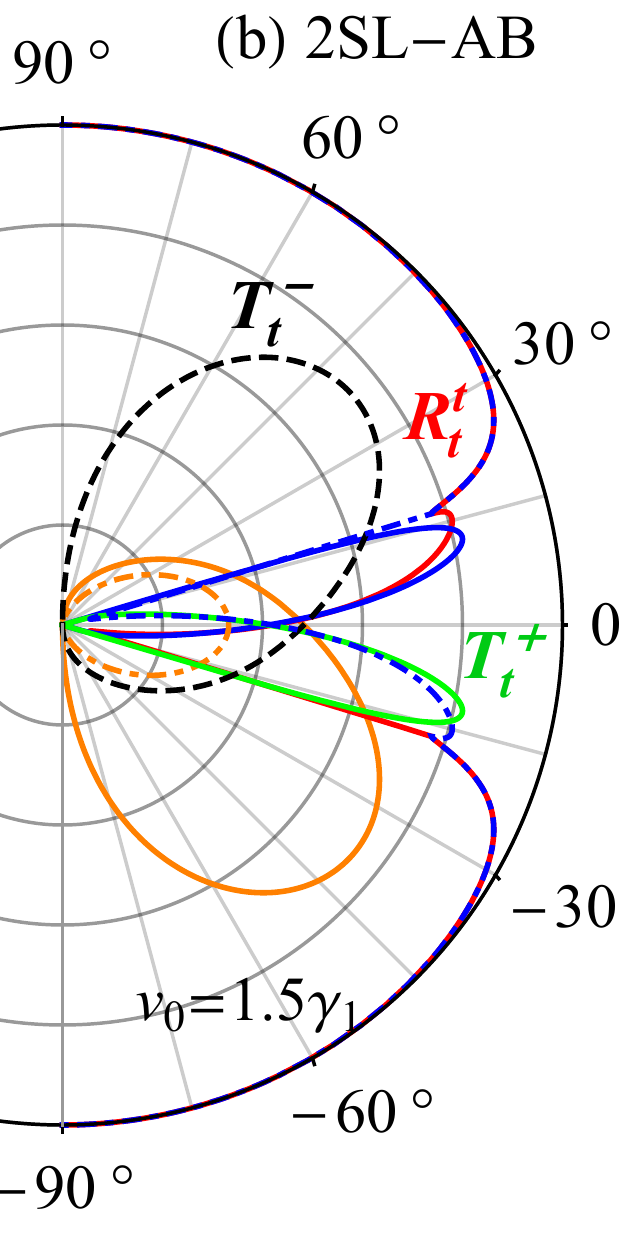}\ \
\includegraphics[width=1.25 in]{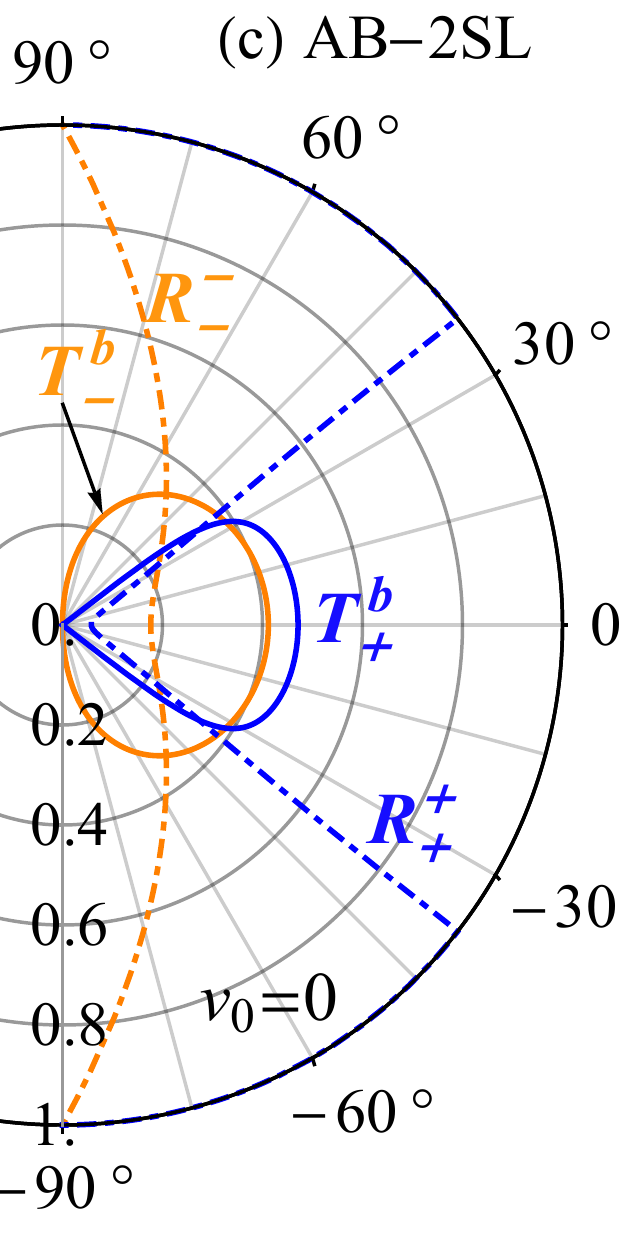}\ \ 
\includegraphics[width=1.25 in]{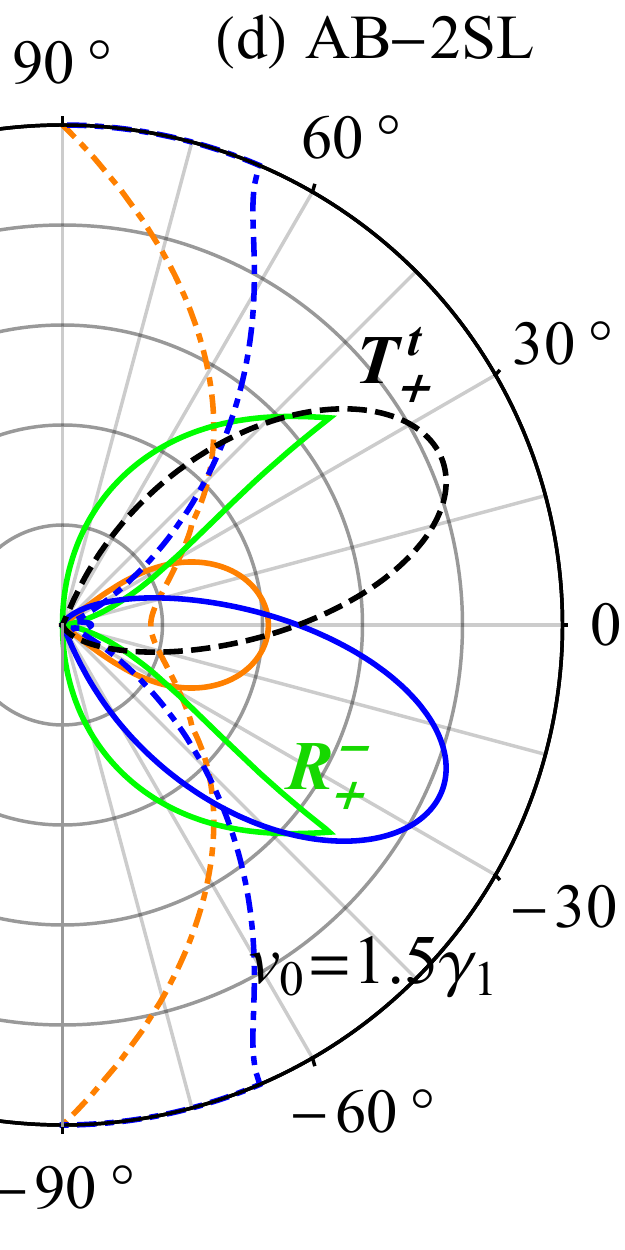}
\caption{(Colour online)   The angle-dependent transmission and reflection probabilities through (a, b) 2SL-AB and (c, d) AB-2SL junctions. The systems in (b, d) are the same as in (a, c), respectively, but where  the right side of the junction is subjected to an electrostatic potential  of strength $ v_0= 1.5\gamma_1. $ In (a) $E=1.2\gamma_1$ for all channels while in (b) $E=1.7\gamma_1$ for $T_{b(t)}^+$ and $E=0.6\ \gamma_1$ for the rest of the channels and in (c, d) $E=(0.6, 1.7)\gamma_1$ for $R_+^+/T_+^{b(t)}$ and $R_-^-/T_-^{b(t)}$, respectively. We choose  energy values in (b, d) such that they  correspond to only one propagating mode in the AB-BL region.      }\label{polar-SL-AB}
\end{figure}

%%%%%%%%%%%%%%%%%%%%%%%%%%%%%%%%%%%%%%%%%%%%%%%%%%%%%%%%%%%%%%%%
In this section, we evaluate how the stacking of the connected region changes the transport properties across a domain wall. The angle-dependent transmission and reflection probabilities for pristine systems 2SL-AB are plotted in Fig. \ref{polar-SL-AB}(a). The charge carriers can be incident from the two layers in the 2SL structure and impinge on  AB-BL where, depending on their energy, they can access only one propagating mode $k^+$ or two $k^\pm$  if the energy is large enough.  Scattering from the top or bottom layer of  2SL into one of these modes is equivalent $T_{t}^\pm=T_{b}^\pm$  as well as backscattering  $R_{t}^{t(b)}=R_{b}^{b(t)}$ and hence, as before, layer symmetry is preserved (see Fig. \ref{polar-SL-AB}(a)). In Fig. \ref{polar-SL-AB}(b) we show results with the AB-BL region subjected to an electrostatic potential of strength $v_0=1.5\gamma_1$. Surprisingly, we see that the layer symmetry is broken and an asymmetric feature with respect to  normal incidence shows up in the transmission and non-scattered reflection  probabilities, see Appendix \ref{Sec:Appendix}, such that $[T/R](\phi)=[T/R](-\phi).$ For example,  $T_{b}^\pm(\phi)=T_t^\pm(-\phi)$ as well as the non-scattered reflection channels $R_{b}^b(\phi)=R_{t}^t(-\phi)$ as discussed in Sec. \ref{Symmetry}. 
This asymmetric feature can be understood by resorting to the bands on both sides of the junction, where due to the electrostatic potential the band alignment of 2SL and AB-BL is altered. In this case, the center  of the AB-BL band is shifted upwards in energy with respect to the crossing of the 2SL\cite{34} energy bands. The origin of such asymmetry is a direct consequence of the asymmetric coupling in AB-BL which leads to  shifting of the bands by $\gamma_1$. Therefore, at low energy $\left\vert E-v_{0} \right\vert< \gamma_1$ there is only one propagating mode $k^{+}$ (i.e one type of charge carrier  ) and consequently only $T_{b(t)}^+$ is available. For larger energy, on the other hand, there are two modes available giving rise to four channels $T_{b(t)}^\pm$.
\begin{figure}[tb]
\vspace{0.4cm}
\centering \graphicspath{{./Figures1/SL-AB-SL/}}
\includegraphics[width=1.5  in]{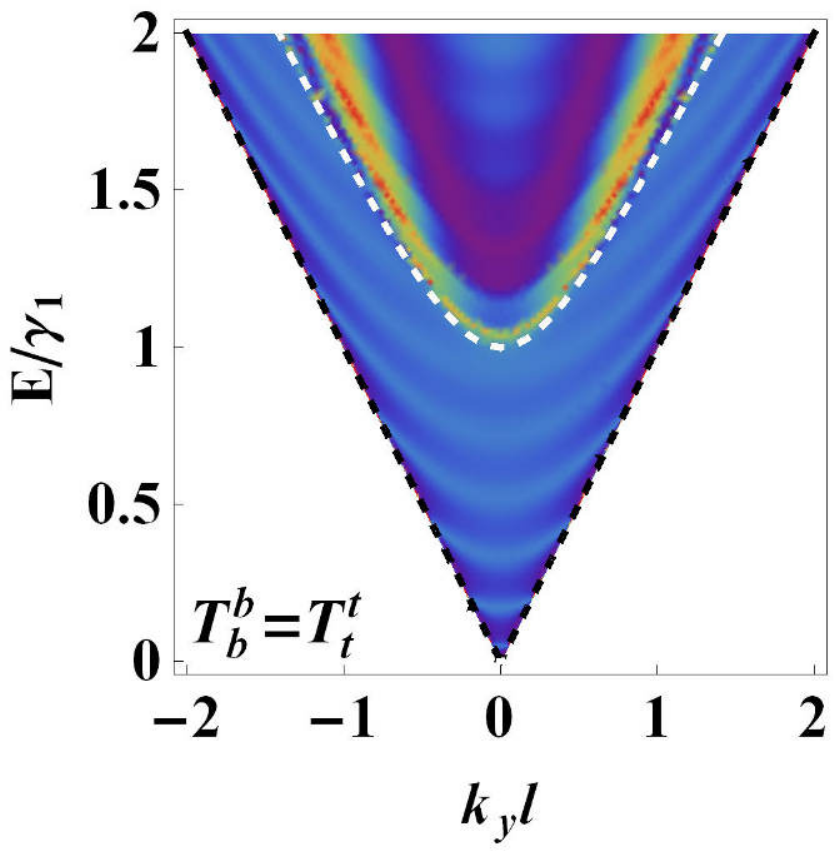}\ \
\includegraphics[width=1.73  in]{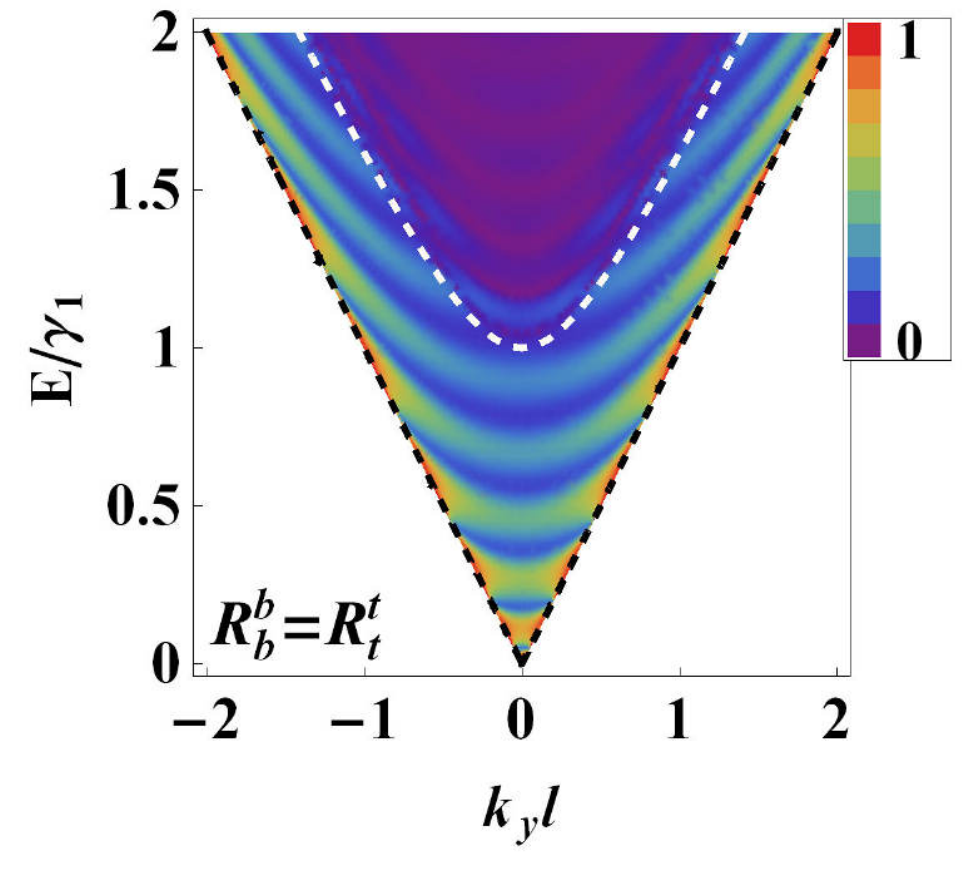}\\
\includegraphics[width=1.5 in]{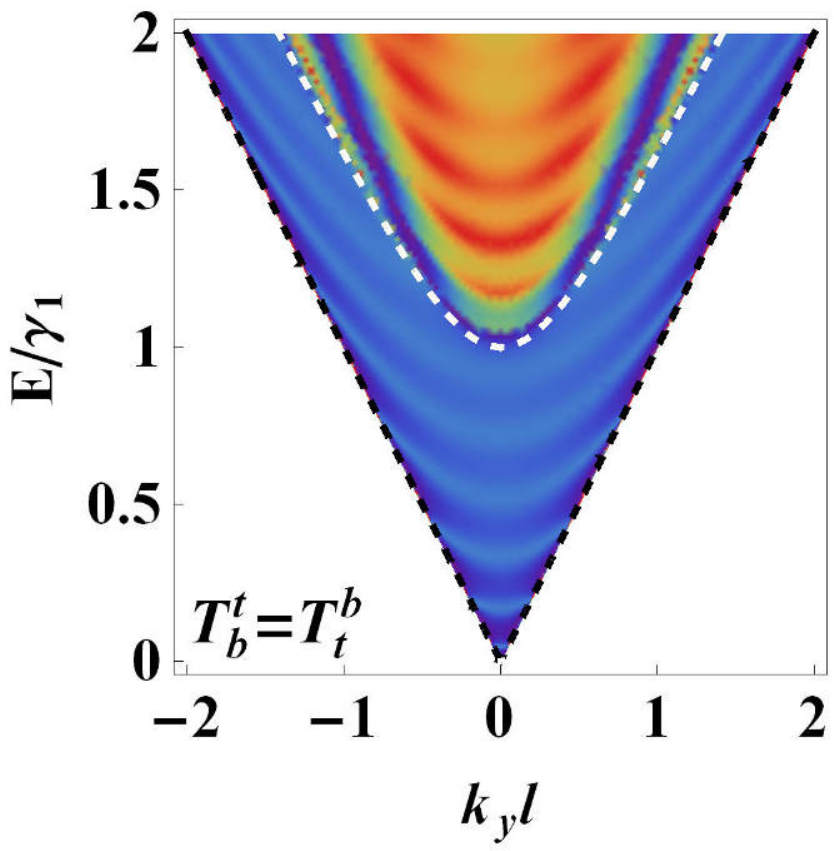}\ \
\includegraphics[width=1.73  in]{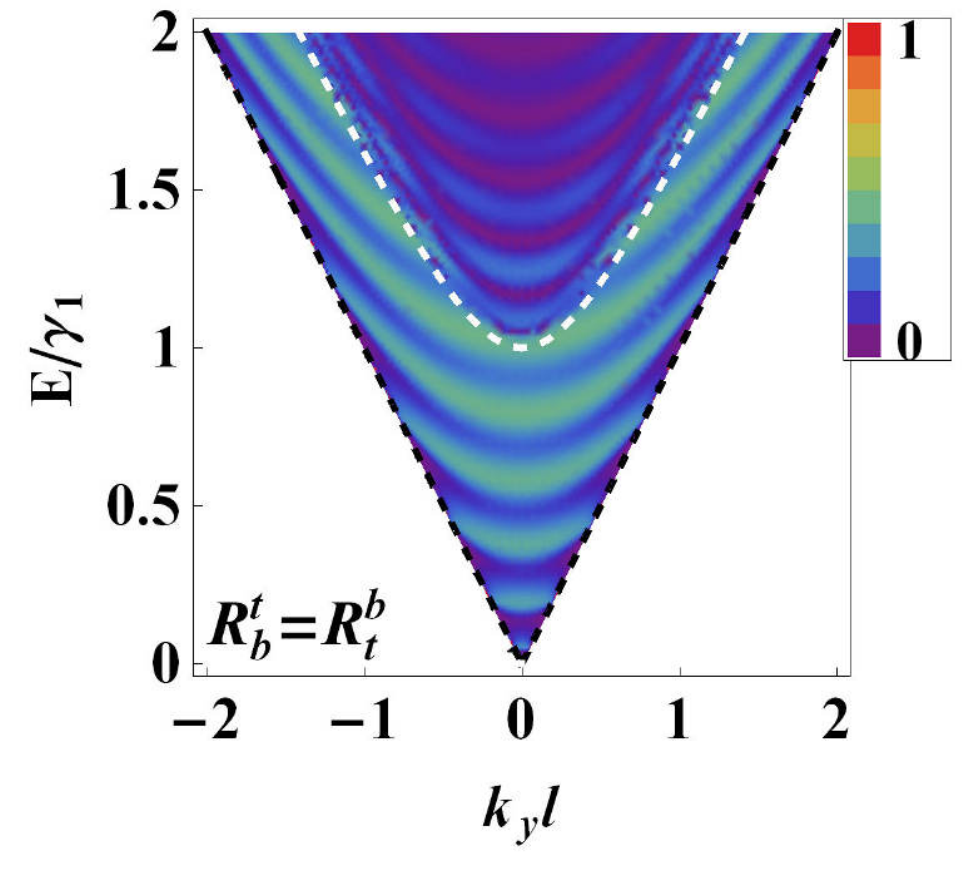}
\caption{(Colour online) Density plot of the  transmission and reflection probabilities through 2SL-AB-2SL as a function of Fermi energy and
transverse wave vector $k_y$ with  $v_0=\ \delta =0.$ }\label{SL-AB-SL}
\end{figure}

The angular asymmetry feature is present only in the region in the $(E,k_y)$-plane where there is only one propagating mode. This can be also understood as a manifestation of the asymmetric amplitude of the wave function in the AB-BL side due to the evanescent modes in this region \cite{24}.  The theory of tunnelling through an interface of monolayer and bilayer was presented earlier\cite{24} and such asymmetry was noticed as well.  Moreover, in our case there are two single layer graphene sheets connected to the bottom and top layers of the bilayer system but the  asymmetric feature in Ref. [\onlinecite{24}] will be recovered when considering only one propagation channel. For instance, the transmission probabilities $T_{t}^\pm$ and $\ T_{b}^\pm$ presented in Fig. \ref{polar-SL-AB}(b) show the same asymmetric features discussed in Ref. [\onlinecite{24}]. This asymmetry feature is reversed in the other valley, so that the total transmission or reflection averaged over both layers is symmetric as can be seen from Fig. \ref{polar-SL-AB}(b).  However, this valley-dependent angular asymmetry could also be used for the basis of a layer-dependent valley-filtering device as proposed in other works\cite{Costa2015,Rycerz2007}. 
 
The above analogy, which is discriminating between the presence of one or two modes,   applies also to the non-scattered reflection probabilities $R_{b}^b$ and $\ R_{t}^t$. These non-scattered currents are carried by the states localized on the disconnected sublattices $\alpha_2$ and $\beta_1$, as seen in Fig. \ref{fig01}. In that case, there is one traveling mode\cite{pelc} and thus, inherently, a layer asymmetric feature will be present.  In contrast, for the scattered channels $R_{b}^t$ and $\ R_{t}^b$ the charge carriers must jump between the layers of AB-BL. This  occurs through the localized states on the connected sublattices $\alpha_1$ and $\beta_2$ where there are two travelling modes and, hence, these probabilities exhibit layer symmetry  as shown in Fig. \ref{polar-SL-AB}(b).  In the AB-2SL configuration, where charge carriers incident from the AB-BL impinge on the 2SL, we show the angle-dependent  transmission and reflection probabilities in Fig. \ref{polar-SL-AB}(c) for pristine 2SL and AB-BL.

Similar to the previous configuration 2SL-AB, the results are symmetric in this case because the Dirac cones of both systems (2SL and AB-BL) are aligned. Furthermore, there is an equivalence in the transmission channels such that  $T_{\pm}^t=T_{\pm}^b$ with partial reflection associated with the non-scattered channels  $R_{-}^-$ and $R_{+}^+$. While for the scattered channels  $R_{+}^-$ and $R_{+}^-$ are almost zero. This is  due to  efficient transmission resulting from the absence of the electrostatic potential in the 2SL. An electrostatic potential   of strength $v_0=1.5 \gamma_1$ induces a scattering between the two modes in the reflection channels so that now $R_{-}^+=R_{+}^-\neq0$ as depicted in Fig. \ref{polar-SL-AB}(d). In addition, it  breaks the band alignment and gives rise to the layer asymmetry feature in the transmission probabilities $T_+^{b(t)}$  where only  one travelling mode exists i.e. $E<\gamma_1$.  Thus,  $T_-^{b(t)}$ always preserves layer symmetry in this case, see Fig. \ref{polar-SL-AB}(d), because the mode $k^-$ exists for  $E>\gamma_1$ where also the mode $k^+$  is available as discussed above. This is also the same reason that configurations consisting of AA-BL always preserve layer symmetry. Indeed, AA-BL does not have a region in the $(E,k_y)$-plane with only one propagating mode, and there are always two travelling modes for  all energies.

%%%%%%%%%%%%%%%%%%%%%%%%%%%%%%%%%%%%%%%%%%%%%%%%%%%%%%%%%%%%%%%%
\subsubsection{ 2SL-AB-2SL}
%%%%%%%%%%%%%%%%%%%%%%%%%%%%%%%%%%%%%%%%%%%%%%%%%%%%%%%%%%%%%%%%
Different configurations have been proposed to connect a single layer to the AB-stacked bilayer graphene\cite{25,34,15,Lima}. Now, two SL are connected to the AB-stacked bilayer, see Fig. \ref{intro-fig02}(a). In Fig. \ref{SL-AB-SL} we show the dependence of the transmission and reflection probabilities on the transverse wave vector $k_y$ and the Fermi energy.
It appears that all channels are symmetric with respect to normal incidence since the Dirac cones of AB and 2SL are aligned. It also implies that  scattered and non-scattered channels of the transmission and reflection  are equivalent such that $(T/R)_{b}^t=(T/R)_{t}^b$ and $(T/R)_{t}^t=(T/R)_{b}^b$ (see Fig. \ref{SL-AB-SL}).
\begin{figure}[t!]
\vspace{0.4cm}
\centering \graphicspath{{./Figures1/SL-AB-SL/}}
\includegraphics[width=1.5  in]{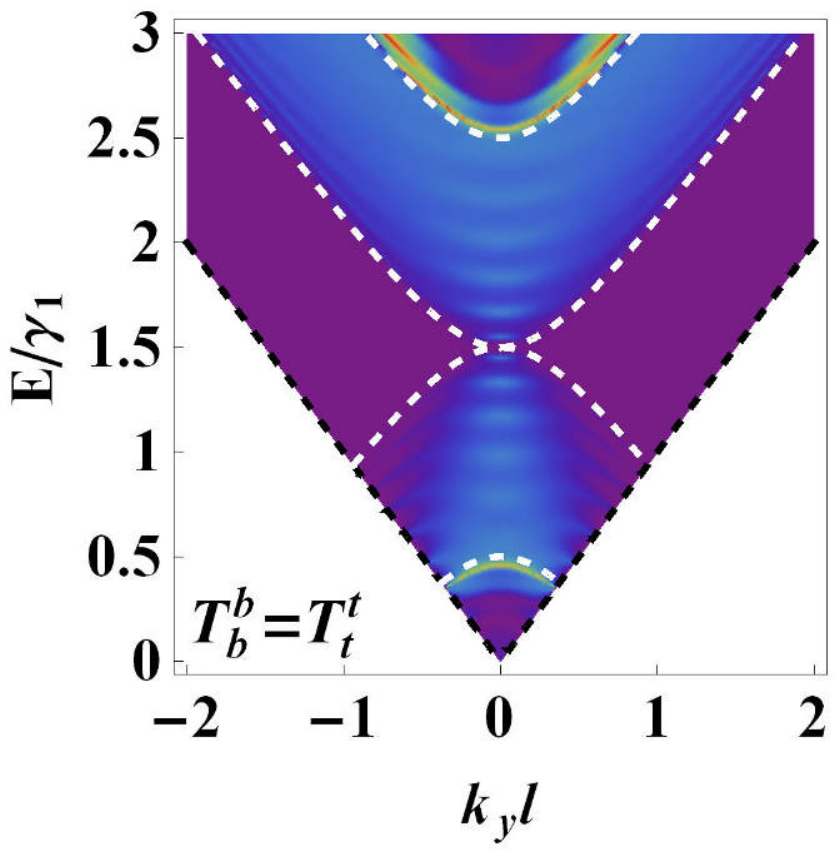}\ \
\includegraphics[width=1.73  in]{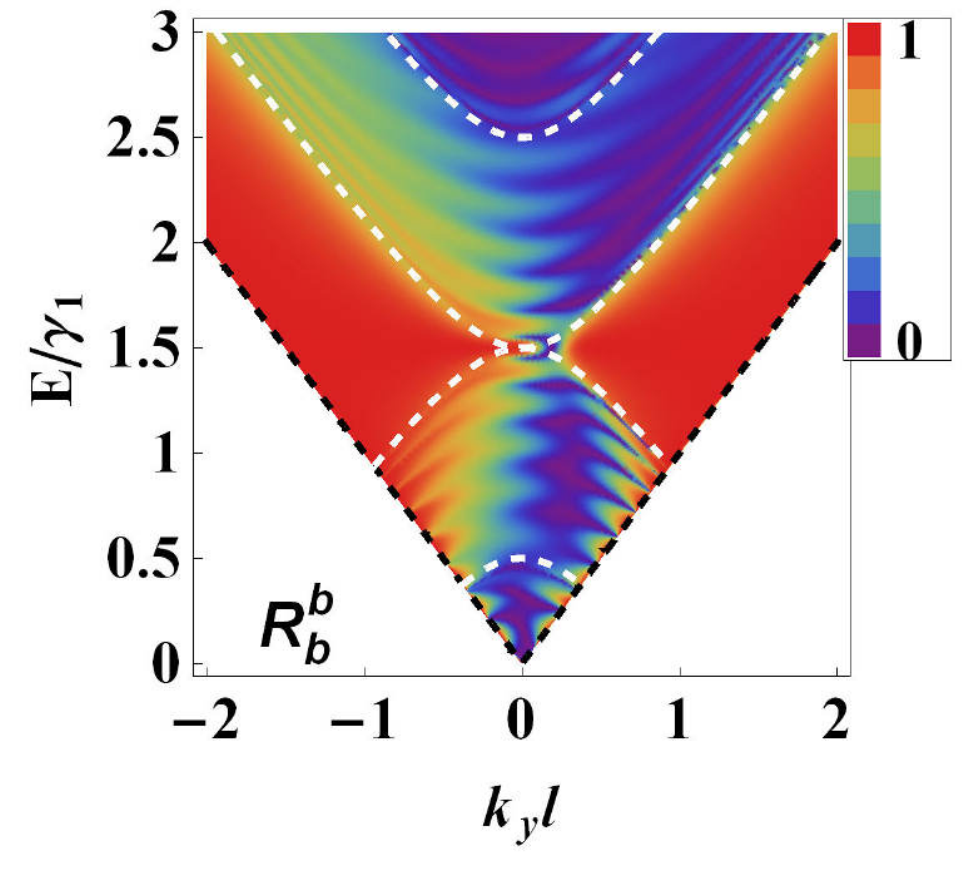}\\
\includegraphics[width=1.5 in]{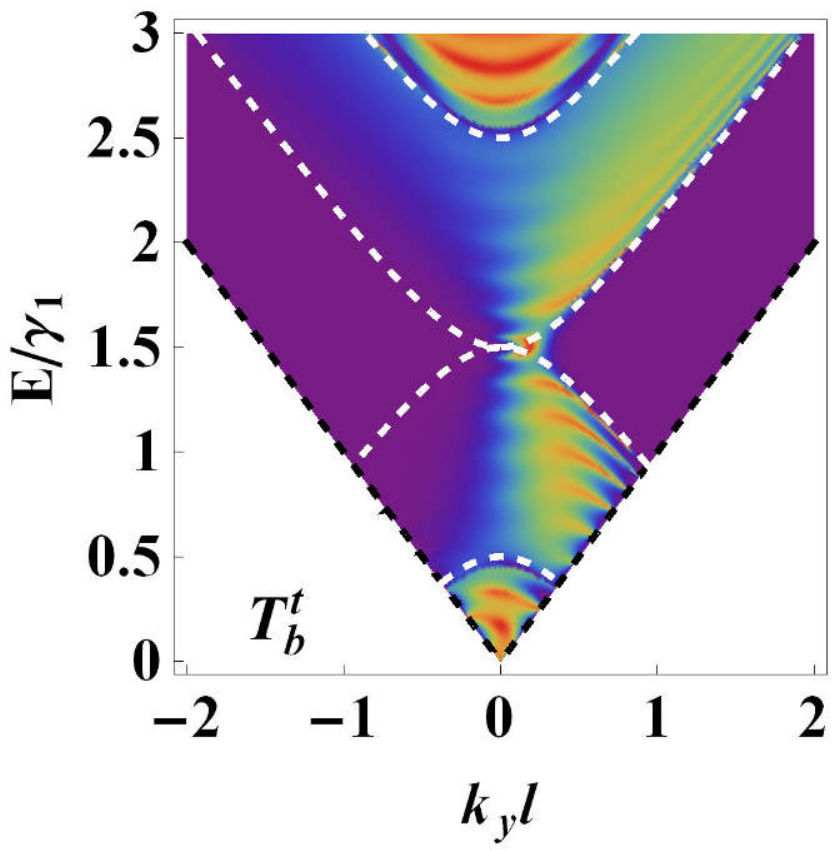}\ \
\includegraphics[width=1.73  in]{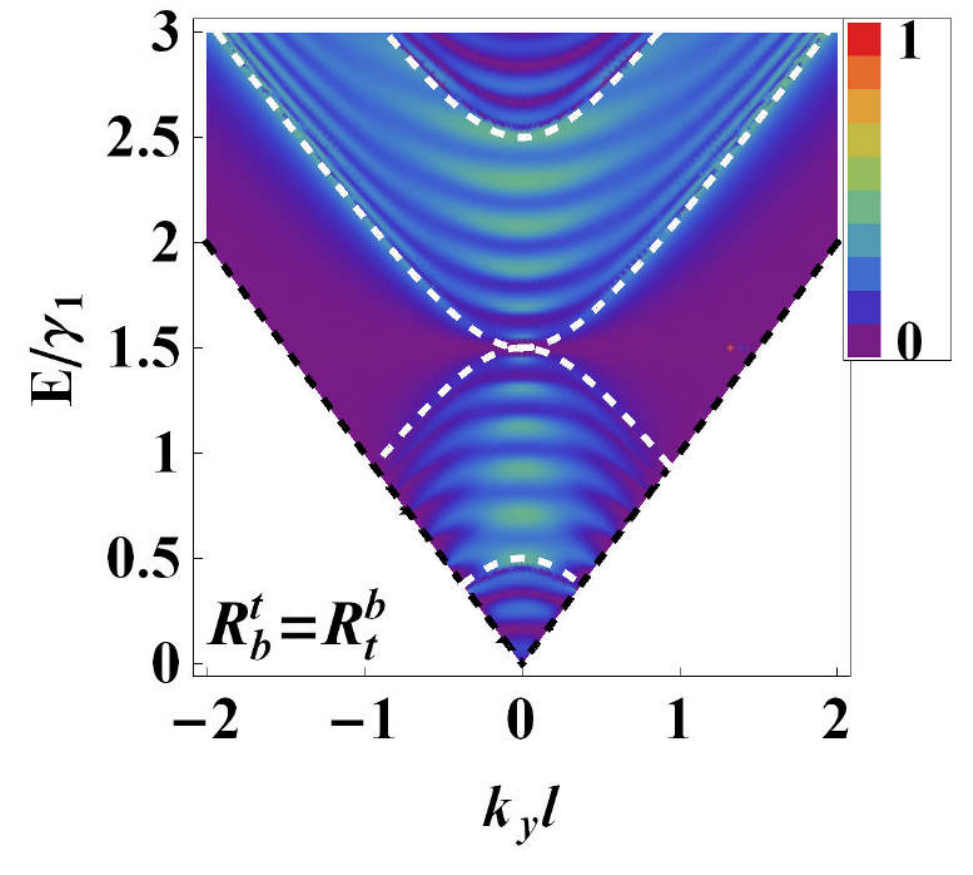}\\
\includegraphics[width=1.5  in]{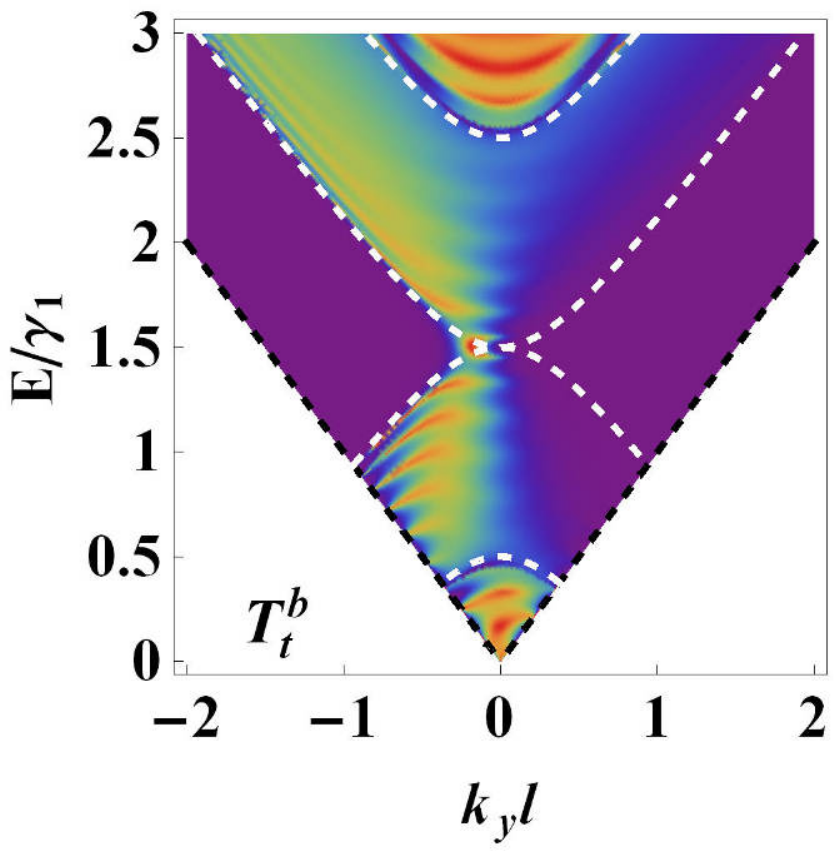}\ \
\includegraphics[width=1.73  in]{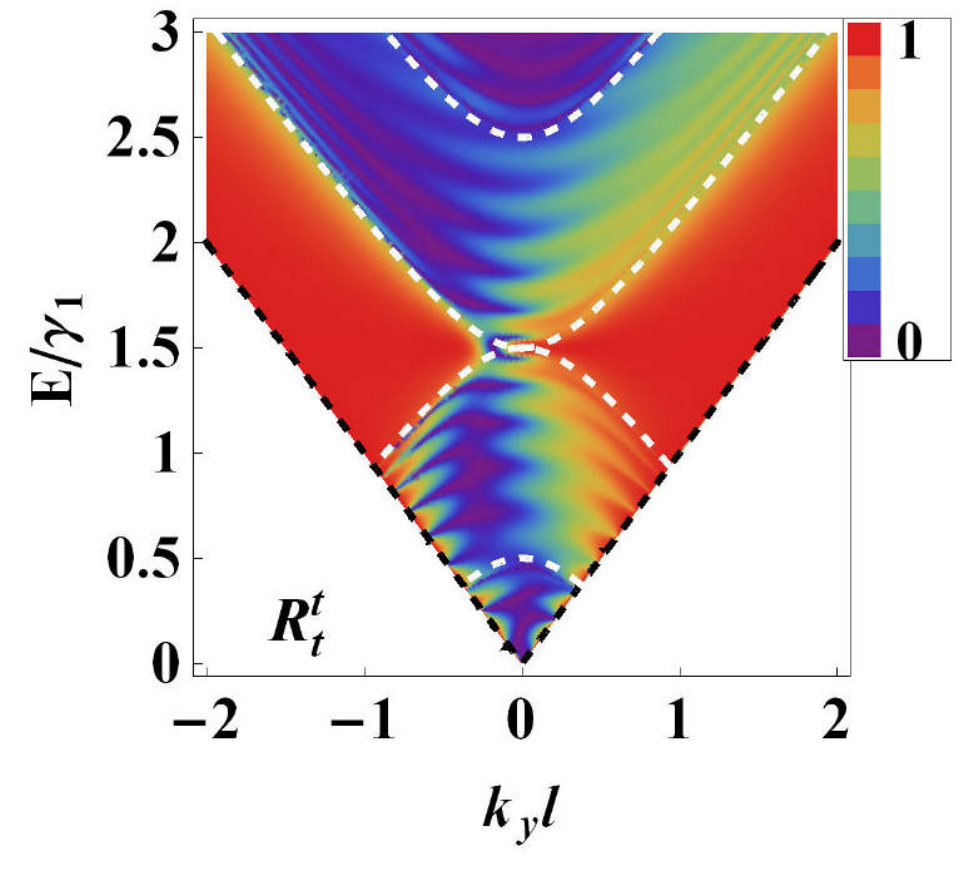}
\caption{(Colour online) The same as in Fig. \ref{SL-AB-SL}, but now with  $v_0=3\gamma_1/2$.  }\label{SL-AB-SLv}
\end{figure}
\begin{figure}[t!]
\vspace{0.4cm}
\centering \graphicspath{{./Figures1/SL-AB-SL/}}
\includegraphics[width=1.5  in]{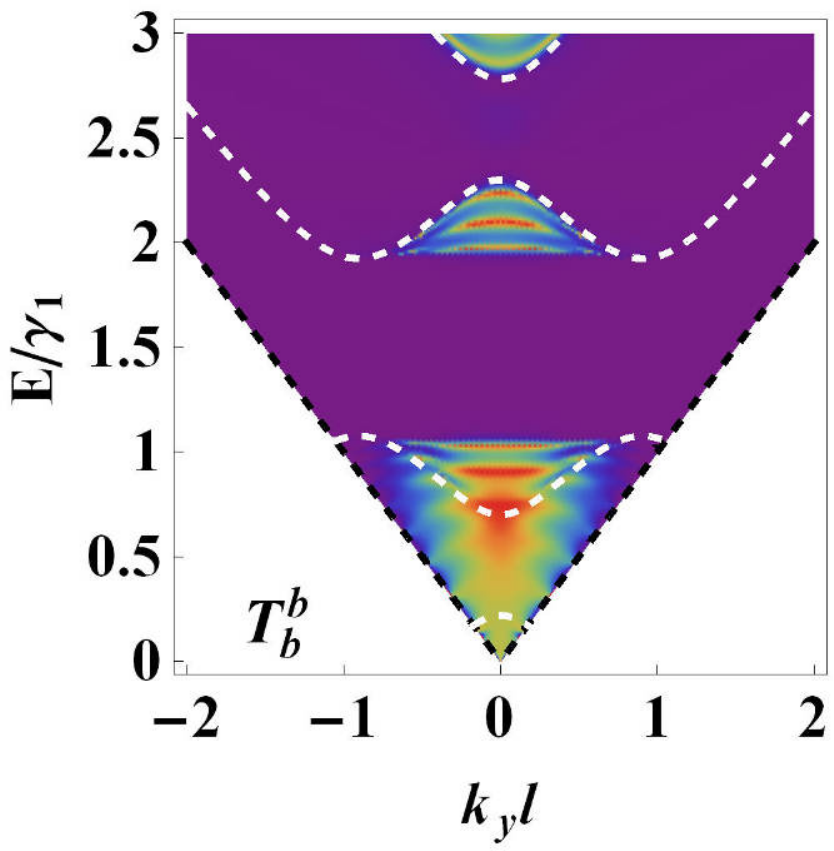}\ \
\includegraphics[width=1.73  in]{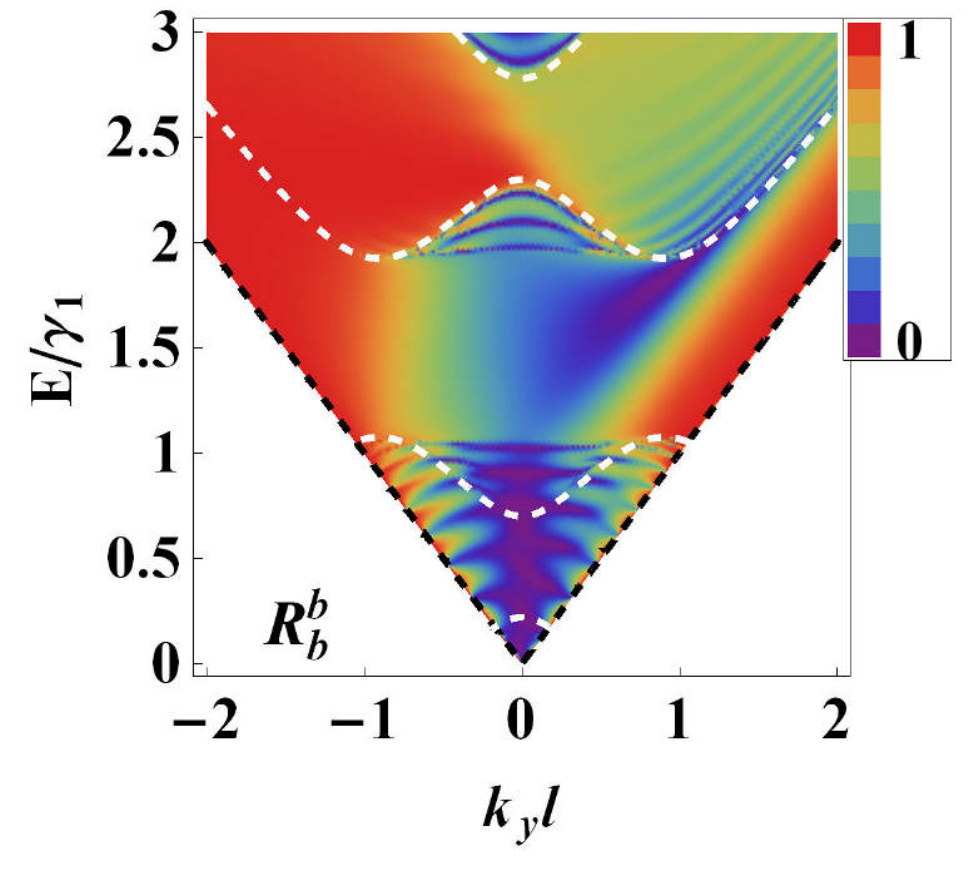}\\
\includegraphics[width=1.5 in]{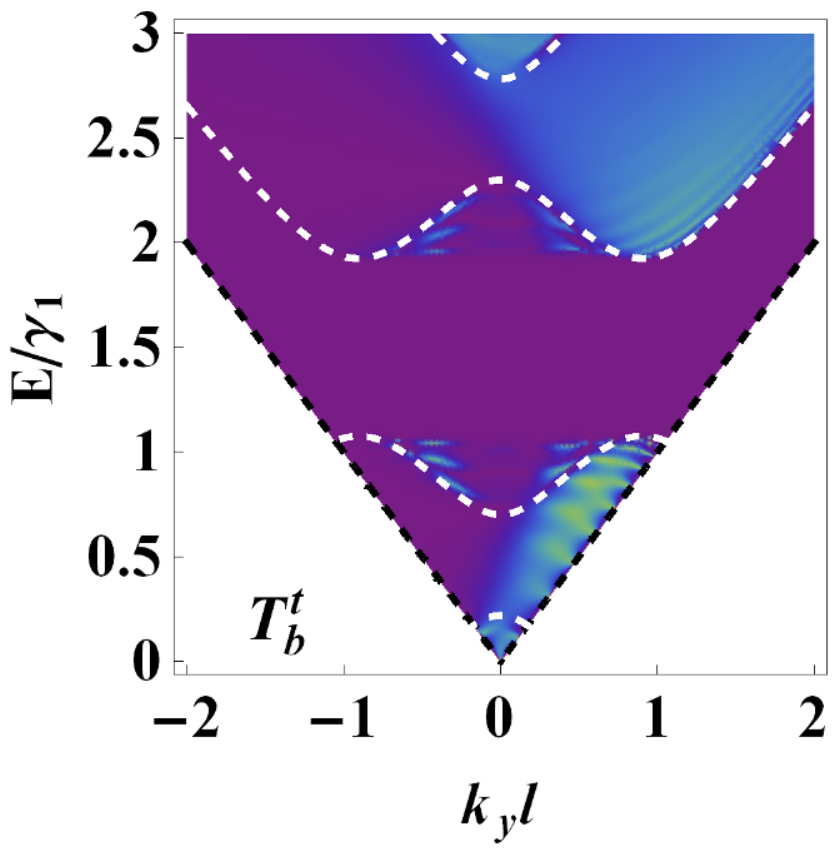}\ \
\includegraphics[width=1.73  in]{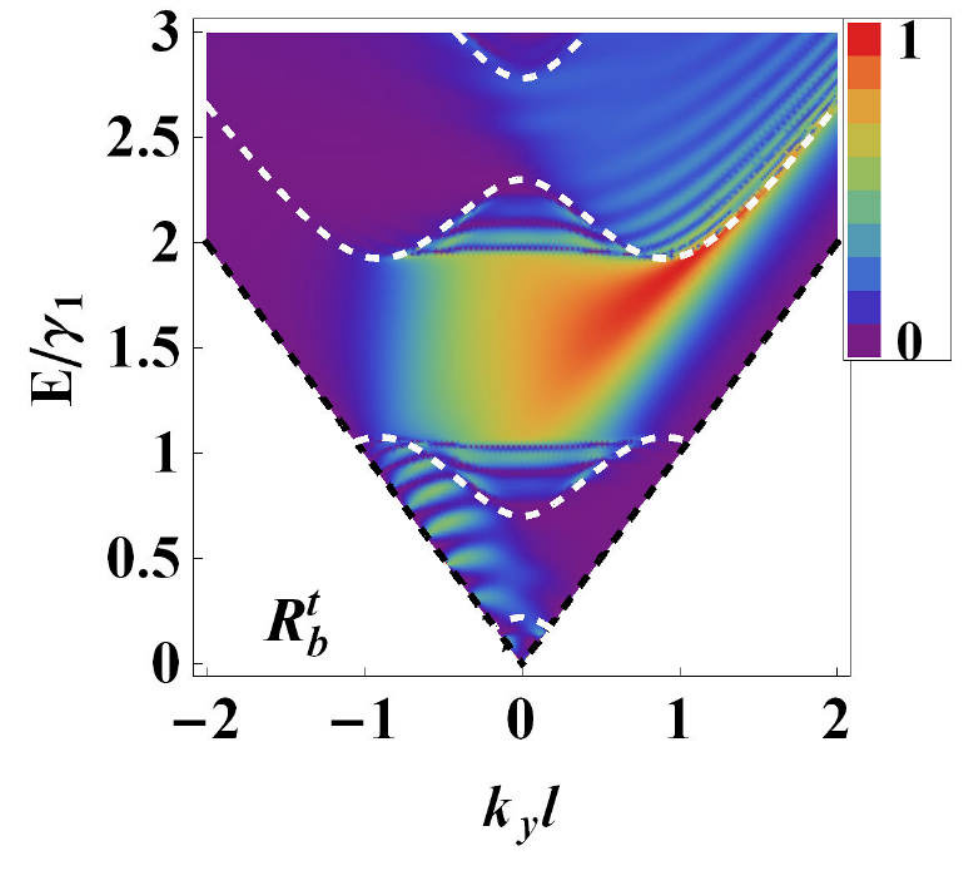}\\
\includegraphics[width=1.5  in]{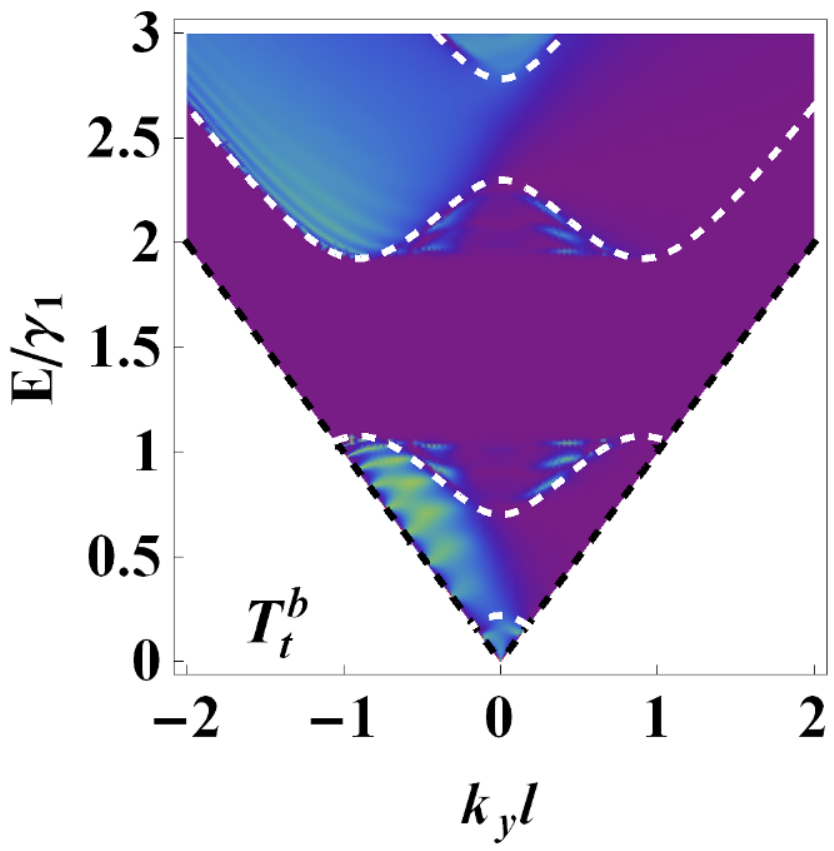}\ \
\includegraphics[width=1.73  in]{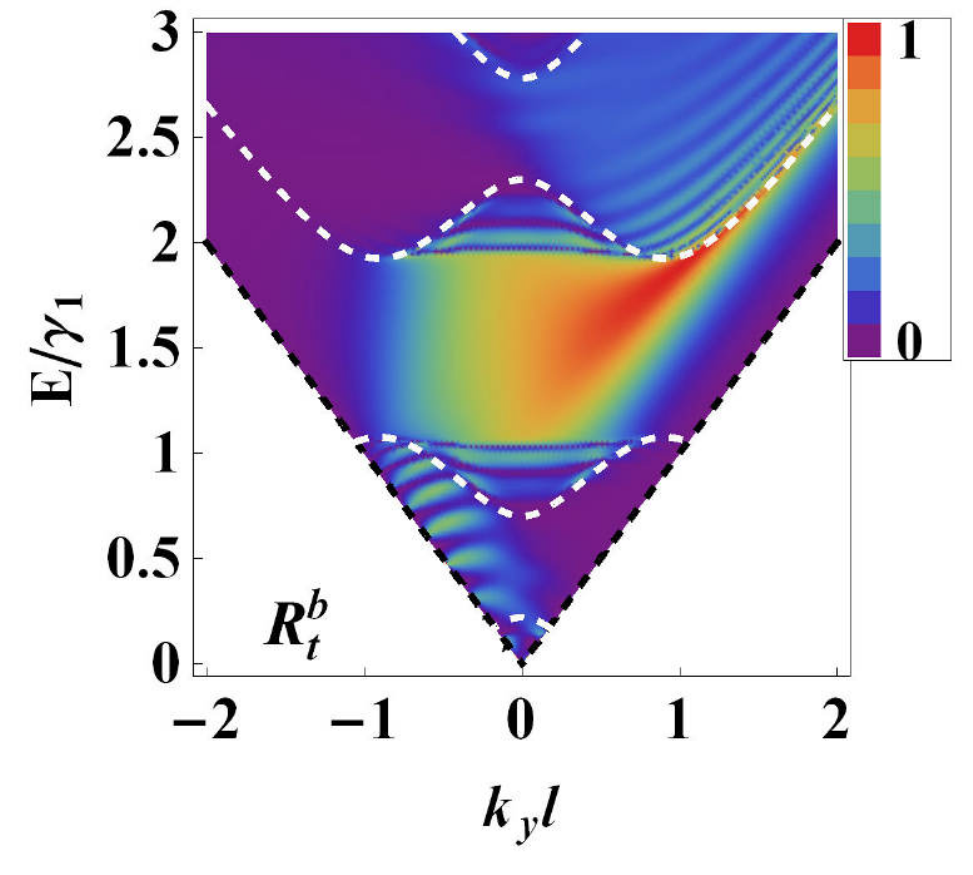}\\
\includegraphics[width=1.5  in]{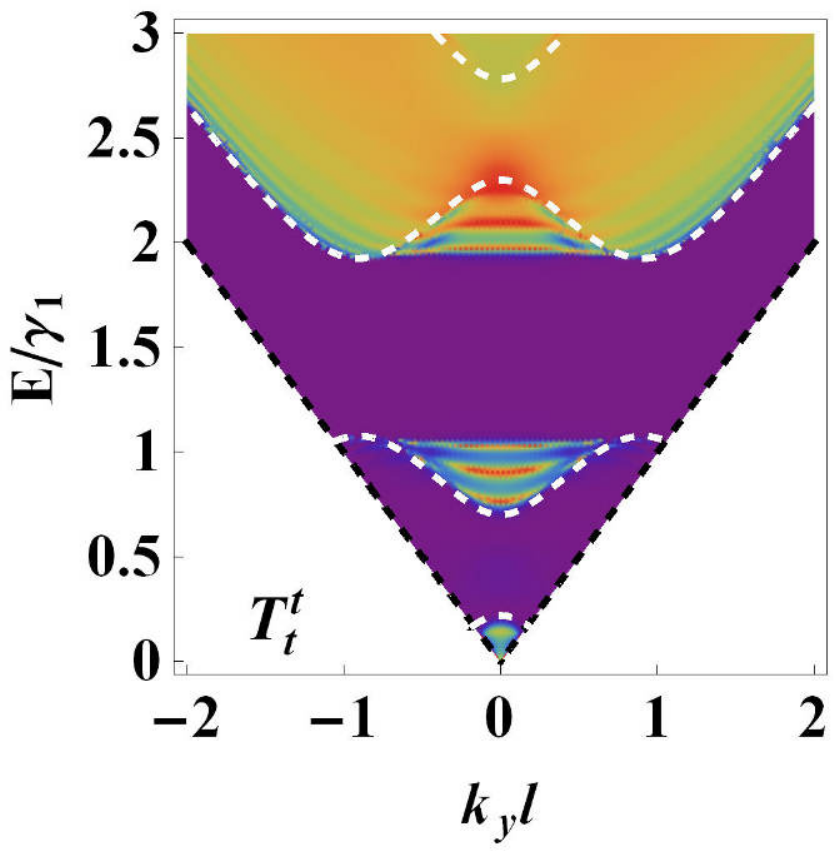}\ \
\includegraphics[width=1.73  in]{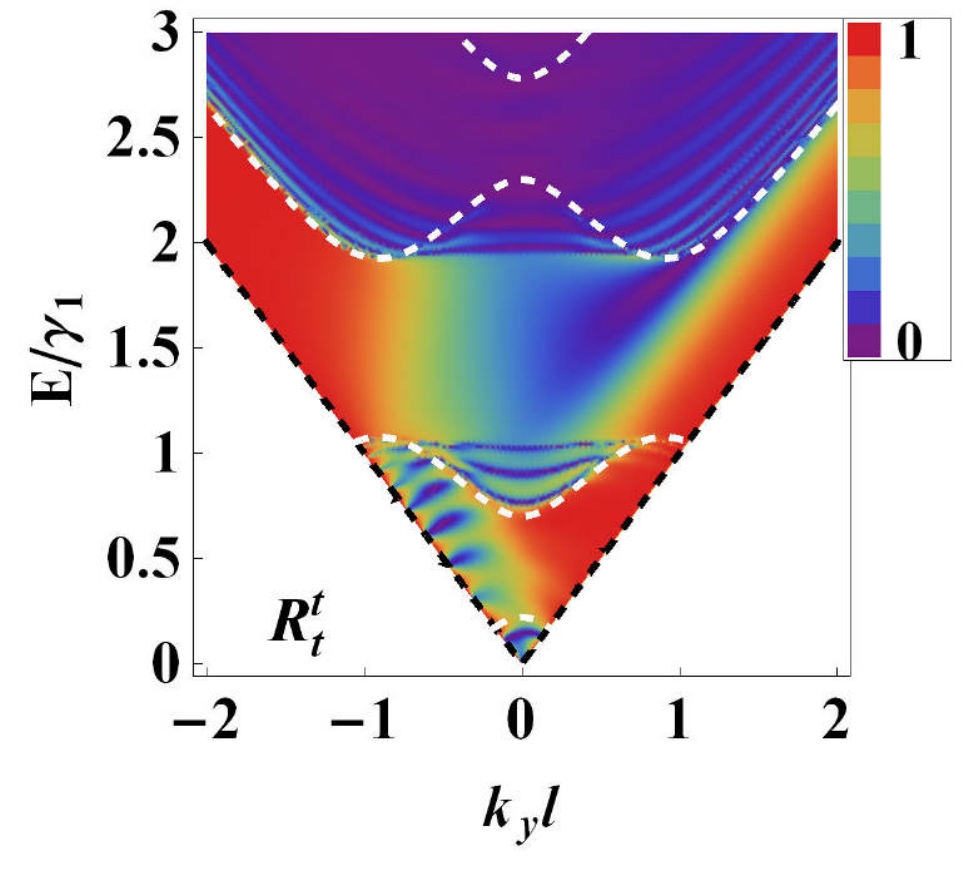}\\
\caption{(Colour online) The same as in Fig. \ref{SL-AB-SL}, but now with   $v_0=3\gamma_1/2$, $\delta =0.8\gamma_1.$ New localized states appear inside the \textit{\textquotedblleft Mexican hat\textquotedblright} shape of the low energy bands of AB-BL due to the strong gate potential. }\label{SL-AB-SLve}
\end{figure}
\begin{figure}[t]
\vspace{0.4cm}
\centering \graphicspath{{./Figures1/SL-AB-SL/}}
\includegraphics[width=1.535 in]{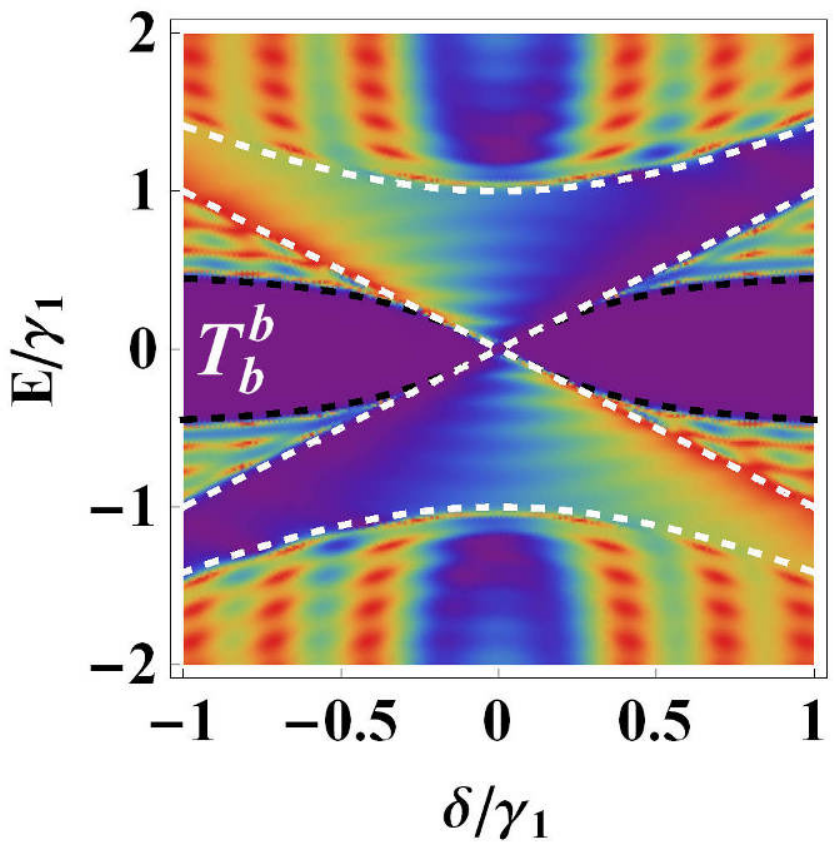}\
\includegraphics[width=1.775  in]{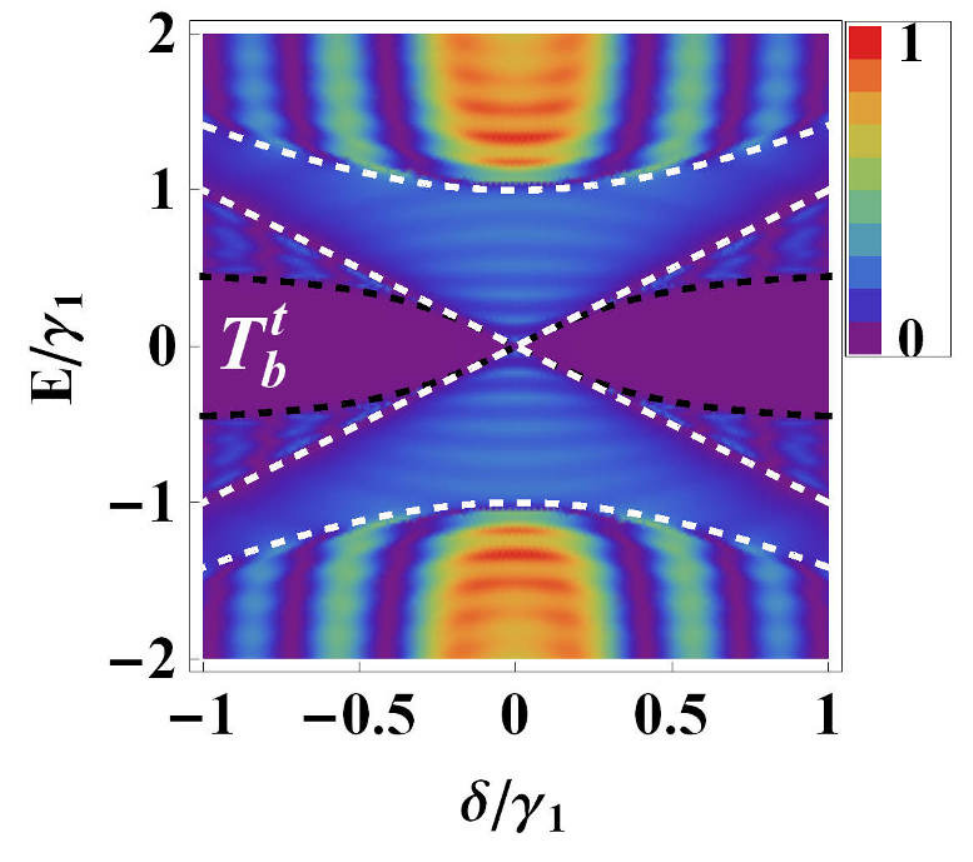}
\caption{(Colour online) Transmission probabilities as function of Fermi energy and bias for normal incidence. }\label{Tbt-Tbb}
\end{figure}

Another interesting feature of this configuration is that for $E<\gamma_1$ the scattered and non-scattered transmissions are equal $T_{i}^{j}=T_{i}^{i}$.  In this energy regime such device can be used as an electronic beam splitter\cite{Brand,Lima}. 

Fig. \ref{SL-AB-SLv} displays the same plot as in Fig. \ref{SL-AB-SL} but with an electrostatic potential  on the AB-BL region.  There is an important difference as compared to the pristine AB-BL case, the layer symmetry   is broken such that $T_{t}^{b}(k_{y})= T_{b}^{t}(-k_y)$ as clarified in Fig. \ref{SL-AB-SLv}. This can be also understood by pointing out that charge carriers  scattered from top to bottom
when moving from left to right in the $K$ valley are equivalent to charge carriers scattering from bottom to top when moving oppositely in the second  valley $K'$. 

Introducing a finite bias ($\delta>0$) to the AB-BL region along with an electrostatic potential ($v_0>0$) will shift the bands and opens a gap in the spectrum. As a result of the presence of a strong electric field, the transmission channels are completely suppressed inside the gap due to the absence of  traveling modes as seen in Fig. \ref{SL-AB-SLve}. Moreover, non-zero asymmetric reflection appears in the gap as well as a violation of the equivalence of non-scattered transmission channels. This is a result of the breaking of inter-layer sublattice equivalence \cite{Ben}. In addition, some localized states appear inside the ``Mexican hat'' of the low energy bands where they are pushed by the strong electric field ($\delta=0.8 \gamma_1$), see Fig. \ref{SL-AB-SLve}.

\begin{figure}[tb]
\vspace{0.4cm}
\centering \graphicspath{{./Figures1/AB-SL-AB/}}
\includegraphics[width=1.5  in]{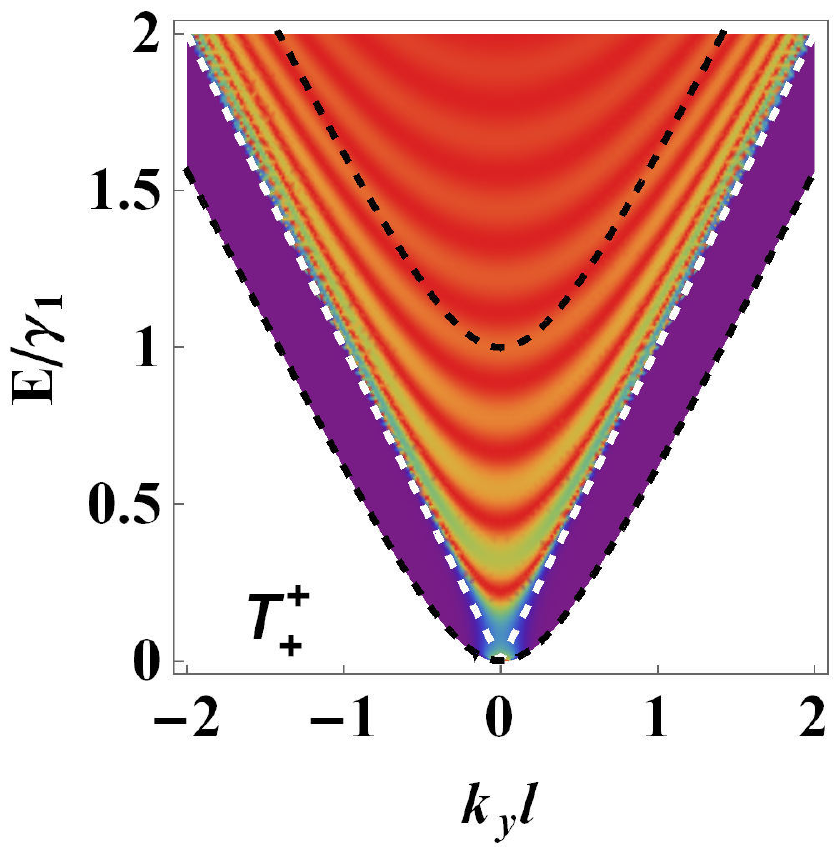}\ \
\includegraphics[width=1.73  in]{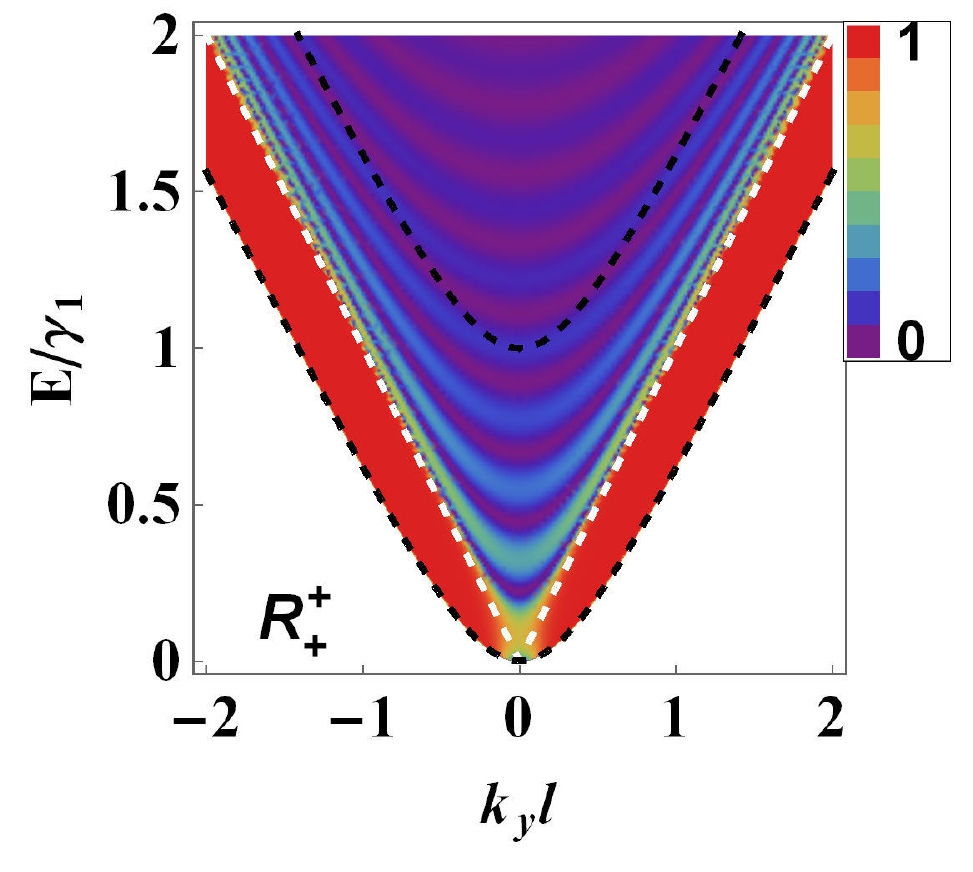}\\
\includegraphics[width=1.5 in]{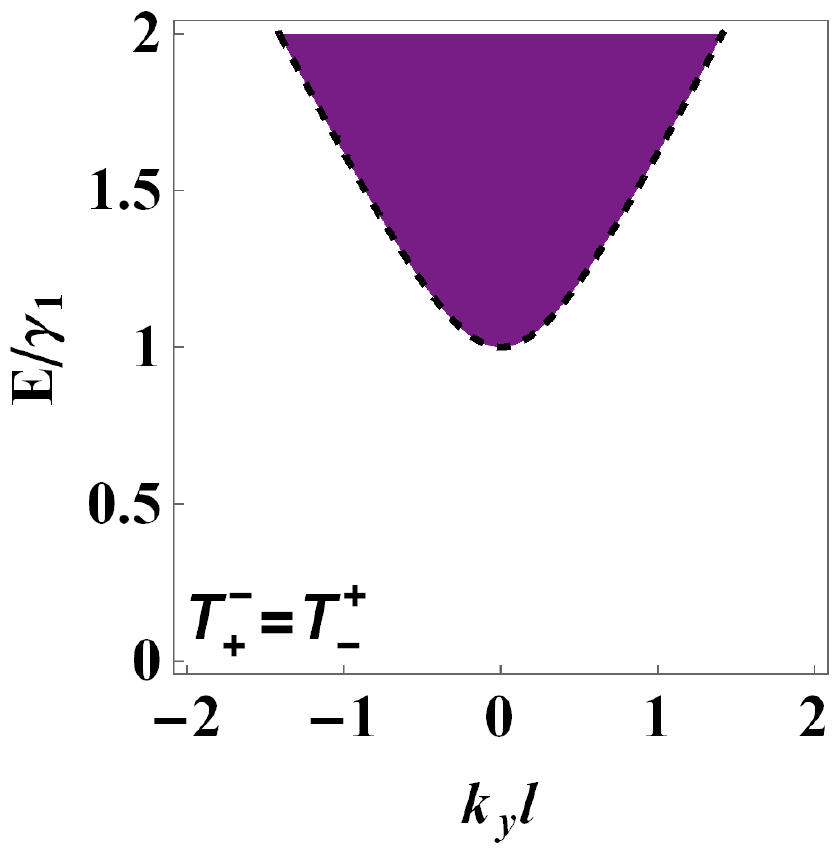}\ \
\includegraphics[width=1.73  in]{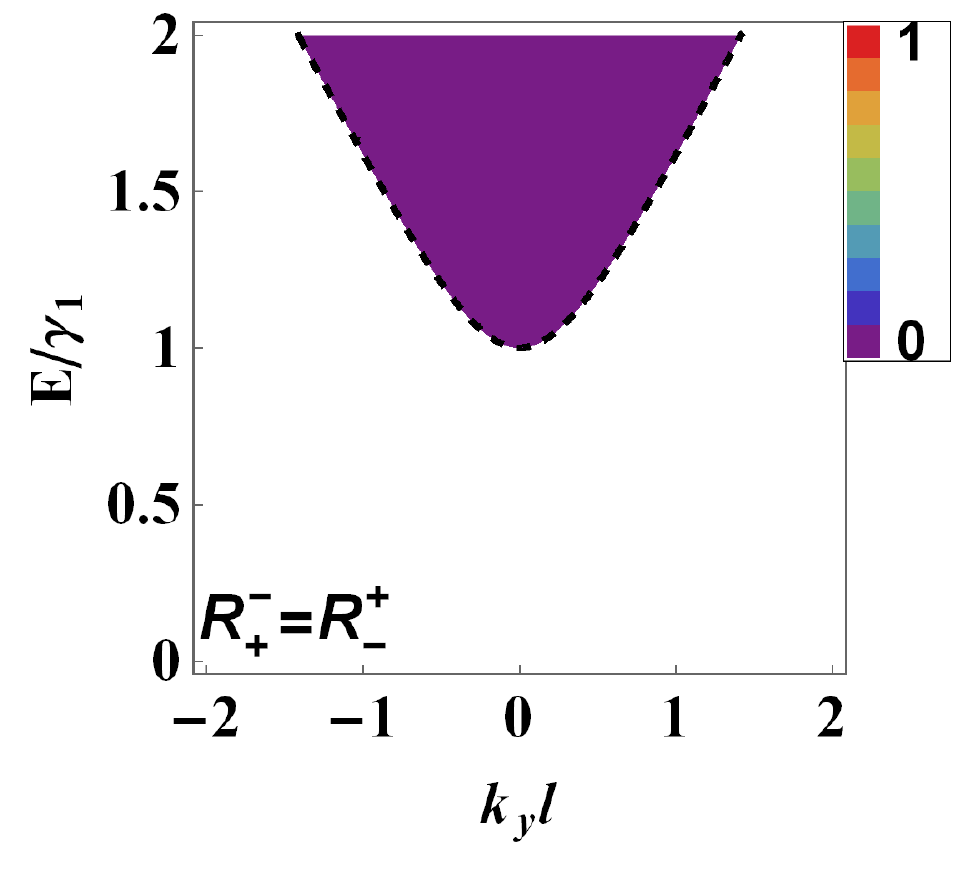}\\
\includegraphics[width=1.5  in]{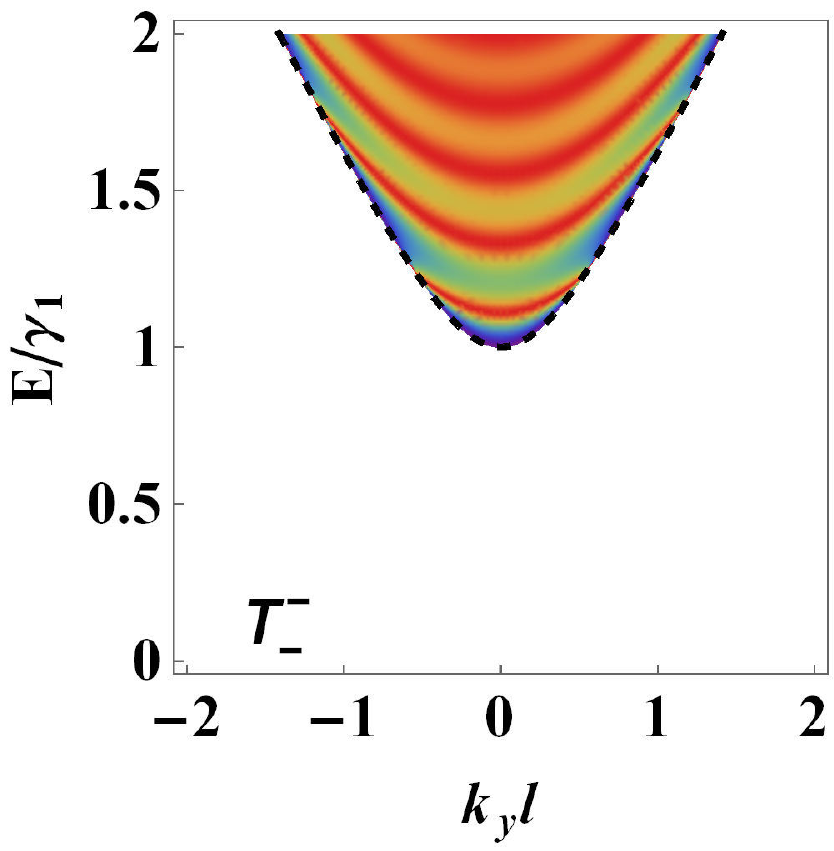}\ \
\includegraphics[width=1.73  in]{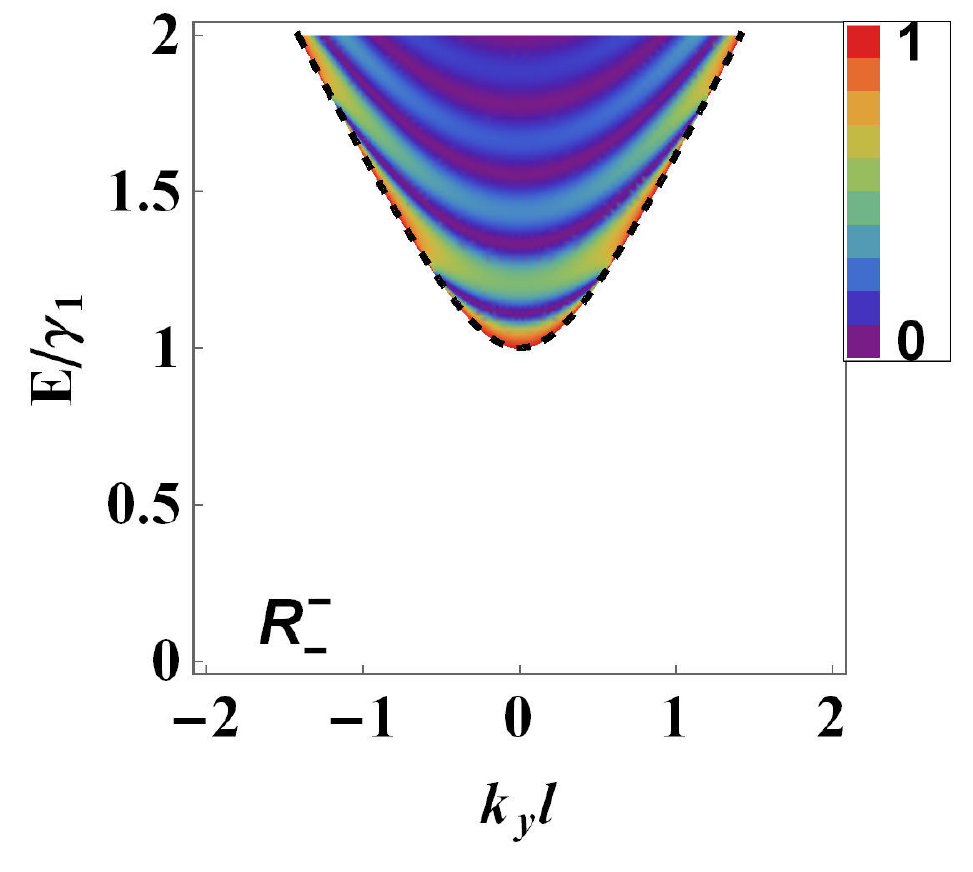}
\caption{(Colour online) Density plot of the  transmission and reflection probabilities through AB-2SL-AB as a function of Fermi energy and
transverse wave vector $k_y$ with  $v_0= \delta =0$ and $d=25$nm. }\label{AB-2SL-AB}
\end{figure}
\begin{figure}[tb]
\vspace{0.4cm}
\centering \graphicspath{{./Figures1/AB-SL-AB/}}
\includegraphics[width=1.5  in]{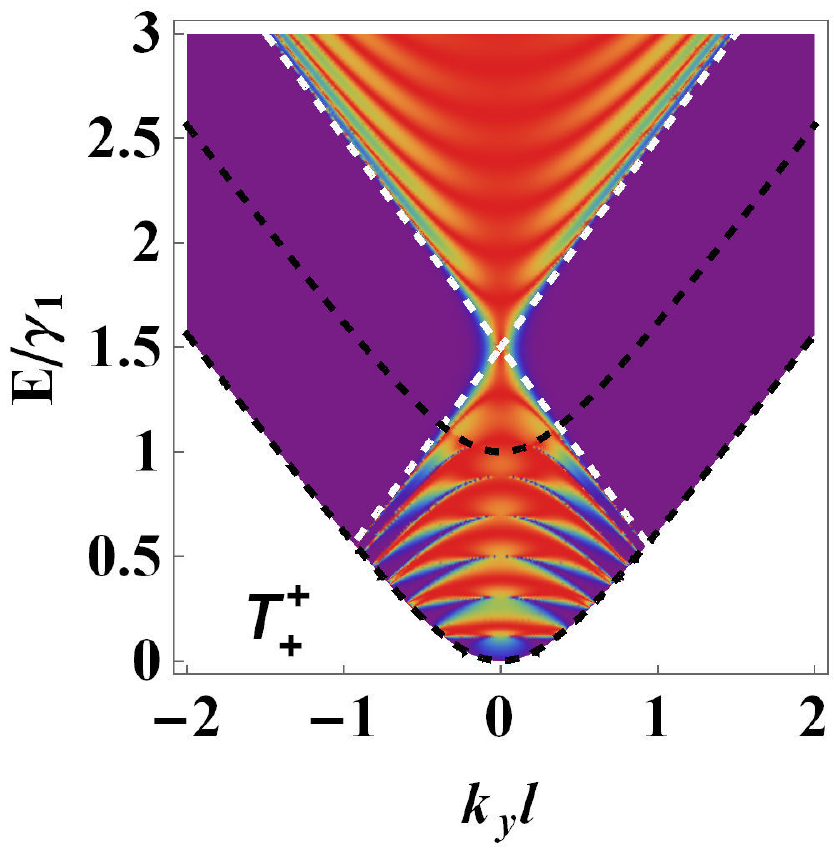}\ \
\includegraphics[width=1.73  in]{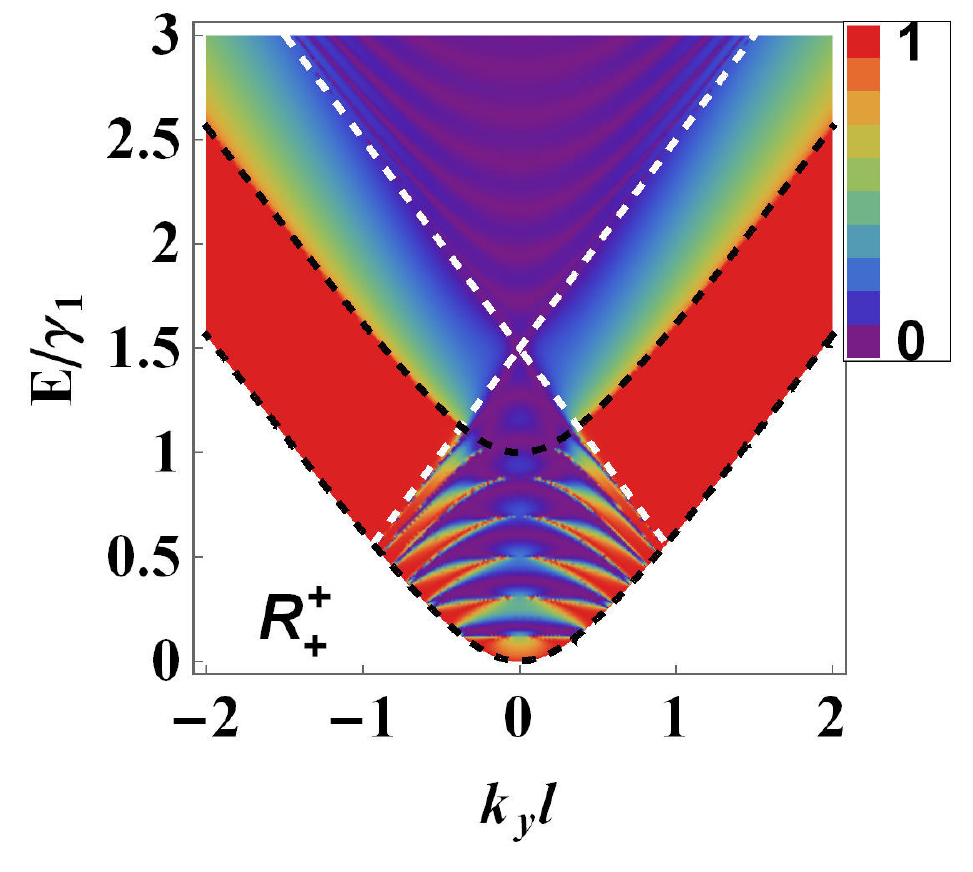}\\
\includegraphics[width=1.5 in]{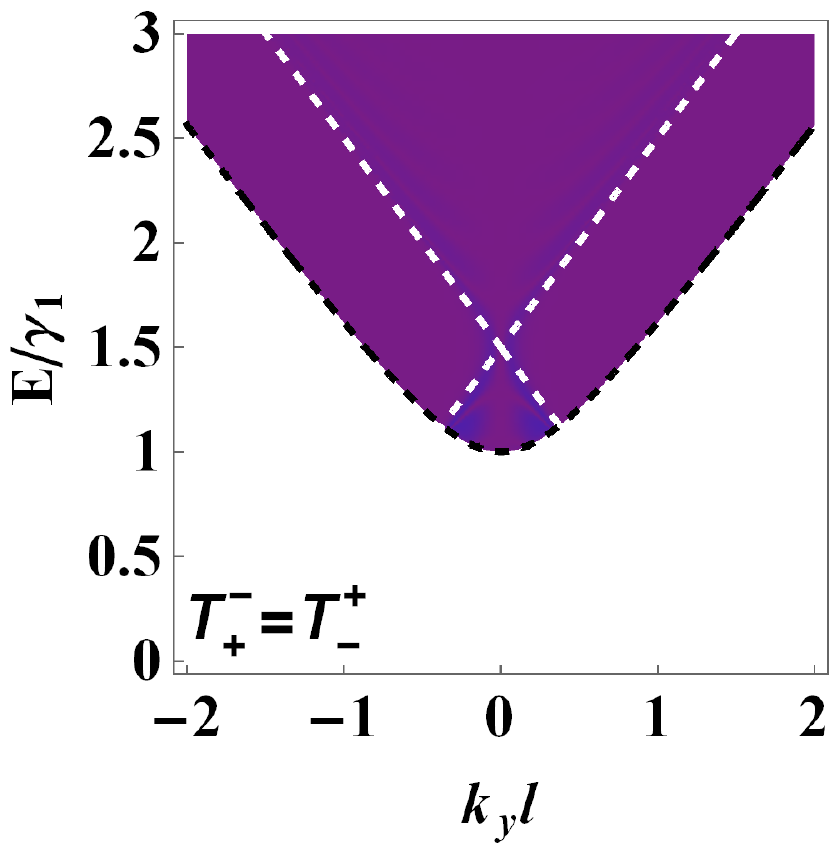}\ \
\includegraphics[width=1.73  in]{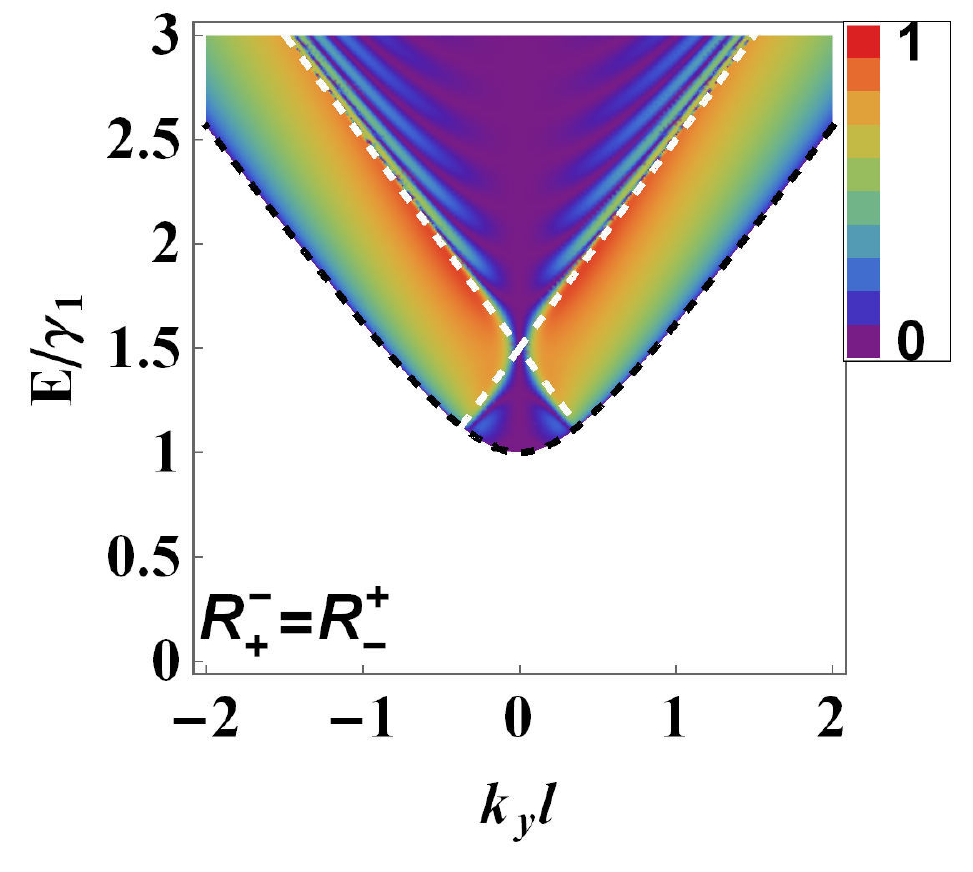}\\
\includegraphics[width=1.5  in]{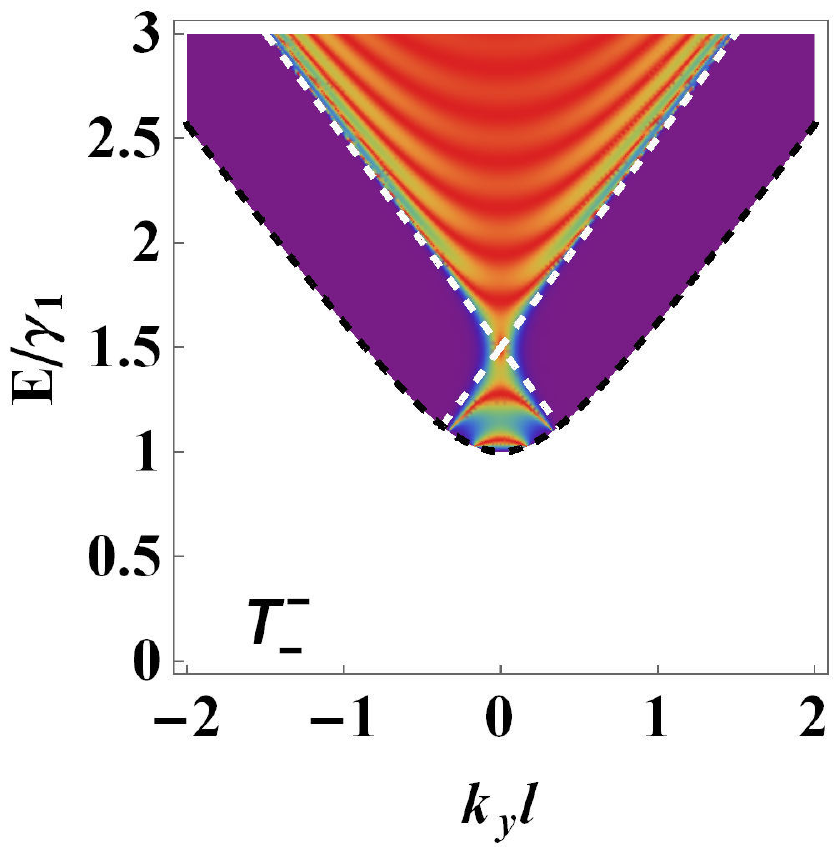}\ \
\includegraphics[width=1.73  in]{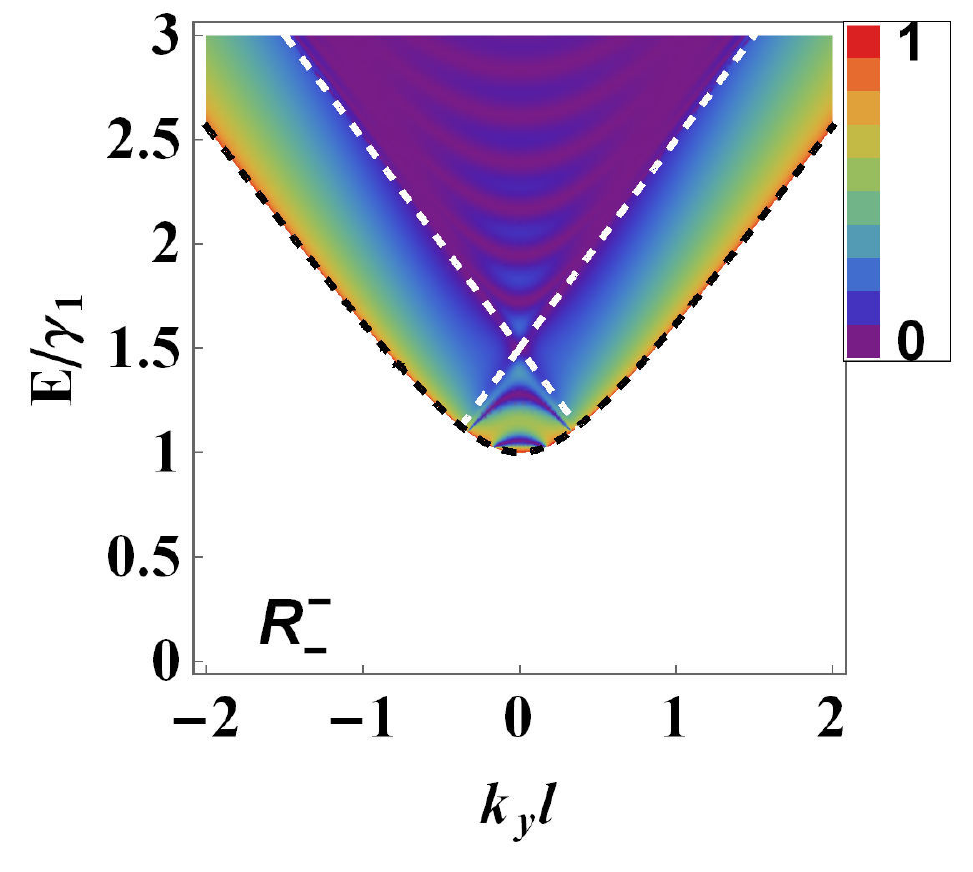}
\caption{(Colour online) The same as in Fig. \ref{AB-2SL-AB}, but here with   $v_0=3\gamma_1/2$ }\label{AB-SL-ABv}
\end{figure}
  There is a link between the transmission probabilities of our system 2SL-AB-2SL and  those investigated  by Gonz\'{a}lez \textit{et al.} \cite{34}. The channels $T_b^b$ and $T_b^t$ are qualitatively equivalent  to those obtained in Ref. [\onlinecite{34}]. For example, $T_b^t$ shows electron-hole  ($e-h$) and $\delta\rightarrow-\delta$ symmetry whereas $T_b^b$ exhibits another symmetry which can be obtained under the exchange $(e,\delta)\leftrightarrow\  (h,-\delta)$. The results in   Fig. \ref{Tbt-Tbb} are in good agreement with those of Ref. [\onlinecite{34}] where we fix $v_0=0$ and $d=25 $ nm.

%%%%%%%%%%%%%%%%%%%%%%%%%%%%%%%%%%%%%%%%%%%%%%%%%%%%%%%%%%%%%%%%   
\subsubsection{ AB-2SL-AB }\label{AB-2SL-AB-T}
%%%%%%%%%%%%%%%%%%%%%%%%%%%%%%%%%%%%%%%%%%%%%%%%%%%%%%%%%%%%%%%%

 For leads composed  of AB-BL where the intermediate region is pristine 2SL, we show the results in Fig. \ref{AB-2SL-AB} for the transmission and reflection probabilities. Now charge carriers will scatter between the different modes of the AB-BL on the left and right leads as shown in Fig. \ref{intro-fig02}(b). As expected,  all channels are symmetric and as a result of the finite size of the 2SL region, resonances appear in  $T$ as shown in Fig. \ref{AB-2SL-AB}. These so-called Fabry-P\'{e}rot resonances appear at quantized energy levels\cite{masir2010} \begin{equation}\label{eq019}
E^n_{SL}(k_y)=\sqrt{k_y^{2}+\left( \frac{n \pi \ }{d}\right)^{2}}.
\end{equation}
This is the dispersion relation for modes confined in the 2SL region with width $d$.

\begin{figure}[tb]
\vspace{0.4cm}
\centering \graphicspath{{./Figures1/AB-SL-AB/}}
\includegraphics[width=1.5  in]{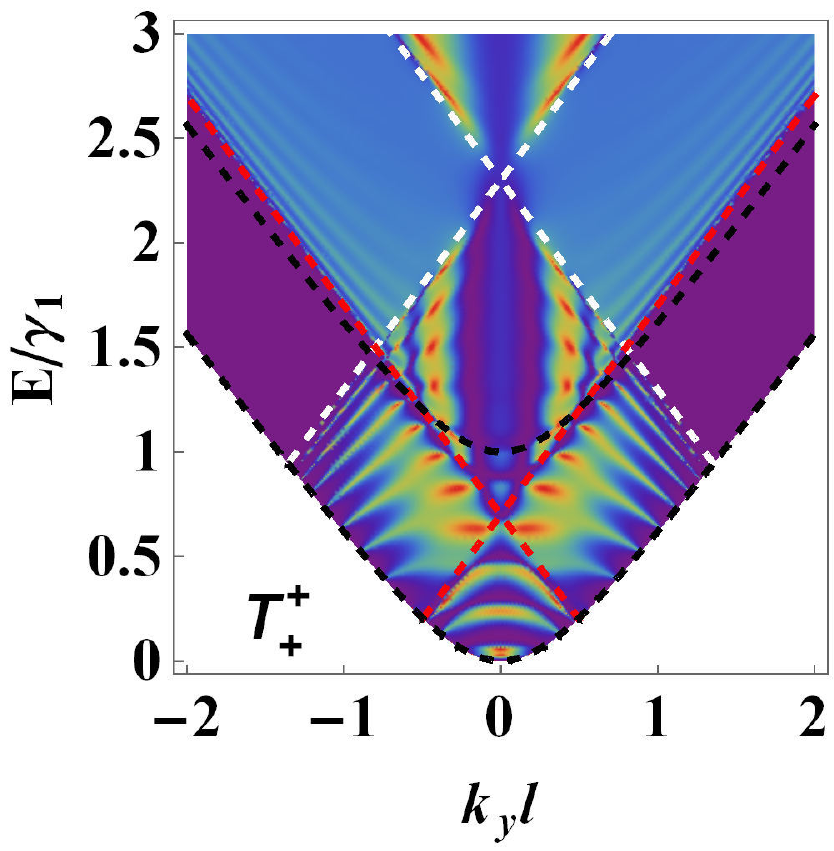}\ \
\includegraphics[width=1.73  in]{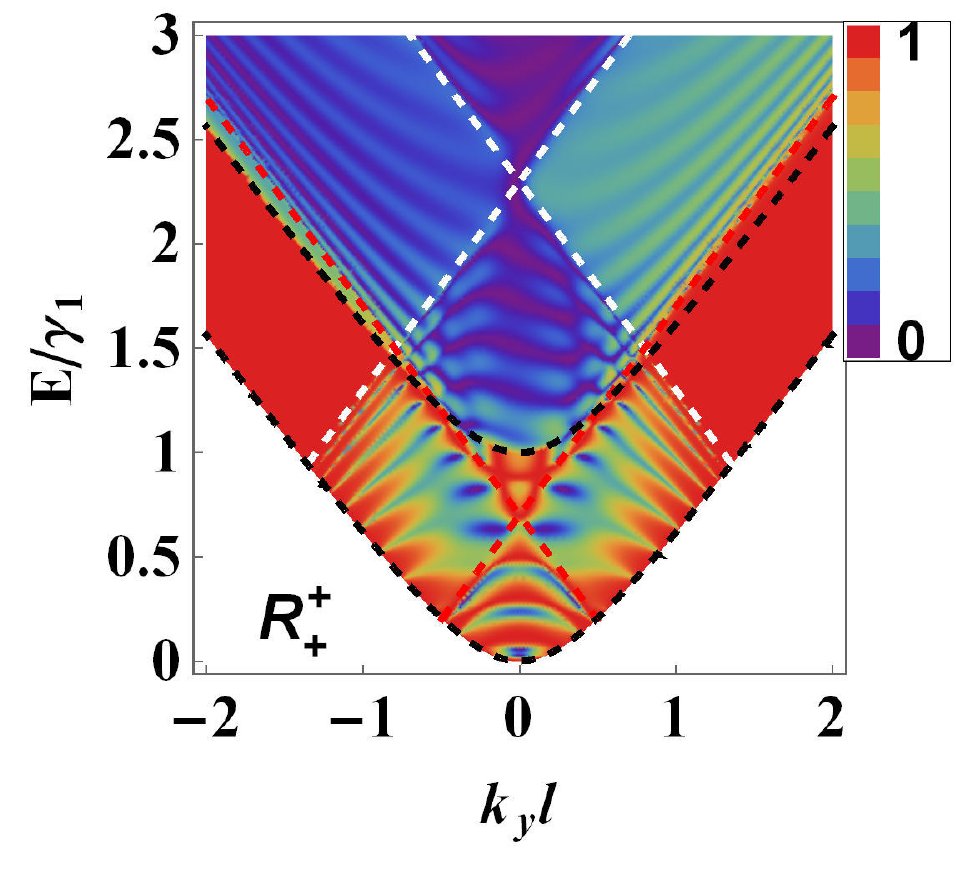}\\
\includegraphics[width=1.5 in]{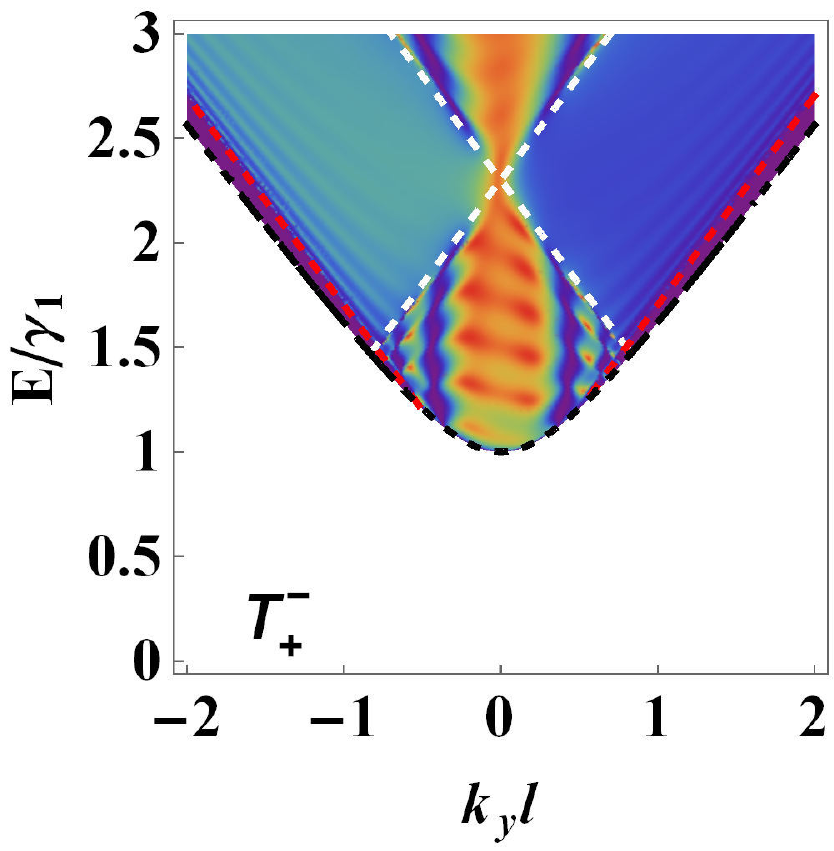}\ \
\includegraphics[width=1.73  in]{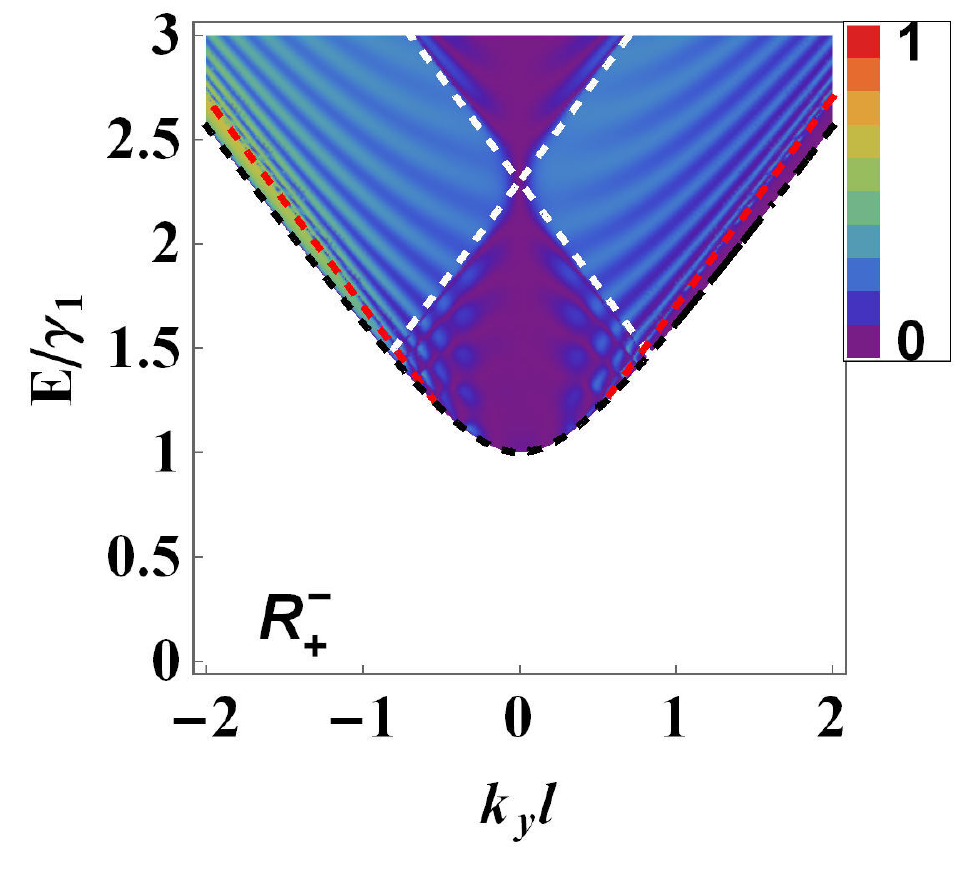}\\
\includegraphics[width=1.5  in]{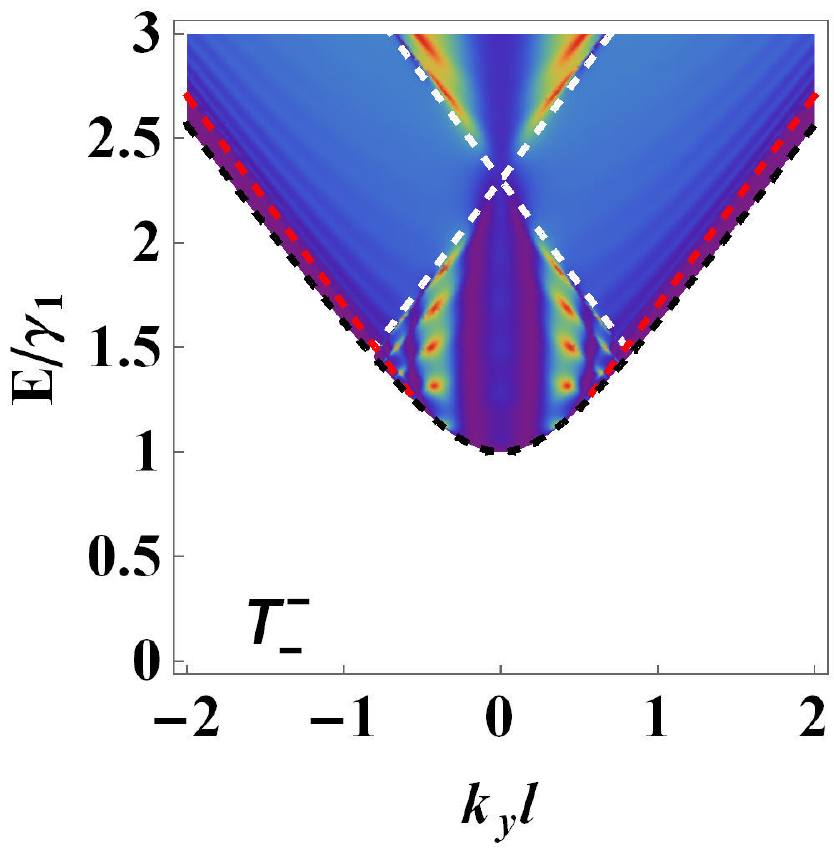}\ \
\includegraphics[width=1.73  in]{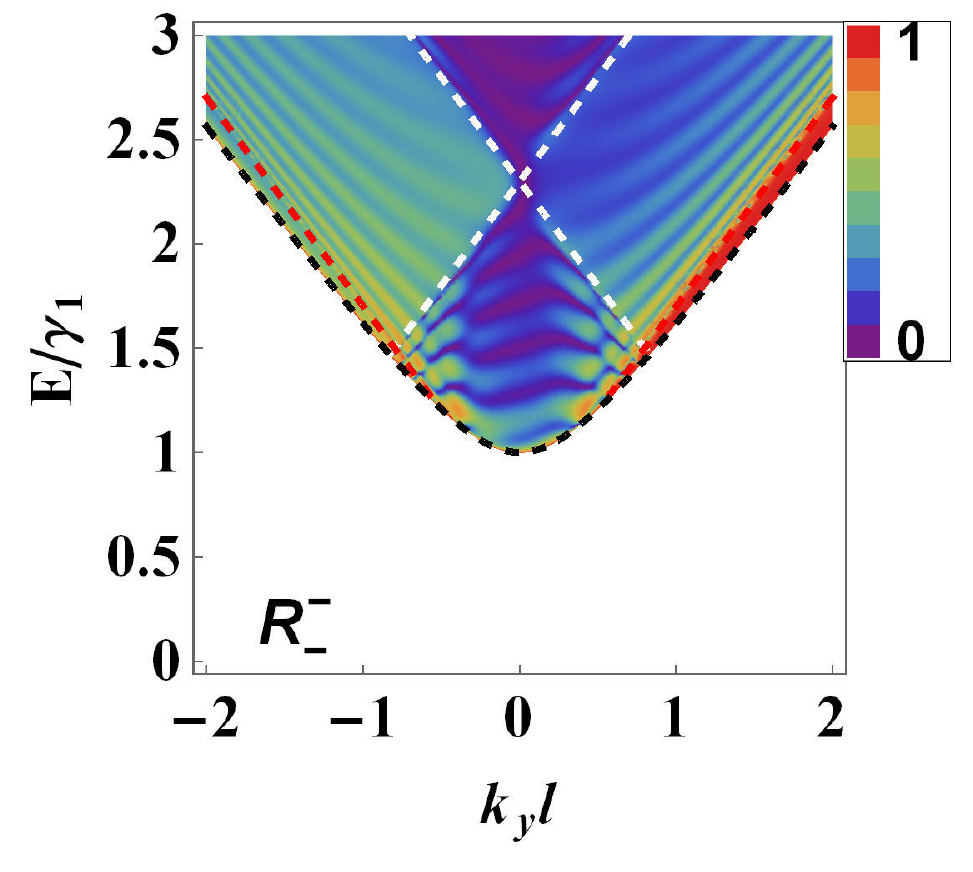}\\
\includegraphics[width=1.5  in]{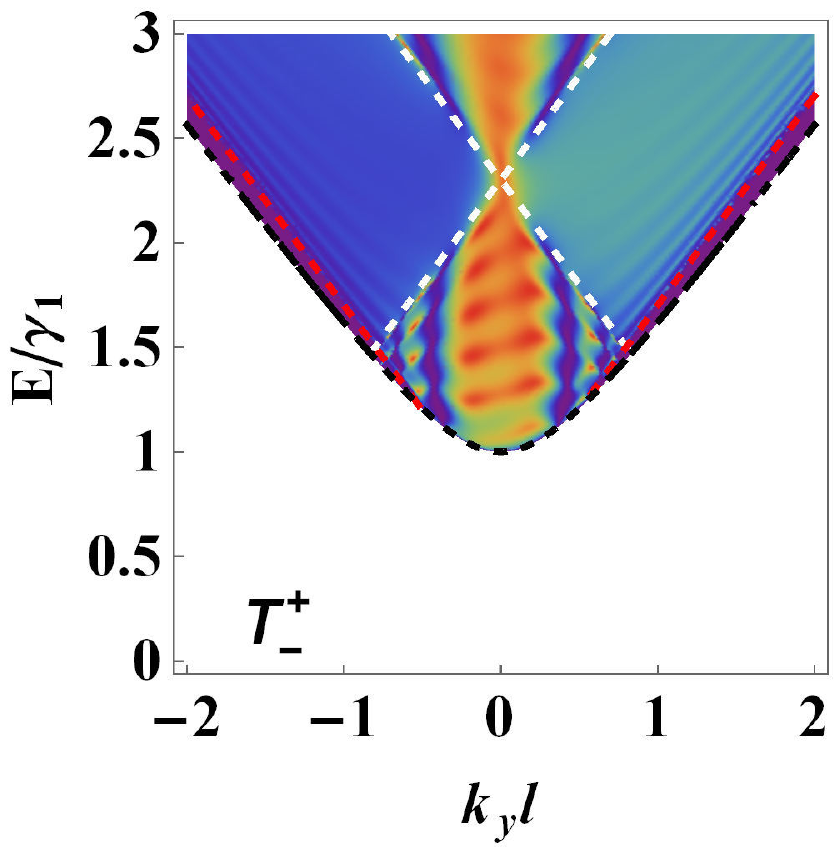}\ \
\includegraphics[width=1.73  in]{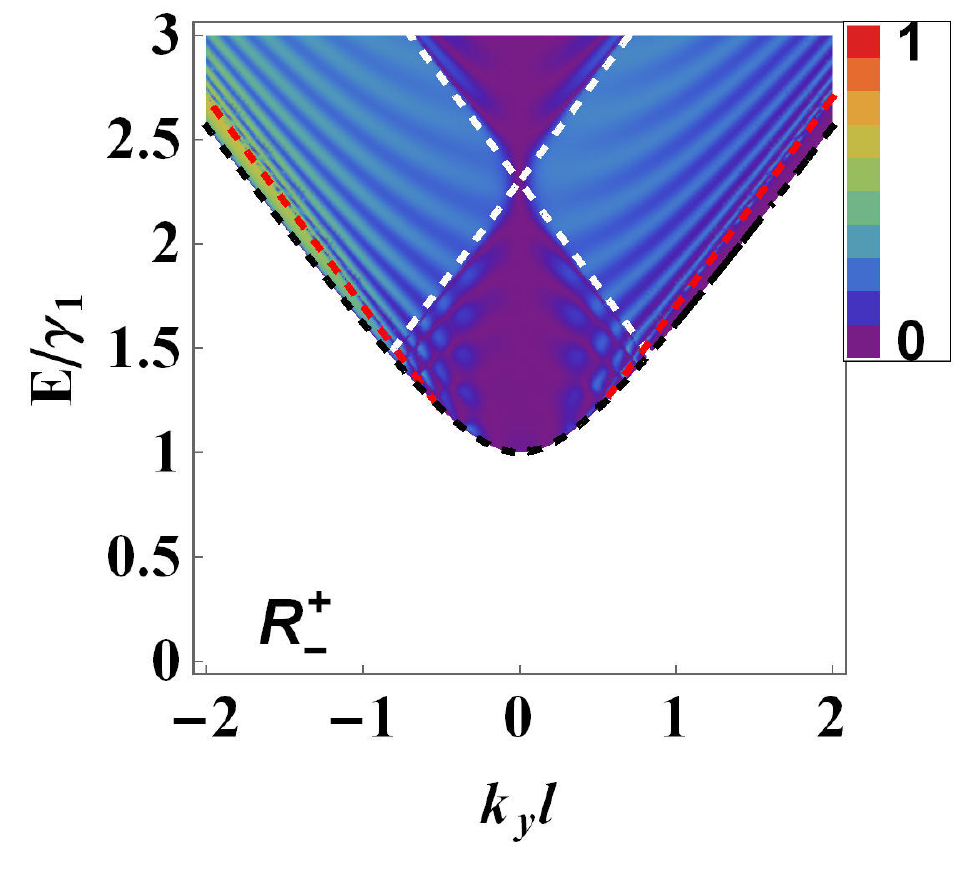}
\caption{(Colour online) The same as in Fig. \ref{AB-2SL-AB}, but here  with $v_0=3\gamma_1/2$, $\delta =0.8\gamma_1$.  Red and white dashed curves  correspond to the bands of bottom and top layers of 2SL while the black dashed curves are the AB-BL bands.   }\label{AB-SL-ABve}
\end{figure}

The results presented in Fig. \ref{AB-2SL-AB} reveal no scattering between the two modes $k^+$ and $k^-$ and charge carriers  are only transmitted or reflected through the same channel from which they come from.  Unexpectedly,  introducing an electrostatic potential induces a strong scattering in the reflection channels ($R_+^-=R_-^+\neq0$) and very weak scattering in the transmission channels  ($T_+^-=T_-^+\neq0$), as seen in Fig. \ref{AB-SL-ABv}.  When the 2SL are biased, the Dirac cones at bottom and top layers will be shifted up (white dashed lines) and down (red dashed lines) in energy, respectively (see Fig. \ref{AB-SL-ABve}).  This bias will strengthen the coupling between the two modes resulting in a strong  scattering between them.
In addition,  the inversion symmetry is broken due to the bias leading to an asymmetry with respect to normal incidence. 

\subsubsection{ Conductance }

\begin{figure}[tb]
\vspace{0.4cm}
\centering \graphicspath{{./Figures1//SL-AB/}}
\includegraphics[width=1.65  in]{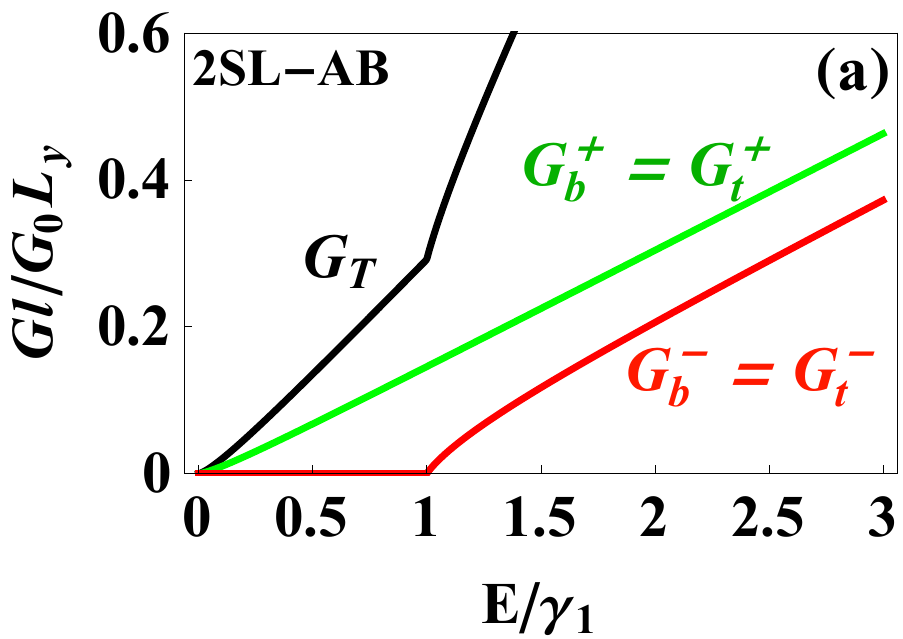}\ \
\includegraphics[width=1.65  in]{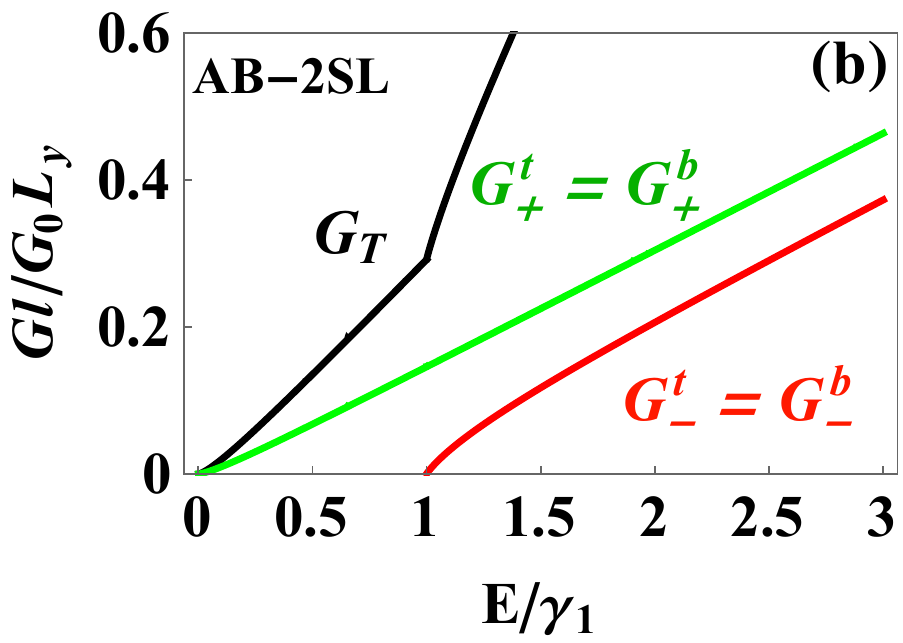}\\
\includegraphics[width=1.65 in]{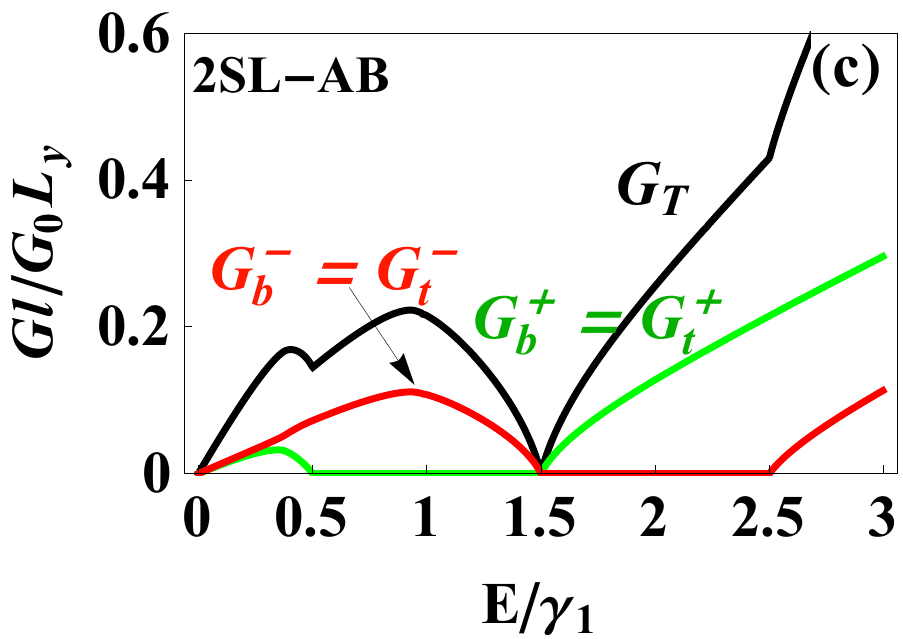}\ \
\includegraphics[width=1.65  in]{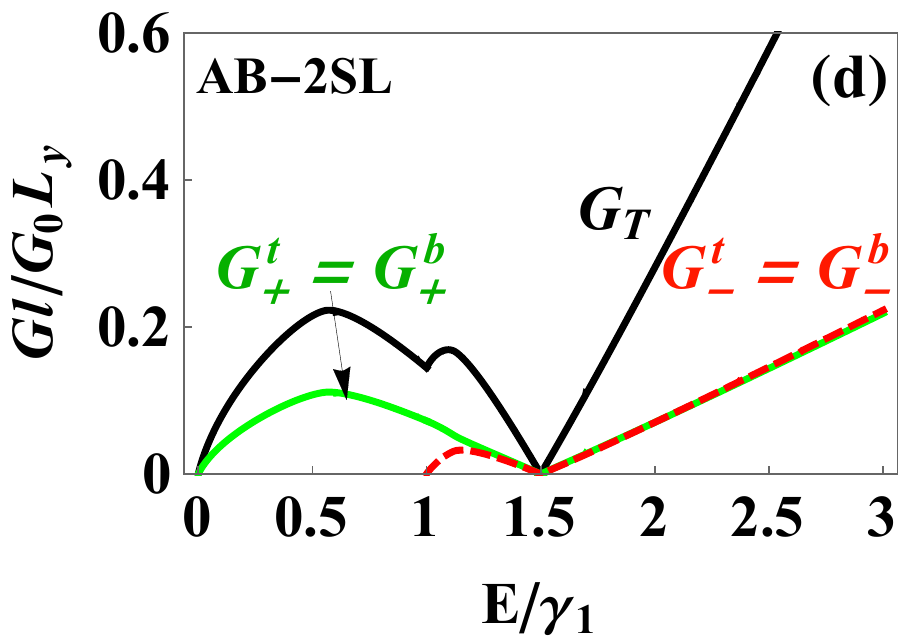}\\
\includegraphics[width=1.65  in]{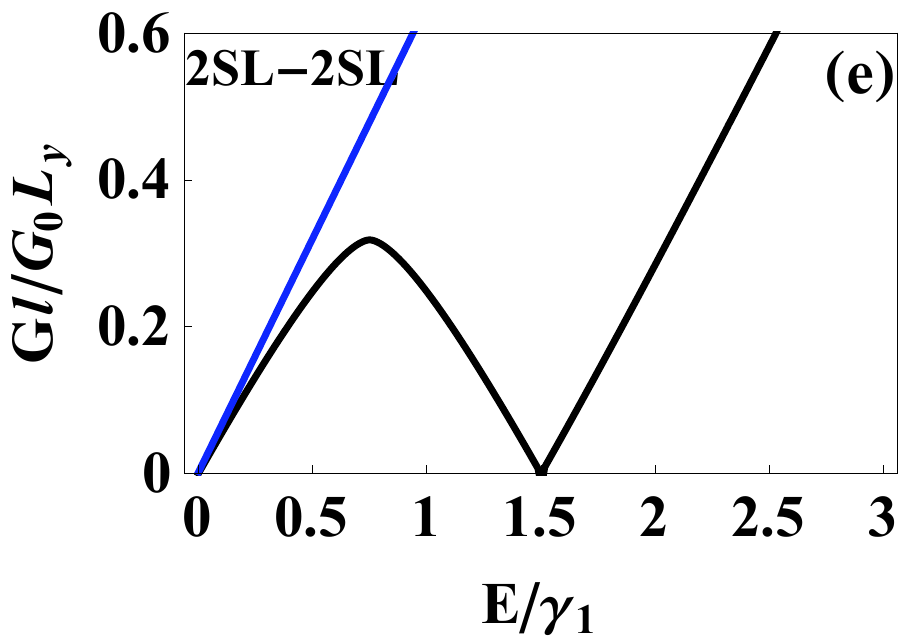}\ \
\includegraphics[width=1.65  in]{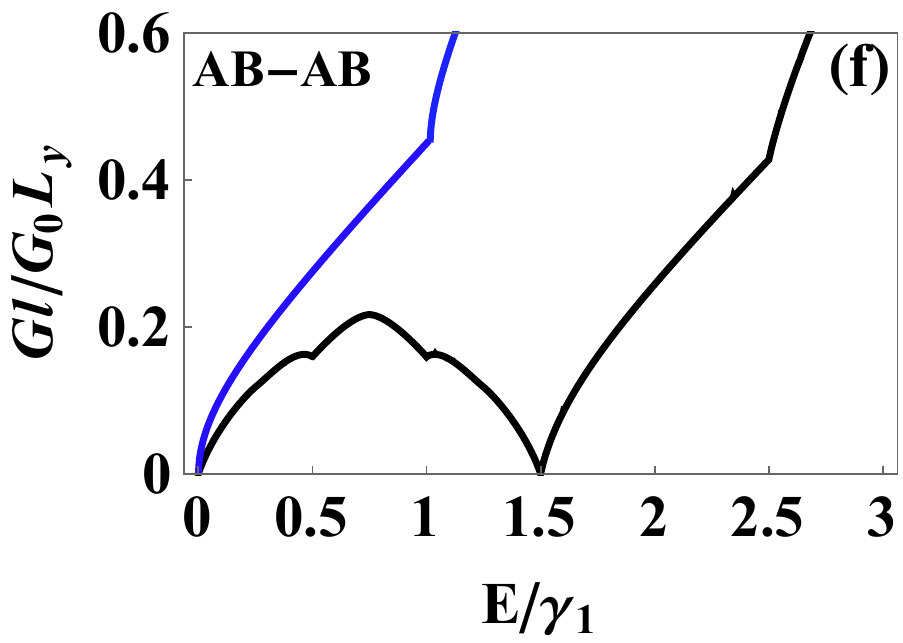}
\caption{(Colour online) Conductance of different junctions for different magnitudes of the applied gate: (a, b) $v_0= \delta =0$, (c, d) $v_0=3\gamma_1/2$, $\delta =0$,  (e, f) the total conductance for 2SL-2SL and AB-AB junctions, respectively, with $v_0=0$(blue curves) and $v_0=1.5\gamma_1$(black curves).  }\label{AB-SL-G}
\end{figure}

The conductance of the two-block system consisting of 2SL and BA-BL is shown in Fig. \ref{AB-SL-G}
for different values of the applied gate voltage. Figs. \ref{AB-SL-G}(a,b) reveal that the system where charge carriers are incident from the 2SL and impinge on AB-BL and vice versa are equivalent to the case when both 2SL and AB-BL are at the same potential.  
As  seen in Figs. \ref{AB-SL-G}(a,b),  $G_+^{t(b)}=G_{t(b)}^+$ are contributing to the total conductance $G_T$ starting from $E=0$ where the $k^+$ mode exists. On the contrary, $G_-^{t(b)}=G_{t(b)}^-$ only contributes when $E>\gamma_1$ where $k^-$ states are available and this appears as a sharp increase in $G_T$ at $E=\gamma_1$. On the other hand, considering an applied electrostatic potential  on the right side of the two-block system will break this equivalence as seen in Figs. \ref{AB-SL-G}(c,d). In addition, as a result of the shift of the Dirac cone in AB-BL (see  Fig. \ref{AB-SL-G}(c)) or 2SL (see Fig. \ref{AB-SL-G}(d)) due to the electrostatic potential, all conductance channels are zero at $E=v_0$. Similar to the AA-BL case,  the conductances of the pristine systems 2SL-AB/AB-2SL (see Figs. \ref{AB-SL-G}(a, b)) \textcolor{blue}{clearly} preserve the time reversal symmetry. Even though, both systems have different transmission probabilities as can be seen from Figs. \ref{polar-SL-AB}(a, c). We also show in Figs. \ref{AB-SL-G}(e, f) the total  conductance in the absence of domain wall in 2SL-SL and AB-AB systems, respectively, for $v_0=0$ (blue curves) and $v_0=1.5\gamma_1$ (black curves). This shows that transport channels in the presence of domain walls are experimentally  recognisable. 
\begin{figure}[tb]
\vspace{0.4cm}
\centering \graphicspath{{./Figures1//SL-AB/}}
\includegraphics[width=1.65  in]{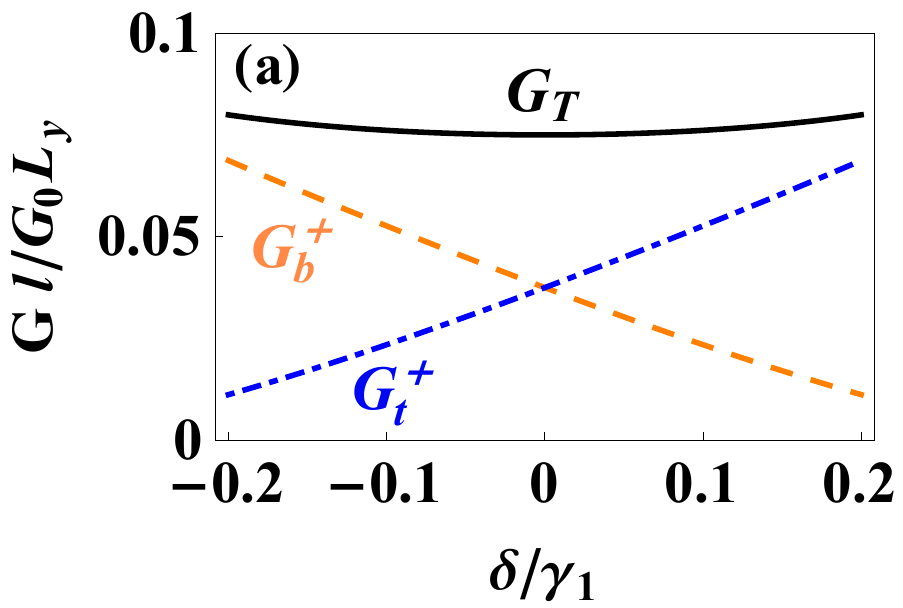}\
\includegraphics[width=1.62  in]{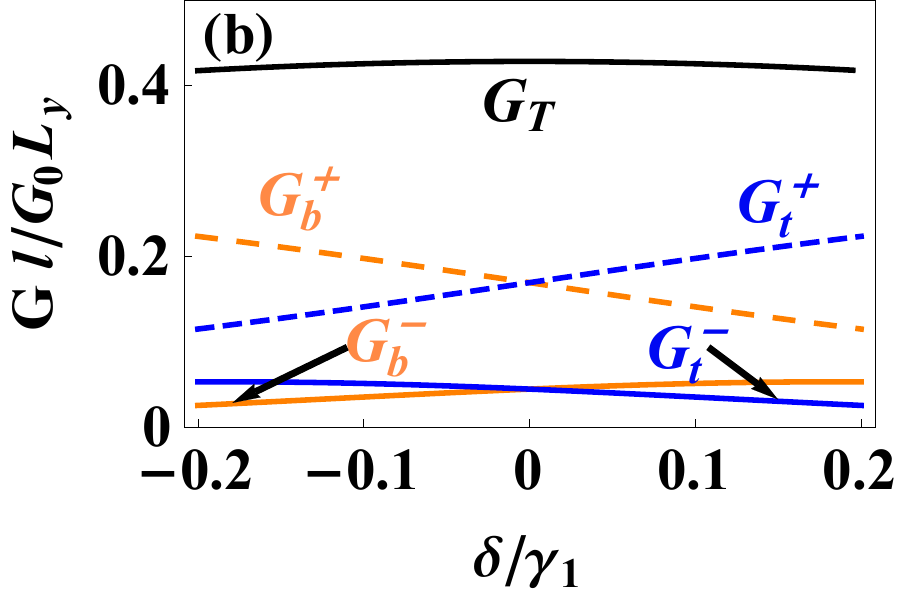}
\caption{(Colour online) Conductance across the 2SL-AB system as a function of the bias on the AB-BL with $v_0=0.$ (a) and (b) correspond to the single and double modes regime  with $E=0.3 \gamma_1$ and $E=1.15 \gamma_1$, respectively.
 With $G^{\pm}_T=G_t^\pm+G_b^\pm.$ \textcolor{red}{}}\label{SL-eAB-G}
\end{figure}
\begin{figure}[tb]
\vspace{0.4cm}
\centering \graphicspath{{./Figures1//SL-AB-SL/}}
\includegraphics[width=1.65  in]{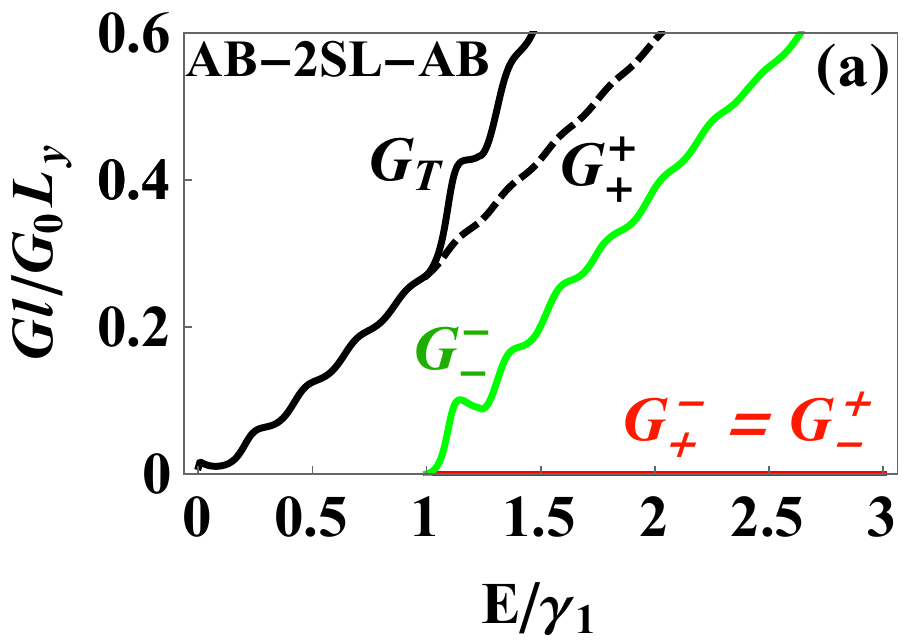}\ \
\includegraphics[width=1.65  in]{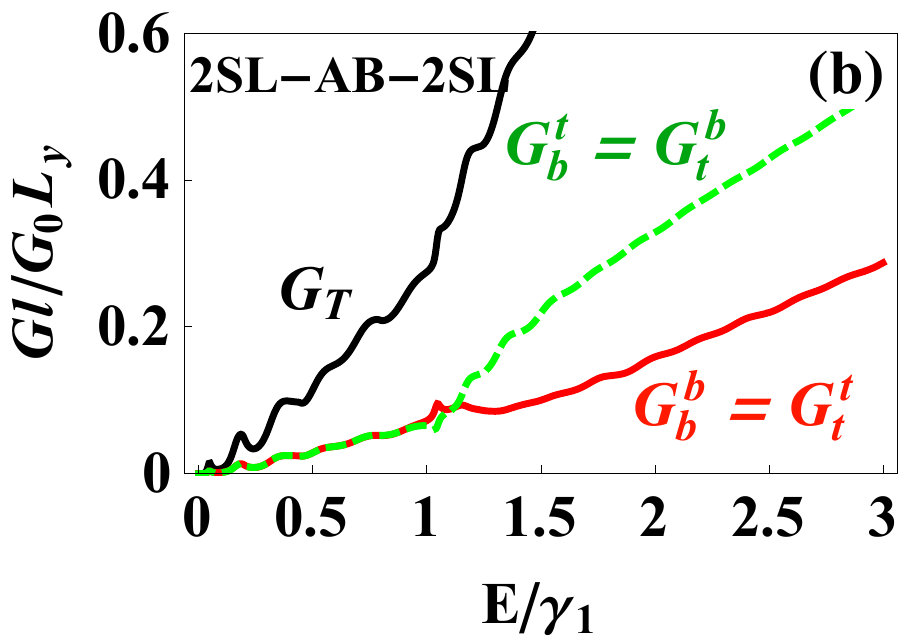}\\
\includegraphics[width=1.65 in]{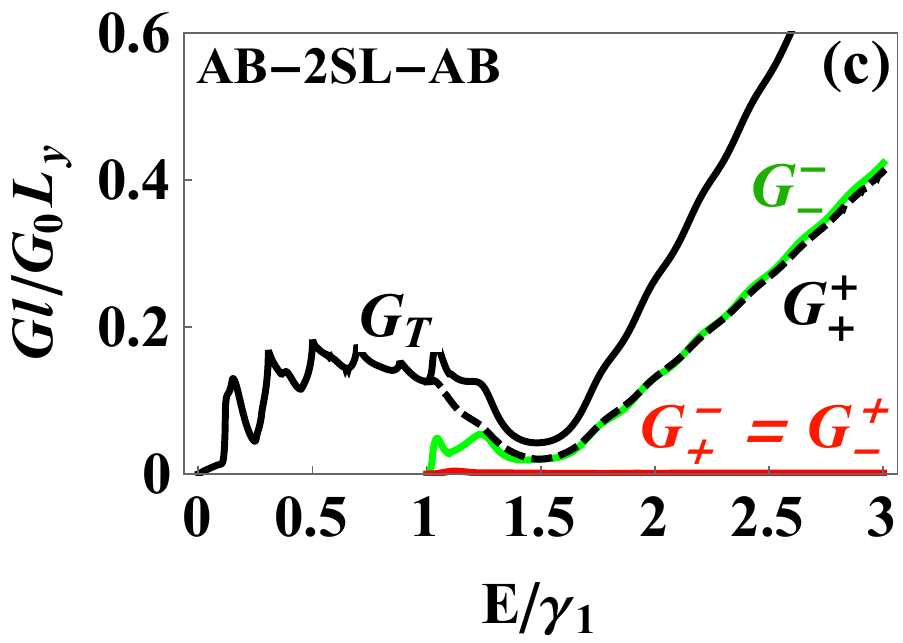}\ \
\includegraphics[width=1.65  in]{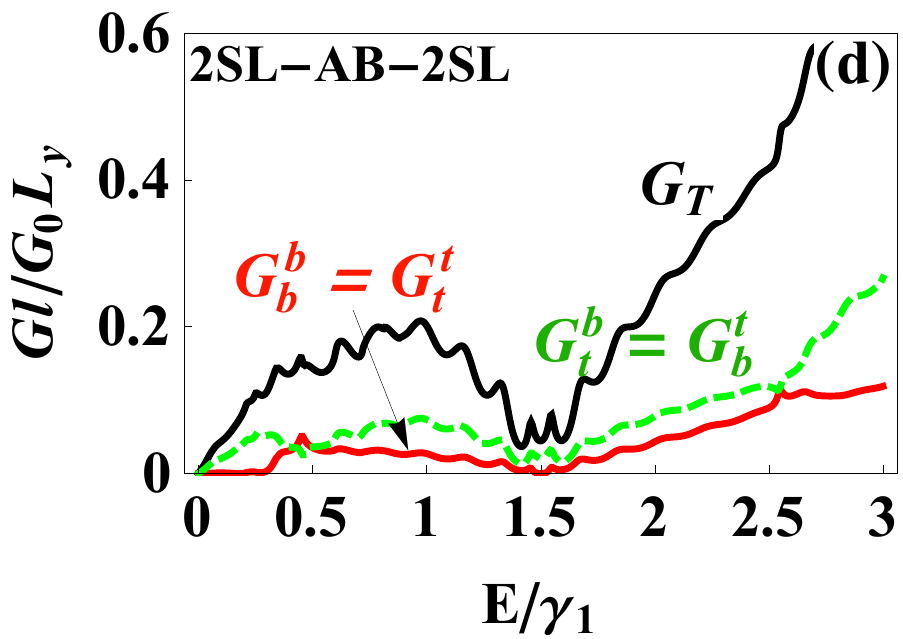}\\
\includegraphics[width=1.65  in]{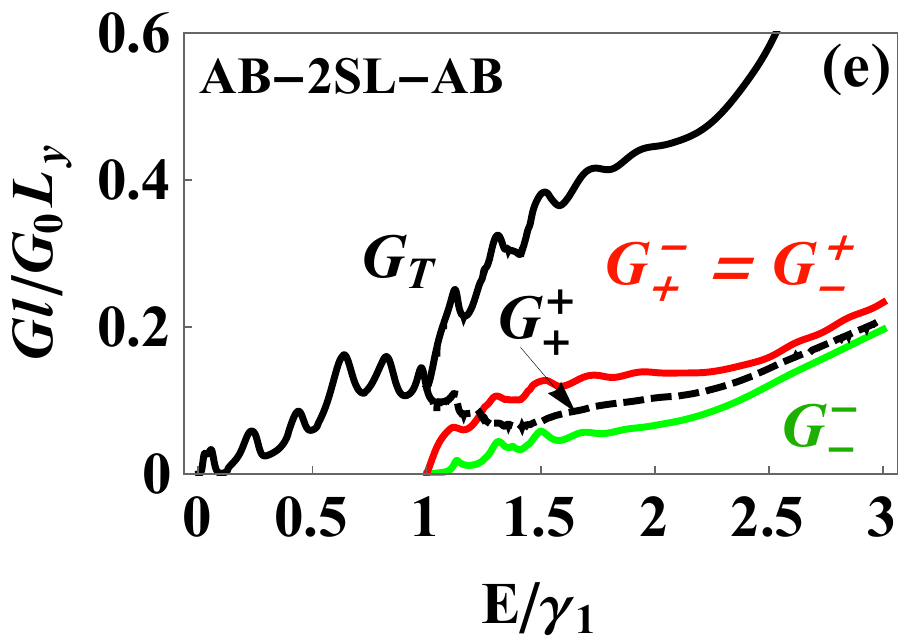}\ \
\includegraphics[width=1.65  in]{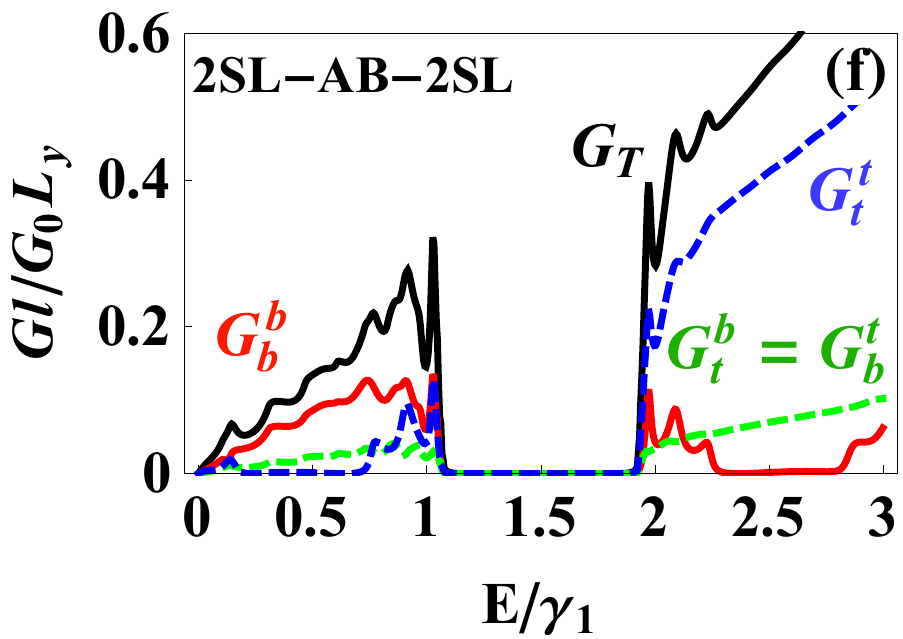}
\caption{(Colour online) Conductance of different junctions for different magnitudes of the applied gate: (a, b) $v_0=\ \delta =0$, (c, d) $v_0=3\gamma_1/2$, $\delta =0$ and (e, f) $v_0=3\gamma_1/2$,\ $\delta =0.8\gamma_1$. }\label{AB-SL-AB-G}
\end{figure}  

In Fig. \ref{SL-eAB-G} we show the conductance in a 2SL-AB system as a function of the bias for transport using a single Fig. \ref{SL-eAB-G}(a) or a double Fig. \ref{SL-eAB-G}(b) mode. The results show that the contribution from the top and bottom layers to the conductances have  opposite behaviours as a function of the inter-layer bias. The total conductance $G_{T}$, however, has a convex form, increasing with the application of an inter-layer bias. From Fig. \ref{SL-eAB-G}(b), on the other hand, we see that when a second mode is available, four channels contribute to the conductance and the total conductance assumes a concave form, i.e. decreasing with increasing inter-layer bias. This is a characteristic experimental feature that can signal the presence of a second mode of propagation.

For  the three-block system we show the conductance of the configuration AB-2SL-AB and 2SL-AB-2SL in the left and right columns of Fig. \ref{AB-SL-AB-G}, respectively.
The resulting conductance of  the first configuration shows only two non-zero channels $G_+^+$ and $G_-^-$, while the scattered ones $G_+^{-}=G_-^{+} =0$ since $T_+^{-}=T_-^{+} =0$ (see Fig. \ref{AB-SL-AB-G}(a)). Furthermore, for low energy $G_T=G_+^+$ since the mode $k^{-}$ is not available in this regime but it starts conducting when $E>\gamma_1$. The applied electrostatic potential on the 2SL keeps the scattered conductance channels  at zero and a minimum in the conductance appears around the shifted Dirac cone $E=v_0$ of the 2SL  as depicted in Fig. \ref{AB-SL-AB-G}(c). As pointed out before, if the Fermi energy approaches the strength of the electrostatic potential, a non-zero minimum is present in the conductance because  charge carriers  can be
transmitted through a width $d$ of 2SL via 
evanescent modes\cite{Snyman}.  In Fig. \ref{AB-SL-AB-G}(f) this minimum disappears and the conductance dramatically increases at $E=\gamma_1$. This is because the bias will couple the two modes and two additional scattered channels $G_+^-$ and $G_-^+$ start conducting. The resonant peaks resulting in the conductance, see Figs.  \ref{AB-SL-AB-G}(a,c,e), are due to the finite size of the intermediate region and hence strictly depend on its width $d$.
\begin{figure}[tb]
\vspace{0.4cm}
\centering \graphicspath{{./Figures1/AA-2SL-AB/}}
\includegraphics[width=1.5  in]{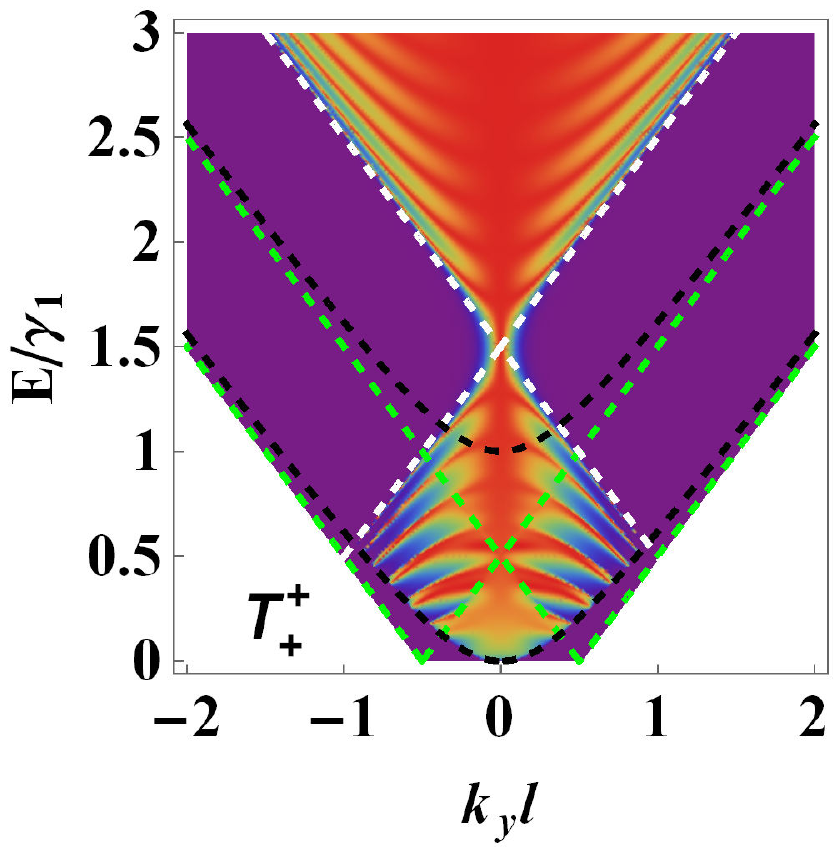}\ \
\includegraphics[width=1.73  in]{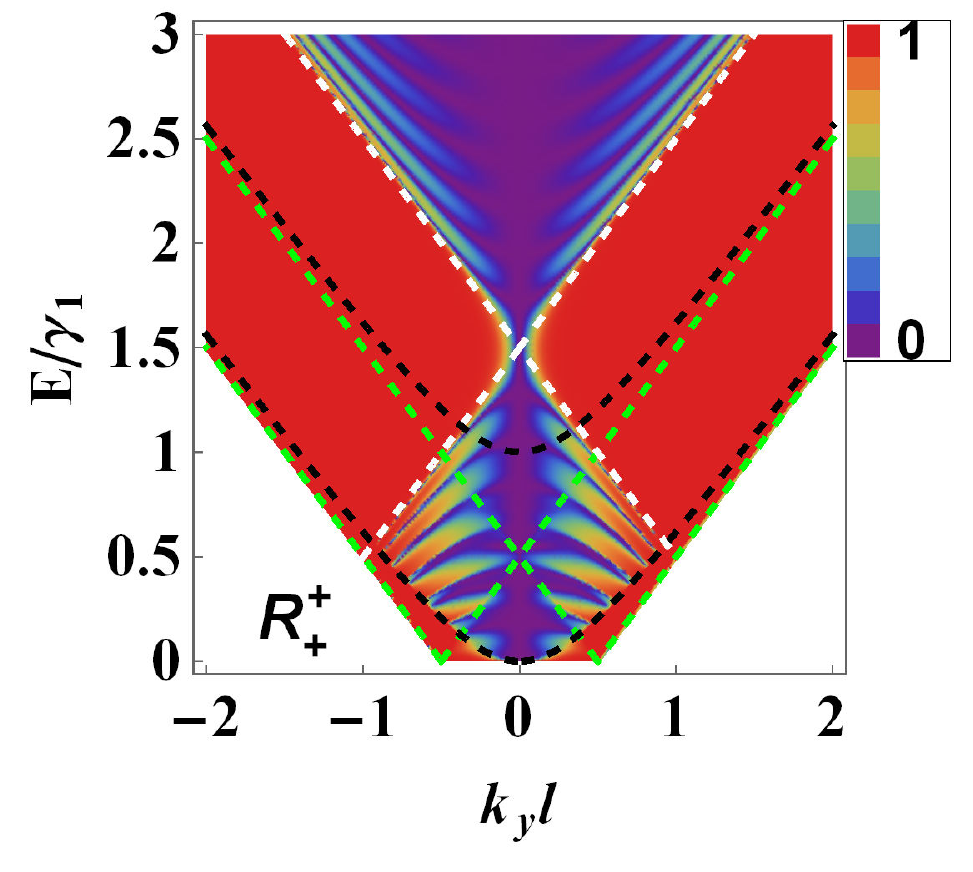}\\
\includegraphics[width=1.5 in]{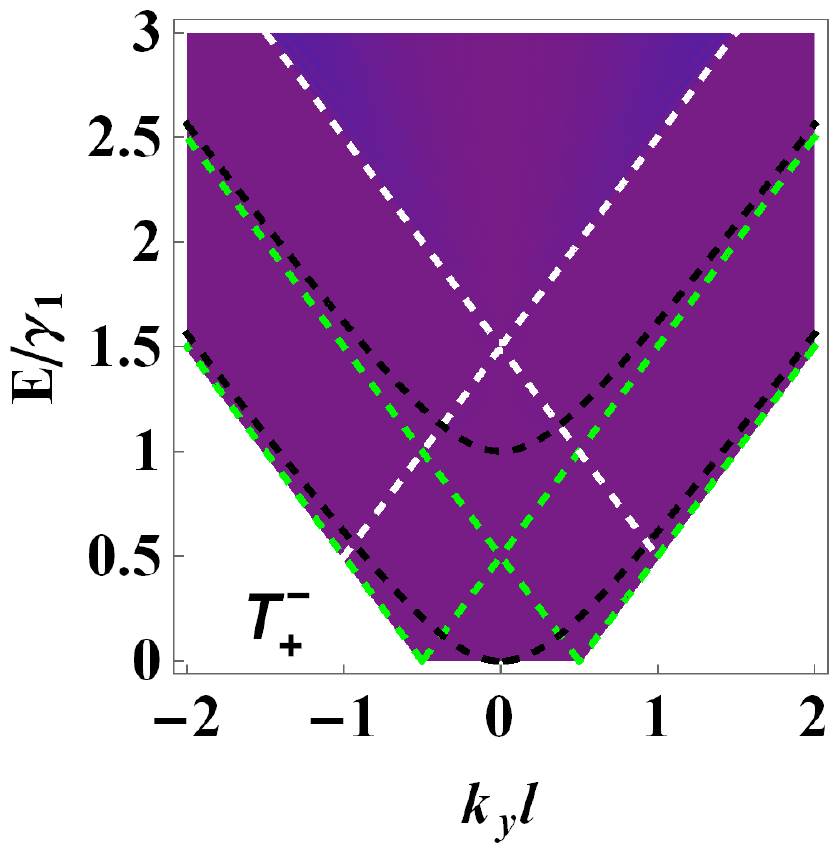}\ \
\includegraphics[width=1.73  in]{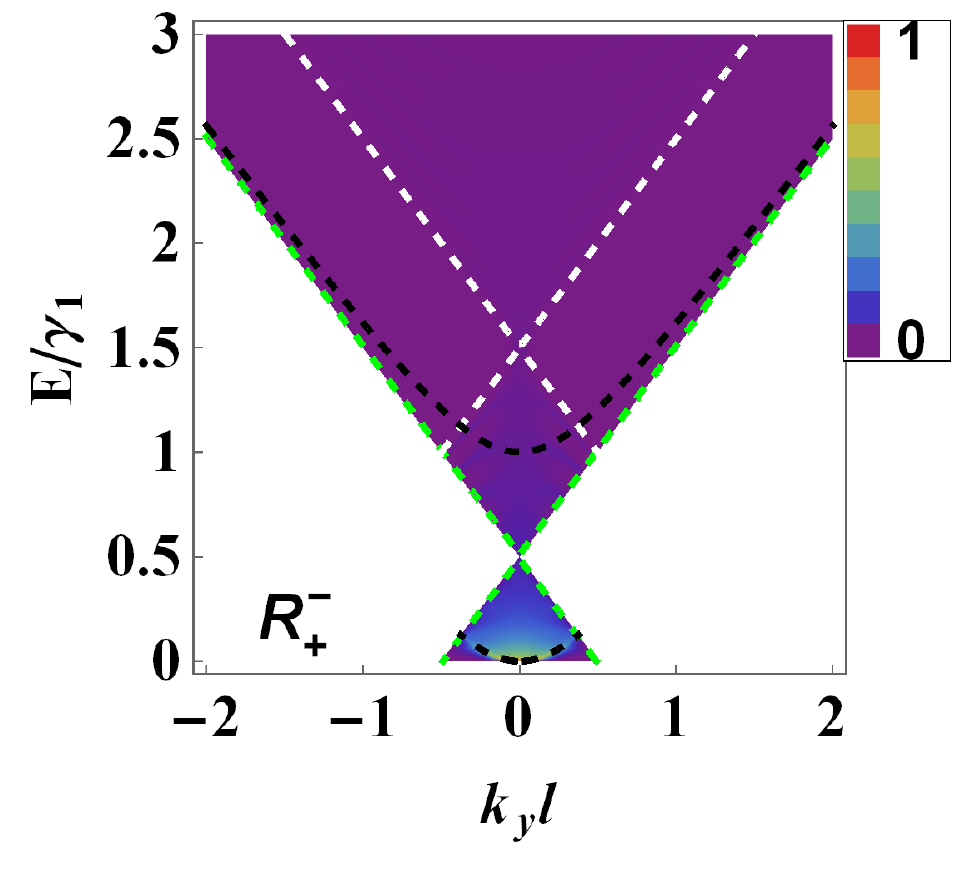}\\
\includegraphics[width=1.5  in]{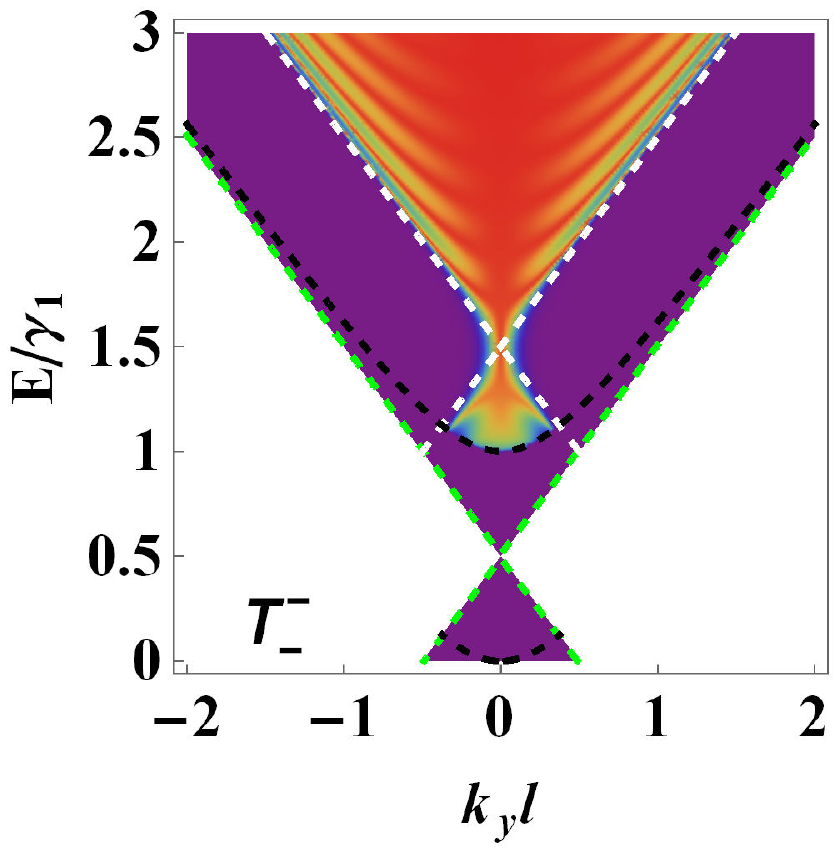}\ \
\includegraphics[width=1.73  in]{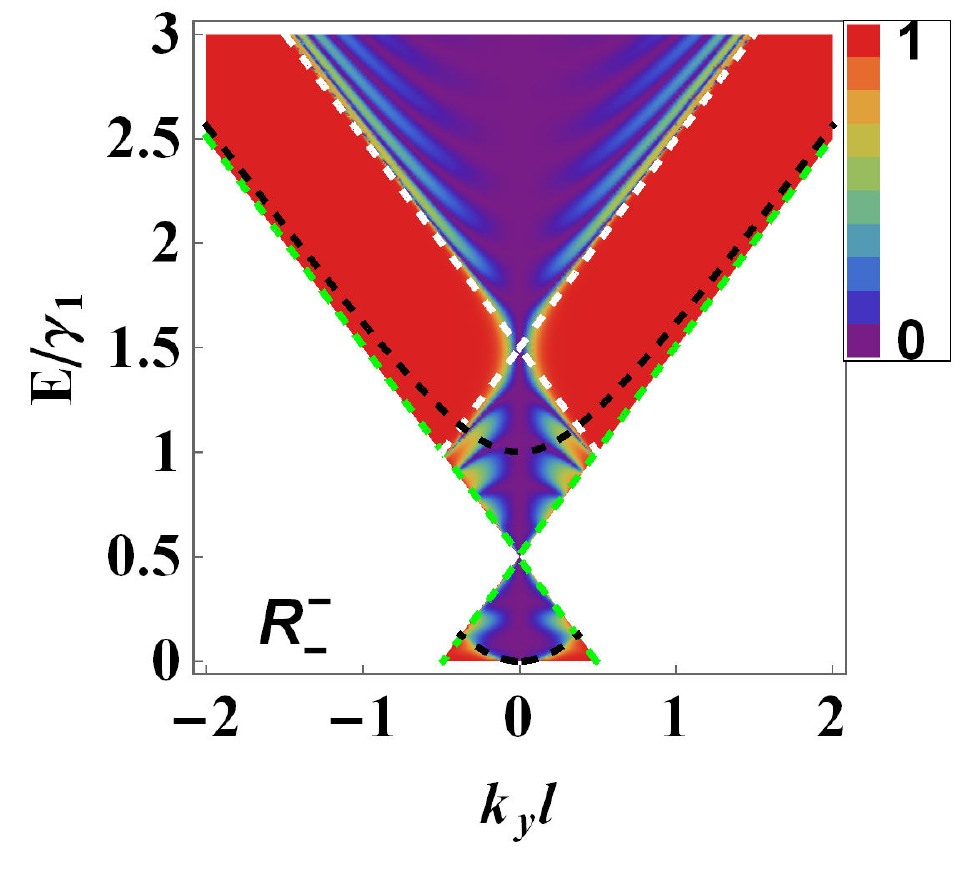}\\
\includegraphics[width=1.5  in]{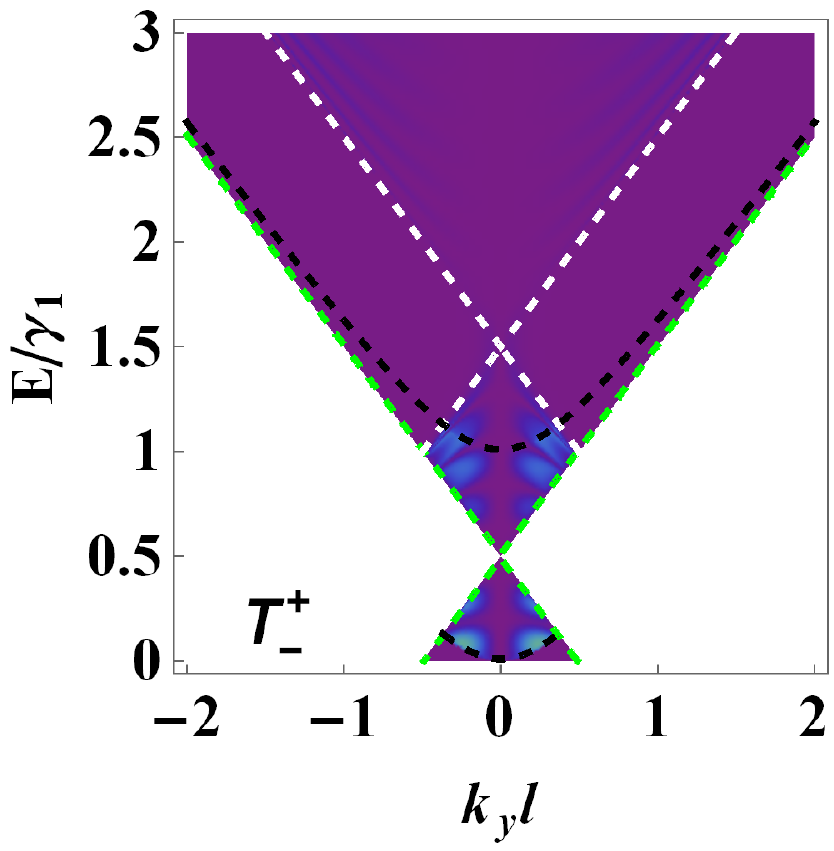}\ \
\includegraphics[width=1.73  in]{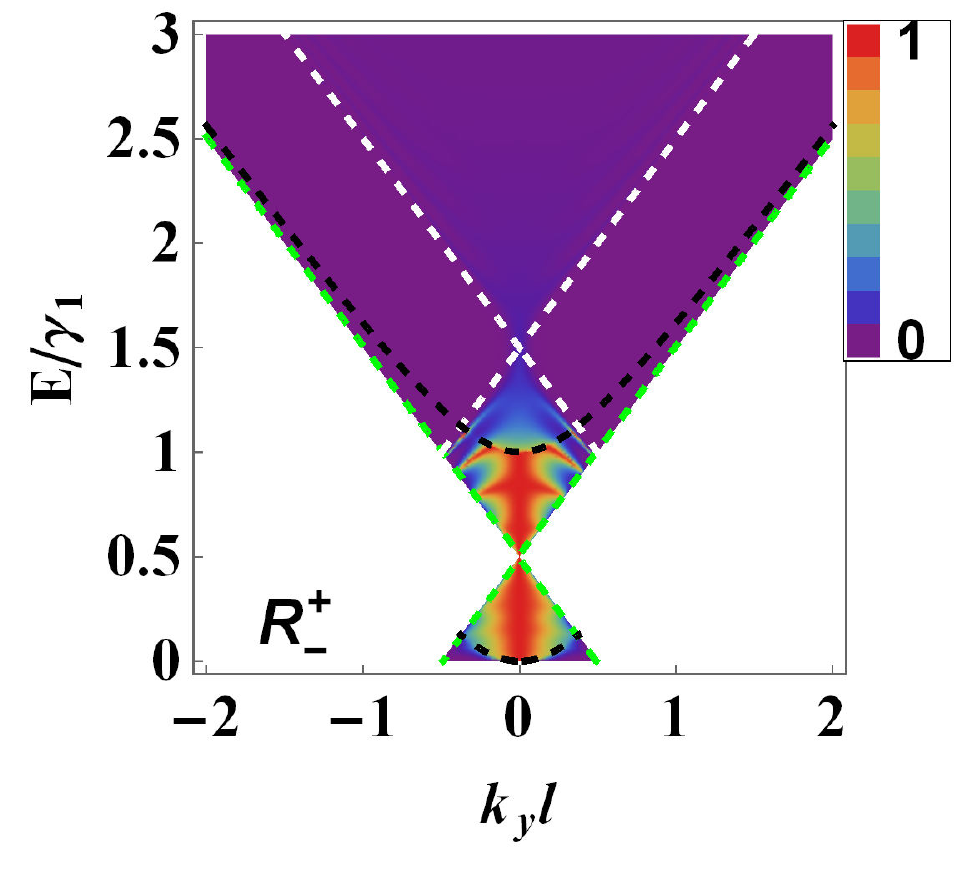}
\caption{(Colour online)Density plot of the  transmission and reflection probabilities through AA-2SL-AB junction as a function of Fermi energy and
transverse wave vector $k_y$ with  $v_0=1.5\gamma_1,\ \delta =0$ and $d=25$nm. The superimposed dashed curves represent the bands of AB-BL(black), AA-BL(green) and 2SL (white), with $\gamma_1$ being the inter-layer coupling of AB-BL.   }\label{T-AA-SL-AB}
\end{figure}
\begin{figure}[tb]
\vspace{0.4cm}
\centering \graphicspath{{./Figures1/AA-2SL-AB/}}
\includegraphics[width=1.65  in]{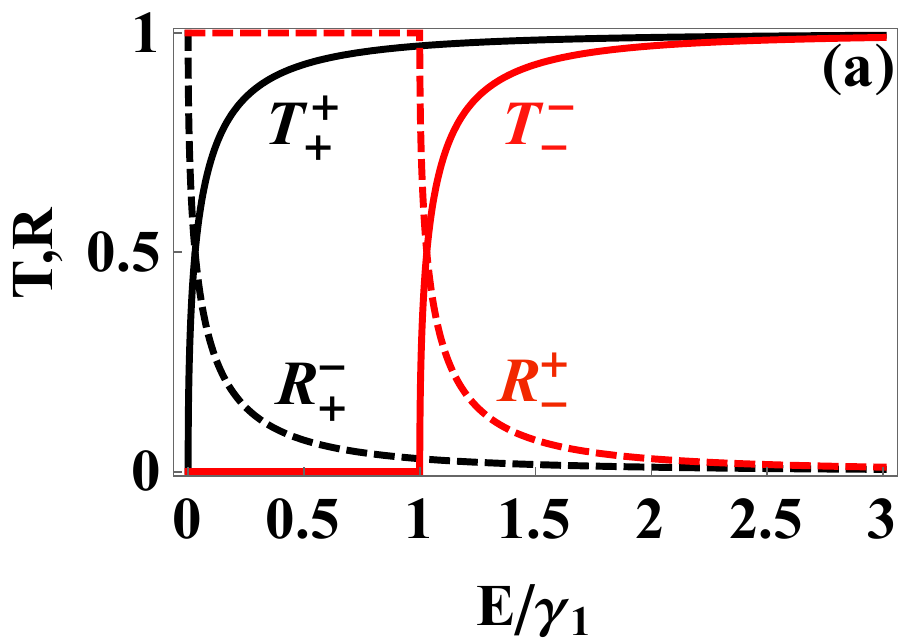}\ \
\includegraphics[width=1.65  in]{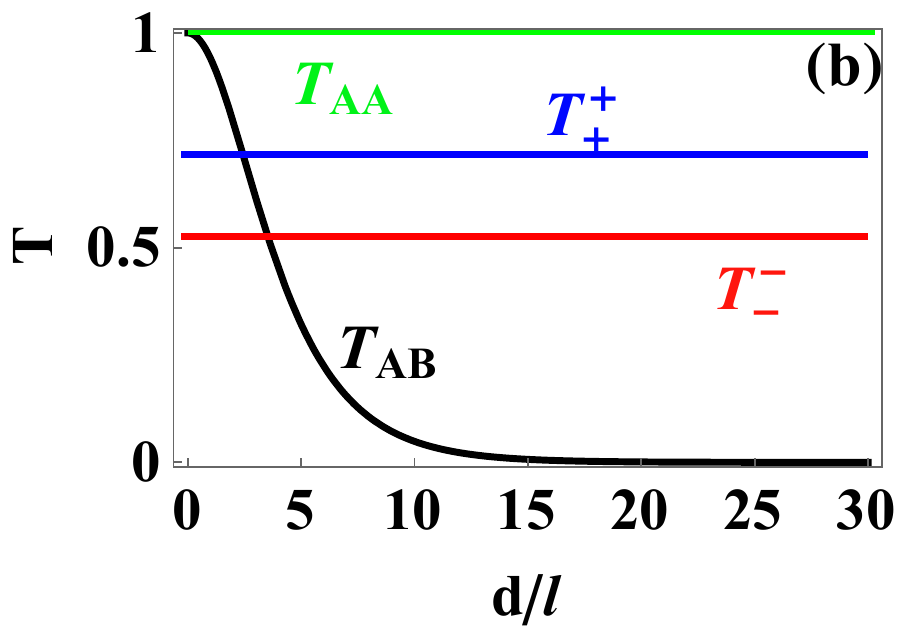}\\
\includegraphics[width=1.65  in]{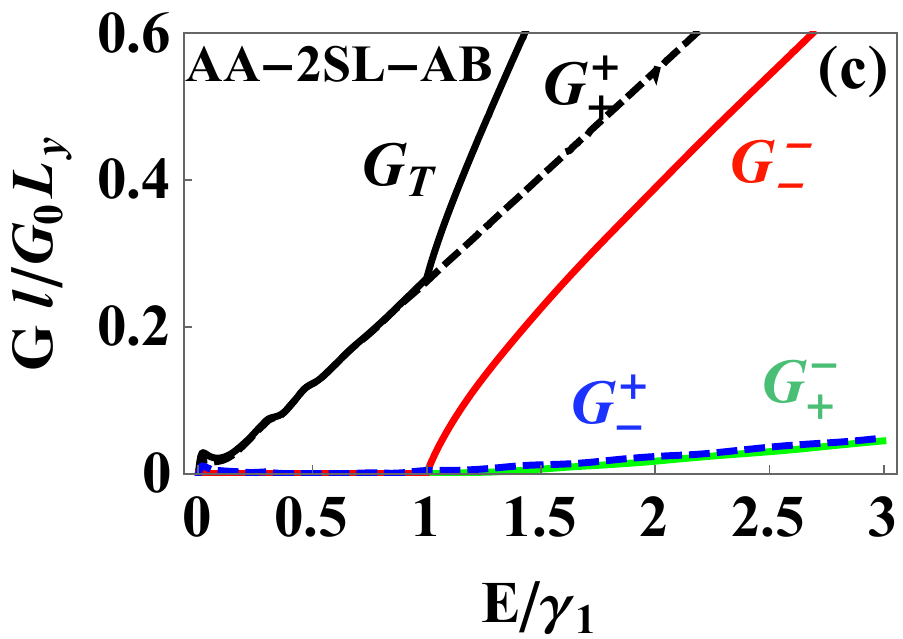}\ \
\includegraphics[width=1.65 in]{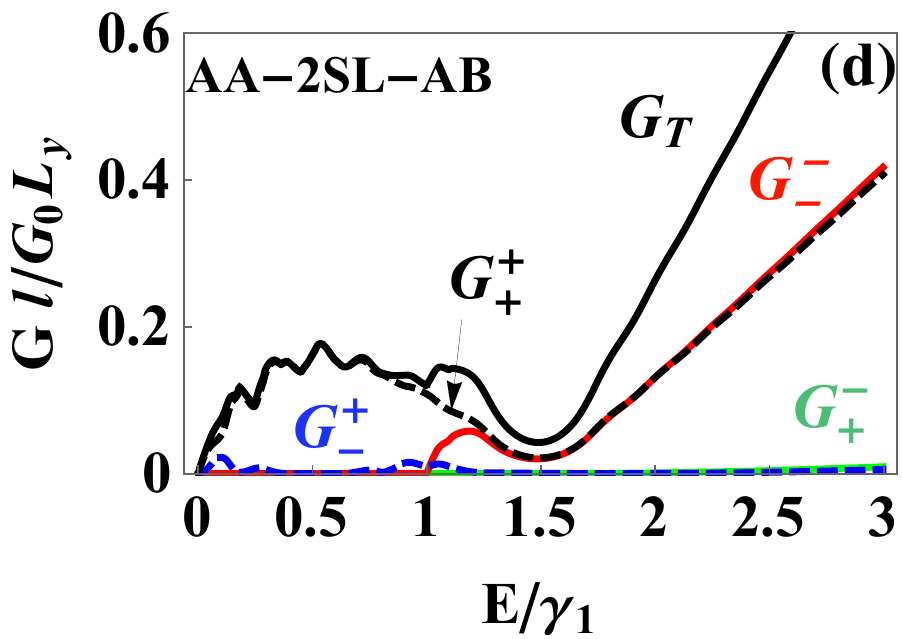}\\
\includegraphics[width=1.65 in]{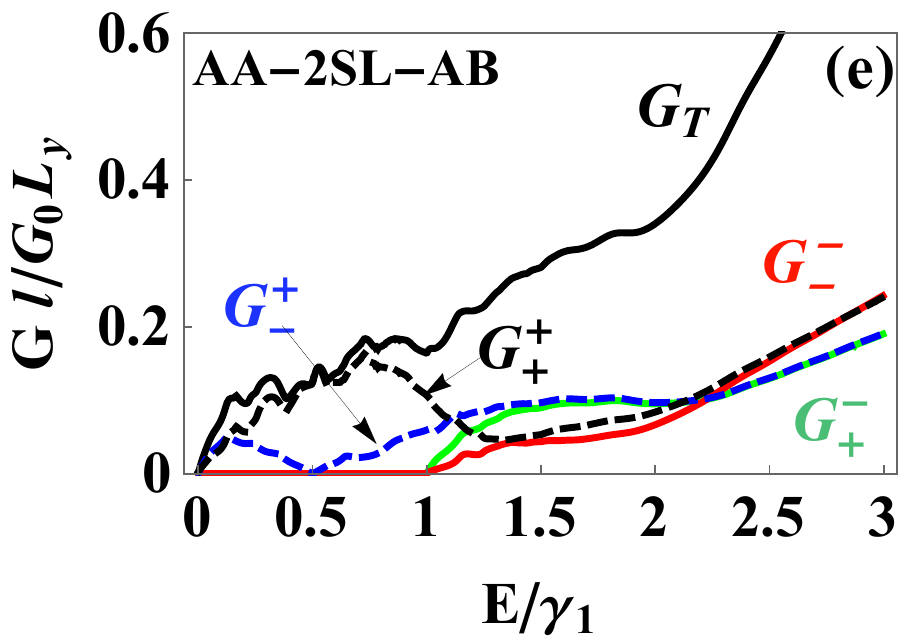}
\caption{(Colour online) (a) Transmission and reflection probabilities for normal incidence for  $v_0=3\gamma_1/2,\ \delta=0$. (b) Transmission probabilities  with normal incidence for AA-BL (AB-BL) n-p-n junction, green (black) curves. Blue  (red) curves are  the non-zero channels $T_+^+$ ($T_-^-$)  in AA-2SL-AB. All   energies are considered to be less than the electrostatic potential strength. Conductance of AA-2SL-AB junction for different magnitudes of the applied gate: (c) $v_0= \delta =0$, (d) $v_0=3\gamma_1/2$, $\delta =0$,  (e) $v_0=3\gamma_1/2$, $\delta =0.6 \gamma_1$, with $\gamma_1$ being the inter-layer coupling of AB-BL.  }\label{cond-AA-SL-AB}
\end{figure}

On the other hand, the conductance of the configuration 2SL-AB-2SL
has different features. In Fig. \ref{AB-SL-AB-G}(b)  the four channels, in contrast to the previous configuration,  start conducting from $E=0$. This possess layer symmetry such that $G_t^t=G_b^b$ and $G_b^t=G_t^b$.
Of particular importance is the equivalence of the four channels for $E<\gamma_1$ while for $E>\gamma_1$ charge carriers  strongly scatter between the layers ($i.e.\ G_i^j>G_i^i$) as shown in Fig. \ref{AB-SL-AB-G}(b). This equivalence of the four channels in the regime $E<\gamma_1$ vanishes when an  electrostatic potential is applied ($v_0>0$) to the intermediate region as seen in Fig. \ref{AB-SL-AB-G}(d). However,  the scattered and non-scattered conducting channels are still equivalent in this case where $G_t^{t(b)}=G_b^{b(t)}$ with $G_i^j>G_i^i$ for all energy ranges, see  Fig. \ref{AB-SL-AB-G}(d).

As discussed before, the most characteristic feature of the inter-layer bias in the AB-BL is  the opening of a gap in the energy spectrum  between $v_0\pm\delta$ which is reflected in the  conductance as seen in Fig. \ref{AB-SL-AB-G}(f).  The resonant sharp peaks in the conductance near the edges of the gap result from the localized states inside the Mexican hat of the low energy bands.  Another consequence of the inter-layer bias is the breaking of the equivalence in the non-scattered conducting channels where now $G_t^{t}\neq G_b^{b}$ as seen in Fig. \ref{AB-SL-AB-G}(f).
 %%%%%%%%%%%%%%%%%%%%%%%%%%%%%%%%%%%%%%%%%%%%%%%%%%%%%%%%%%%%%%%
\subsection{AA-2SL-AB}\label{Concl}
%%%%%%%%%%%%%%%%%%%%%%%%%%%%%%%%%%%%%%%%%%%%%%%%%%%%%%%%%%%%%%%

Here we consider the case where the leads  consist of BL with different stackings  separated by two uncoupled  graphene sheets. Such a structure can be formed if in the decoupled region one of the graphene sheets has larger lattice constant, e.g. due to strain, leading to     an inter-layer shift when the two layers couple.

Notice that the inter-layer coupling strength $\gamma_{1}$ differs for the two bilayer structures. Their ratio is $\gamma_1^{AA}/\gamma_1^{AB}\approx1/2$ \cite{li2009band,AA-gamma1,AA-Yuehua2010}. To account for this difference  the energy is normalized to $\gamma_1^{AB}$ such that the upper Dirac cone of pristine AA-BL is now located at $E=1/2$  instead of  $E=1$ as in the previous sections.  In the junction AA-2SL-AB the  charge carriers incident from  AA-BL and transmitted through 2SL into AB-BL. The results for the transmission and reflection  probabilities of this junction are shown in Fig. \ref{T-AA-SL-AB} for $v_0=1.5 \gamma_1,\ \delta=0$ and $ d=25$ nm. The  carriers  incident from lower($k^+$)/upper($k^-$) Dirac cones in AA-BL can be transmitted into   one of the modes ($k^+$ or $k^-$) in the AB-BL, see Fig. \ref{intro-fig02}(e). On the other hand, the reflection process occurs between the intra- or inter-cone in the AA-BL.

 Remarkably, Fig. \ref{T-AA-SL-AB} shows that the scattered transmission probabilities are very small and that almost all transmission is carried by the non-scattered channels. This is not immediately expected since a priori the $k^{+}$-mode in AA-BL is not related to the $k^{+}$-mode in AB-BL. However, both modes have the same parity under in-plane inversion, showing that this feature is robust against variations in the inter-layer coupling.

In contrast to the AA-2SL-AA junction where the scattering between lower and upper cones is forbidden in case of zero bias, here the two cones are coupled even without  bias. This results in non-zero reflection in the  scattered  channels $R_+^-$ and $R_-^+$.  

For normal incidence, the scattered transmission ($T_+^-$ and $T_-^+$) and the non-scattered reflection ($R_+^+$ and $R_-^-$) channels are zero (see  Fig. \ref{T-AA-SL-AB})  because in that case both the AA and AB Hamiltonian are block diagonal in the even and odd modes basis. Now, we can investigate Klein tunnelling when transitioning in-between the two types of stacking. For this, we show the non-zero channels of transmission and reflection for normal incidence in Fig. \ref{cond-AA-SL-AB}(a). We find that in contrast with the AA-2SL-AA case, perfect Klein tunnelling does not occur in the junction AA-2SL-AB. However, as shown in Fig. \ref{cond-AA-SL-AB}(b), we do find that the transmission probability does not depend on the length or even presence of the 2SL region, in contrast to the previous cases with two domain walls. 

For  $\delta\neq0$ the coupling between the different modes is strengthened and, hence,  strong scattering in the transmission and reflection channels occurs. Furthermore, the symmetry with respect to normal incidence in the reflection and transmission channels is broken. 
  
The conductance for the discussed structure is shown in Figs. \ref{cond-AA-SL-AB}(c, d, e) for $(v_0=\delta=0),\ (v_0=1.5\gamma_1,\delta=0)$ and ($v_0=1.5\gamma_1, \delta=0.6\gamma_1$), respectively. For pristine 2SL, the dominant channels are $G_+^+$ and $G_-^-$ . Notice that the latter one starts conducting only when $E>\gamma_1$ and this  shows up as a rapid increase in the total conductance $G_T$ at $E=\gamma_1$. The scattered channels $G_+^-$ and $G_-^+$ are only weakly contributing to the total conductance as a result of weak coupling of the modes.  In contrast to the junctions AA(AB)-2SL-AA(AB), in this case the scattered  channels of   the conductance are not equivalent $G_+^-\neq G_-^+$, see Fig. \ref{cond-AA-SL-AB}(c,d).  This is because  the scattering occurs between modes in bilayer graphene of different stackings.      \ The electrostatic potential introduces a minimum at $E=v_0$ in the total conductance due to the absence of propagating states at this energy in the 2SL, see Fig. \ref{cond-AA-SL-AB}(d). Biasing the intermediate region (2SL) of the junction AA-2SL-AB provides propagating states at $E=v_0$, and hence  removing the minima in $G_T$ as shown in Fig. \ref{cond-AA-SL-AB}(e). In addition, the contribution of the scattered channels  $G_+^-$ and $ G_-^+$ becomes more pronounced as  a result of the strong coupling between the modes induced by the bias. 

 Finally, notice that the counterpart junction AB-2SL-AA, represents the time-reversal case of the system discussed above. We have verified that the transmission channels are equivalent in the absence of a bias. In the presence of a bias, the angular symmetry is broken and, consequently, the reversed junction features the opposite angular asymmetry, preserving time-reversal invariance.

%%%%%%%%%%%%%%%%%%%%%%%%%%%%%%%%%%%%%%%%%%%%%%%%%%%%%%%%%%%%%%%
\section{summary and conclusion}\label{Concl}
%%%%%%%%%%%%%%%%%%%%%%%%%%%%%%%%%%%%%%%%%%%%%%%%%%%%%%%%%%%%%%%
Using the four-band model we  obtained the conductance, transmission  and reflection probabilities through single and double domain walls separating  two single layers and AA/AB-stacked bilayer graphene. We discussed in detail the scattering mechanism from detached layers to bilayer graphene and presented compact analytical formulae for the transmission probabilities. These results showed that  one can find the inter-layer coupling strength solely through measuring the conductance.

We found that an electrostatic potential applied to AB-BL, in an 2SL-AB junction, breaks the layer symmetry in the single-valley transmission probability channels. Such asymmetry  originates from the asymmetric coupling in AB-BL and arises as a consequence of the mismatch in energy between the  2SL and AB-BL Dirac cones caused by the electrostatic potentials applied to the AB-BL region. Layer asymmetry  exists when only one propagating mode  is present and hence is  not seen in configurations consisting of AA-BL where the entire energy range is associated with two transport channels.

We have also evaluated the robustness of chirality-induced properties, such as Klein tunnelling and anti-Klein tunnelling, to scattering on domains without inter-layer coupling. We found that in domain walls separating 2SL and AA-BL, Klein tunnelling is still preserved. On the other hand, for domain walls separating 2SL and AB-BL, the well known anti-Klein tunnelling in AB-BL is not preserved any more, but neither is Klein tunnelling itself. Moreover, in two domain walls separating three regions whose interlayer coupling is all different, i.e. the AA-2SL-AB case, we find that although perfect Klein tunnelling does not hold,  the tunnelling does not depend on the thickness of the 2SL region either. This remarkable effect is attributed to a conservation of parity of the modes.

Furthermore, we have found that a strong gate potential difference allows some states to be localized inside the Mexican hat of the low energy bands in the AB-BL. Those states contribute to the conductance  and appear as  sharp peaks  at the two edges of the gap. We showed that   scattering between these modes, in the transmission channels, is not allowed in the configuration (AA/AB)-2SL-(AA/AB). However, such scattering can be induced by applying an inter-layer bias on the 2SL which  in addition to shifting the bands of the top and bottom layers of 2SL,  also couples the modes. In contrast, we showed that the two modes  of AA-BL are coupled even without biasing the system in the junction AA-2SL-AB and revealed that  the latter junction is equivalent to the AB-2SL-AA. 

In order to limit the number of parameters, through this article we only considered abrupt domain walls, however, the results are robust against smoothness of the domain walls.\cite{Hasan1}

Our study reveals that the presence of the local domain  wall in bilayer graphene samples change the transport properties significantly. Our results  may shed  light on the design of electronic devices based on bilayer graphene.   Finally, we  showed that for a given sample with unknown sizes of local stacking domains, the average inter-layer coupling can be estimated through quantum transport measurements. 

%\vspace*{-0.3 cm}
%%%%%%%%%%%%%%%%%%%%%%%%%%%%%%%%%%%%%%%%%%%%%%%%%%%%%%%%%%%%%%%%
\section*{Acknowledgments}
%%%%%%%%%%%%%%%%%%%%%%%%%%%%%%%%%%%%%%%%%%%%%%%%%%%%%%%%%%%%%%%%
HMA and HB acknowledge  the Saudi
Center for Theoretical Physics (SCTP) for  their generous support and the support of KFUPM under physics research group projects  RG1502-1 and RG1502-2. This work is supported by the Flemish Science Foundation (FWO-Vl) by a post-doctoral fellowship   (BVD).
%\vspace*{-0.3 cm}
\appendix
%%%%%%%%%%%%%%%%%%%%%%%%%%%%%%%%%%%%%%%%%%%%%%%%%%%%%%%%%%%%%%%%
\section{Functions definitions}\label{Sec:Appendix}
%%%%%%%%%%%%%%%%%%%%%%%%%%%%%%%%%%%%%%%%%%%%%%%%%%%%%%%%%%%%%%%%
The transmission probabilities are  calculated by applying appropriate boundary conditions at the 2SL-BL interfaces together with the transfer matrix. After some cumbersome algebra,  we obtain  for  2SL-AB 
\begin{eqnarray}
T_j^\pm=4 \textrm{Re}(k^{\pm}) \frac{\eta\left[ \eta^2+\left(\textrm{Im}(k^{\mp})+\kappa_{j}\ v_0\ \sin\phi\right)^2\ \right]}{C_{0 }+\sum_{m=1}^4C_m\ \cos(m\phi)},
\end{eqnarray}
where\\ 
$C_0=2\left( \textrm{Im}(k^\mp)\textrm{Re}(k^\pm) \right)^2+\epsilon^2\left( \textrm{Im}^{2}(k^\mp)+\textrm{Re}^{2}(k^\pm) \right)
+\Gamma_{1}$,\\ \\ $\Gamma_{1}=2v_{0}^{4}-4v_0^3E+5v_0^{2}E^{2}-3v_0E^{3}+\frac{3}{4}E^4$,\\ \\$C_1=-\epsilon \textrm{Re}(k^\pm)\left[ 4\left( v_0^2+\textrm{Im}^2(k^\mp) \right)-6v_0E+3E^2 \right],$\\ \\$C_2=\epsilon^2\left( \textrm{Im}^{2}(k^\mp)+\textrm{Re}^{2}(k^\pm) \right)+\Gamma _{2}$,\\ \\$\Gamma _{2}=E\left( -4v_0^3+6v_0^2E-4v_{0}E^2+E^3\right)$,\\ \\$C_3=\textrm{Re}(k^\pm)E\left( 2v_0^2-3v_0E+E^2 \right)$,\\ \\$C_4=\frac{1}{4}E^2(E-2v_0)^2$.\\

Similarly, the transmission probabilities for the AB-2SL system are obtained as
\begin{eqnarray}
T_\pm^j=4 \textrm{Re}(k_j)k^{\pm} \frac{\lambda\left[  \mu^{\pm}+\kappa_{j} v_0 \sin\phi\ \textrm{Im}(k^{\mp}) \right] }{\left\vert Q^{\pm} \right\vert^2},
\end{eqnarray}
\begin{eqnarray*}
\mu^{\pm}=\frac{\epsilon\left( \textrm{Im}^{2}(k^{^{\mp}})+E^{2} \right)-
E(\pm1+E)(E+v_0)\sin^{2}\phi }{2\sqrt{E(\pm1+E)}},\\
\end{eqnarray*}
$ \lambda=E\sqrt{E(\pm1+E)},$\\
$Q^{\pm}=\frac{1}{2}[z_{0}-z_{1}\left( k^{\pm}+i\textrm{Im}(k^\mp) \right)+z_{2}k^{\pm}\textrm{Im}(k^\mp)],$\\ with\\ 
$z_0=2i\left[ v_0\alpha\ -ik_jE \right]\left[ \alpha \left( -ik_{j} +\alpha\right)+\epsilon  E\right]$,\\ \\
$z_1=E\left[  \left( ik_{j} +\alpha\right)^{2}-\epsilon^{2}\right]$,\\ and finally\\ 
$z_2=2\epsilon  \left[ik_{j} +\alpha\right]$,\\ where $\alpha=\sqrt{E^2\pm
E}\sin\phi.$\\ \\ \\
%\vspace*{-0.2 cm}
%%%%%%%%%%%%%%%%%%%%%%%%%%%%%%%%%%%%%%%%%%%%%%%%%%%%%%%%%%%%%%%%%%%%%%%%%%%%%%%%%%
%%%%%%%%%%%%%%%%%%%%%%%%%%%%%%%%%%%%%%%%%%%%%%%%%%%%%%%%%%%%%%%%%%%%%%%%%%%%%%%%%%

\end{document}